\documentclass[12pt,letterpaper]{article}

\usepackage[T1]{fontenc}
\usepackage[margin=1in]{geometry}

\usepackage[titletoc]{appendix}
\usepackage{titling}
\usepackage[authoryear]{natbib}

\usepackage[dvipsnames]{xcolor}
\usepackage{float}
\usepackage[colorlinks=true,
            linkcolor=Blue,
            citecolor=Blue,
            urlcolor=blue,
            hyperfootnotes=false,
            pdftitle={A Locally Robust Semiparametric Approach to Examiner IV Designs},
            pdfauthor={Lonjezo Sithole}]{hyperref}

\usepackage[bottom]{footmisc}

\usepackage[onehalfspacing]{setspace}

\usepackage[final]{microtype}
\usepackage{lmodern}
\usepackage[subtle]{savetrees}
\usepackage[small,compact]{titlesec}
\titlespacing{\paragraph}{0pt}{2.25ex plus 1ex minus .2ex}{0.4em}

\usepackage{graphicx,csquotes,enumitem,booktabs,threeparttablex,array}
\usepackage{mathtools,amssymb,amsthm,bm,dsfont,etoolbox}
\allowdisplaybreaks

\theoremstyle{definition}

\newcommand{\continuation}{??}

\newtheorem*{example*}{Example}

\newtheorem*{remark*}{Remark}

\makeatletter
\newcommand{\exampleqedsymbol}{$\triangle$}
\AtBeginEnvironment{example}{%
  \pushQED{\qed}%
  \let\example@oldsym\qedsymbol
  \renewcommand{\qedsymbol}{\exampleqedsymbol}%
}
\AtEndEnvironment{example}{%
  \popQED%
  \let\qedsymbol\example@oldsym
}
\AtBeginEnvironment{example*}{%
  \pushQED{\qed}%
  \let\example@oldsym\qedsymbol
  \renewcommand{\qedsymbol}{\exampleqedsymbol}%
}
\AtEndEnvironment{example*}{%
  \popQED%
  \let\qedsymbol\example@oldsym
}
\newcommand{\remarkqedsymbol}{$\triangle$}
\AtBeginEnvironment{remark}{%
  \pushQED{\qed}%
  \let\remark@oldsym\qedsymbol
  \renewcommand{\qedsymbol}{\remarkqedsymbol}%
}
\AtEndEnvironment{remark}{%
  \popQED%
  \let\qedsymbol\remark@oldsym
}
\AtBeginEnvironment{remark*}{%
  \pushQED{\qed}%
  \let\remark@oldsym\qedsymbol
  \renewcommand{\qedsymbol}{\remarkqedsymbol}%
}
\AtEndEnvironment{remark*}{%
  \popQED%
  \let\qedsymbol\remark@oldsym
}
\makeatother

\theoremstyle{plain}
\newtheorem{theorem}{Theorem}[section]
\newtheorem{assumption}{Assumption}[section]
\newtheorem{assumptionalternative}{Assumption}[section]

\newtheorem{proposition}{Proposition}[section]
\newtheorem{corollary}{Corollary}[section]
\newtheorem{lemma}{Lemma}[section]
\numberwithin{equation}{section}

\title{A Locally Robust Semiparametric Approach to Examiner IV Designs}
\author{Lonjezo Sithole\thanks{Department of Economics, University of Michigan, lsithole@umich.edu. I thank my dissertation committee (led by Florian Gunsilius and Kaspar Wuthrich), Mel Stephens and David Van Dijcke for helpful comments and suggestions. I also gratefully acknowledge financial support from the Rackham Graduate School, University of Michigan via the Rackham Predoctoral Fellowship. All errors in this work are mine.}}
\date{15 September, 2026}

\begin{document}

\maketitle
\vspace{-0.85cm}

\begin{abstract}
I propose a locally robust semiparametric framework for estimating causal effects using examiner IV designs when adjustment for many or continuous covariates makes saturation infeasible or produces sparse cells. The key ingredient of this approach is an orthogonal moment function that removes the first-order effect of estimation errors in the two treatment regressions defining the examiner IV. I derive the orthogonal moment function and show that it remains valid under misspecification when, for each treatment regression, either that regression or the corresponding Riesz representer in the influence function adjustment is correctly specified. The proposed framework not only allows estimation of the examiner IV using a wide range of nonparametric and machine learning techniques, including LASSO, neural networks and random forests, but also delivers root-\(n\) consistent estimation and valid inference under suitable regularity conditions. I examine the finite-sample performance of the estimator through Monte Carlo simulations. I also apply the method to US patent examiners, using regularized regressions to account for differences in technology and other application characteristics.
\end{abstract}
\vspace{0.05in}

\begin{small}
\noindent \textit{JEL Numbers:} C14, C21, C26, C31, C45\\[0.25ex]
\noindent \textit{Keywords:} examiner IV, semiparametric estimation, orthogonal moment function, generated instruments, leniency designs, machine learning.
\end{small}
\clearpage

\section{Introduction \label{sec:Introduction}}

A growing number of empirical studies in economics and other social sciences exploit quasi-random assignment of decision makers to construct instruments for endogenous
treatment or policy variables. For example, differences in treatment rates across judges, examiners,
caseworkers, and other decisionmakers have been used to estimate the effects of
incarceration, foster care placement, bankruptcy protection, disability
benefits, and other interventions on an array of socioeconomic outcomes
\citep{10.1257/aer.96.3.863,mueller2015criminal,gross2022temporary,bald2022economics,dobbie2015debt,black2018effect}.
The key feature of these empirical designs is that an institution assigns cases to decision makers who differ in
their propensity to treat.
In most of these settings, however, assignment of decision makers (hereinafter, examiners) is typically random only after conditioning on observed
characteristics. Judges, for example, may choose courts, shifts, or working days, and patent
applications may be routed within technology classes and filing cohorts.

Let $X$ collect the characteristics that govern assignment, let $T$ denote a binary
treatment, and let $Z$ identify the assigned examiner. The examiner instrument is
\begin{equation*}
\gamma_0(X,Z)
=
\mathbb{E}[T\mid X,Z]-\mathbb{E}[T\mid X].
\end{equation*}
The first term is the treatment rate among cases with characteristics $X$
assigned to examiner $Z$. The second averages this rate over the examiners who
receive cases with those characteristics. Therefore, the examiner instrument is a measure of \emph{relative leniency} of examiner, $Z$, within a given covariate cell. Notice that
$\mathbb{E}[\gamma_0(X,Z)\mid X]=0$. I refer to this zero conditional expectation property as
\emph{conditional centering} and call $\gamma_0$ the \emph{oracle instrument}.
The IV estimand studied below is
\begin{equation*}
\theta_0
=
\frac{\mathbb{E}[Y\gamma_0(X,Z)]}
{\mathbb{E}[T\gamma_0(X,Z)]}.
\end{equation*}

\citet{kolesarcowles} develops the unbiased jackknife IV estimator (UJIVE),
which replaces the conditional expectations in $\gamma_0$ with leave-one-out
estimates of two linear projections. The population instrument is
\[
\gamma^L(X,Z)
=
L[T\mid X,Z]-L[T\mid X],
\]
where $L[\cdot\mid\cdot]$ denotes population least squares. I call the projection
on $(X,Z)$ the long regression and the projection on $X$ the short regression.
At the population level,
\begin{align*}
\gamma^L(X,Z)-\gamma_0(X,Z)
={}&
\{L[T\mid X,Z]-\mathbb{E}[T\mid X,Z]\}\\
&-
\{L[T\mid X]-\mathbb{E}[T\mid X]\}.
\end{align*}
The two instruments agree when the long and short projections have the same
error. Correct specification of both regressions is sufficient for this cancelation but it is not necessary as the
errors can also cancel when neither regression is correct.

When all covariates are discrete, saturated long and short regressions recover
the two conditional expectations.\footnote{The ``saturate and weight'' approach of
\citet[Section~4.5.1]{angristpischke2009} includes indicators for all covariate
cells and their interactions with the excluded instruments. In an examiner IV
design, the long regression saturates the cells defined by examiner and $X$,
and the short regression saturates the cells defined by $X$.} With continuous covariates, saturation is not available though a finite-dimensional
parametric model may still be correct. In general, however, parametric models raise the spectre of misspecification.  More crucially, in this setting, approximating the conditional expectations by linear projections does not in general
preserve conditional centering. A
nonsaturated linear specification satisfies
\[
\mathbb{E}[p(X)\gamma^L(X,Z)]=0
\]
for the controls $p(X)$ included in the regression. Conditional centering
requires orthogonality to \emph{every} square-integrable function of $X$.
\citet{blandhol2025tsls} call this stronger requirement \emph{rich covariates}
and show why it matters for the causal interpretation of linear IV estimands.

Conditional centering has a direct causal implication: it removes the term
involving the untreated potential outcome from the numerator. Under conditional assignment,
\[
\mathbb{E}[Y(0)\gamma_0(X,Z)]
=
\mathbb{E}\!\left[
\mathbb{E}[Y(0)\mid X]\mathbb{E}[\gamma_0(X,Z)\mid X]
\right]
=0.
\]
In this paper, I combine this cancellation with the average monotonicity condition of
\citet{frandsen2023judging}. I show that, under that condition, $\theta_0$ is a weighted
average of individual treatment effects with nonnegative weights. Under the stronger pairwise
monotonicity condition, the ratio is a weighted average of LATEs for pairs of
examiners, with weights proportional to squared differences in their treatment
propensities. The oracle ratio therefore satisfies the two conditions in
\citet{blandhol2025tsls} for a weak causal interpretation: its numerator
contains no term involving $Y(0)$, and it assigns nonnegative weights to
treatment effects.

Because the two conditional expectations in $\gamma_0$ are unknown, their estimation introduces first-order bias in the IV moment. To correct for this bias, I provide one influence function adjustment for each conditional expectation. The augmented
score, formed from adding these influence function adjustments to the original identifying moment function for $\theta_0$, is Neyman orthogonal, and its explicit Riesz representers can be estimated
from four regressions: the conditional expectations of $T$ and $Y$ given
$(X,Z)$ and their conditional expectations given $X$. A direct calculation shows that
the remaining bias consists of products of the four regression errors. With
cross-fitting, the estimator has a root-$n$ expansion when these products are
$o_p(n^{-1/2})$. For i.i.d. observations from a fixed distribution, the
normalized score is the efficient influence function and the estimator attains
the semiparametric efficiency bound.

I also derive conditions for valid inference as the number of examiners and the
number of regressors used to approximate the conditional expectations increase
with sample size. Under the regularity conditions in Corollary~\ref{cor:series},
unregularized series regressions with at most $K_n$ regressors in each equation
satisfy the product condition when $K_n=o(\sqrt n)$ and their approximation error
is $o(n^{-1/4})$. Including unrestricted examiner indicators implies
$K_n\geq J_n$, so these sufficient conditions require $J_n=o(\sqrt n)$. I also calculate rates for partially penalized LASSO, Post-LASSO and
ridge regressions, in which examiner main effects are left unpenalized. For
clustered observations, I provide additional conditions for asymptotic normality
and consistent variance estimation that allow dependence among observations
within a cluster.

The Monte Carlo experiments separate sampling error, error from estimating the
regressions, approximation error, and the difference between the estimands
defined by conditional expectations and linear projections. In the
design with interactions between examiners and covariates, linear UJIVE
converges to 1.40, whereas the ratio based on conditional expectations is 1.48. This
difference remains as the sample grows.

In an application to US patent examiners, I compare two linear UJIVE estimates with a flexible
orthogonal estimate. The sample, treatment, and outcome are the same across the different specifications.
The treatment regressions have out-of-fold mean squared errors that differ by at
most 1.7 percent, but the IV estimates range from 0.042 to 0.129. The comparison
illustrates the distinction in Section~2: predictive accuracy alone does not
establish that the linear projections adequately approximate the conditional
expectations defining the instrument.

\subsection*{Related Literature}

When instrument validity holds only after conditioning on covariates, the
covariate specification can determine the IV estimand.
\citet{sloczynski2026practical} show how covariate specifications change the
weights on conditional LATEs and recommend flexible specifications when
saturation is infeasible. \citet{blandhol2025tsls} show that a weakly causal
linear IV estimand requires
$\mathbb{E}[\gamma^L(X,Z)\mid X]=0$. In examiner IV designs, the instrument
\[
\gamma_0(X,Z)
=
\mathbb{E}[T\mid X,Z]-\mathbb{E}[T\mid X],
\]
satisfies this condition by construction. Both conditional expectations are unknown. Using nonparametric or regularized
estimates of them in the IV moment requires accounting for estimation error
in each.

The closest antecedent to this work is \citet{kolesarcowles}, whose UJIVE uses leave-one-out
estimates of the long and short least squares projections to remove the bias
caused by using an observation to fit its own instrument. \citet{kolesarcowles}
allows growing linear dictionaries under approximation and regularity conditions
that make the average squared approximation errors in the two treatment
regressions vanish.
Equation~(\ref{eq:projection_error_gap}) implies the population condition:
the projection instrument equals $\gamma_0$ when the long and short projection
errors cancel. Under the conditions in \citet{kolesarcowles}, the difference between the
projection and oracle ratios vanishes, so UJIVE estimates $\theta_0$ even along
sequences with many instruments, many covariates, and heteroskedasticity. For an
estimator based on the projections to have the oracle first-order expansion in
Theorem~\ref{thm1}, the effect of the approximation on the IV moment must also
be $o(n^{-1/2})$. \citet{evdokimovkolesar2018} derive asymptotic distributions
and consistent variance estimators for linear jackknife IV estimators with
heterogeneous treatment effects and many instruments or covariates. They
assume a linear representation of the reduced-form conditional expectations
in the chosen instrument and control dictionaries.

My estimation and inference results complement those of \citet{kolesarcowles}
and \citet{evdokimovkolesar2018} by allowing estimation of the conditional
expectations by regularized or machine learning methods when saturation is not
available or produces sparse cells. In this
case, naive plug-in estimation leaves first-order bias from each of the two
treatment regressions. I derive a correction by treating $\theta_0$ as a
nonparametric functional of the conditional expectations. The orthogonal score,
constructed by adding this correction to the original identifying moment function,
permits general estimators of the treatment and outcome regressions.

Although motivated by examiner designs, the estimation results in the sequel apply to the IV
estimand in (\ref{eq:estimand}) with binary treatment and a general instrument vector $Z$.
Estimation requires the same four regressions: the conditional expectations of
treatment and outcome given $(X,Z)$ and given $X$. Under the conditions in
Section~3, the estimator is asymptotically normal even under treatment effect heterogeneity.
In work subsequent to the previous versions of this paper (April 2024 and April 2026),
\citet{yu2026machine} extends the analysis to a correlated random coefficients
model allowing continuous treatments. He also considers IV estimands based on
general functions of the instruments and examines how first-stage estimation
error changes the weights on treatment effects. His orthogonal score is,
however, algebraically equivalent to mine, up to relabelling.\footnote{His
treatment $X$, covariates $W$, and parameter $\beta$ correspond to $T$, $X$,
and $\theta$ here. His $(g_X,m_X,q_Y,l_Y)$ are the four regressions
$(\gamma_1,\gamma_2,m_1,m_2)$ in Section~3.4. Suppressing their arguments,
equations~(15)--(17) of his 13 September 2026 version expand to
\[
\begin{aligned}
\psi_{\mathrm{CRC}}
={}&[Y-m_2-\theta(T-\gamma_2)](\gamma_1-\gamma_2)\\
&+(T-\gamma_1)[m_1-m_2-\theta(\gamma_1-\gamma_2)],
\end{aligned}
\]
which is (\ref{eq:residualized_score}), obtained by substituting the
representers in (\ref{eq:alpha_feasible_closed}) into my score. The equality
holds pointwise for every $\theta$ and every choice of the four regression
functions.}

Flexible prediction methods can serve different purposes in an IV analysis, from
constructing an instrument to estimating a nonparametric structural function.
For a nonparametric structural function, \citet{hartford2017deep} estimate the
conditional treatment distribution and use it in the second-stage loss, while
\citet{brunssmith2025twostage} learns an instrument basis from the reduced form.
For a linear IV coefficient, \citet{chen2021harmless} estimate an optimal
instrument by combining flexible prediction with sample splitting. They
restrict the fitted instrument to depend linearly on the covariates, thereby
excluding nonlinear functions of the controls as a source of identifying
variation.
I study the examiner IV ratio whose instrument is the difference between the
long and short conditional expectations of treatment. Because
$\mathbb{E}[T\mid X]$ removes the part of $\mathbb{E}[T\mid X,Z]$
explained by $X$ alone, nonlinear dependence on $X$ in either regression
does not supply identifying variation: $\gamma_0$ compares treatment
propensities across examiners conditional on $X$. I account for estimation
of both expectations through a separate influence function adjustment for
each.

The influence function adjustments for the examiner IV ratio build on the literature on inference with estimated nuisance
functions. In high-dimensional linear models, debiased estimators correct
regularization bias in low-dimensional coefficients
\citep{vandegeer2014,javanmard2014,zhangzhang2014}. For semiparametric
functionals, influence function calculations and locally robust estimation
provide corrections for estimated regression functions
\citep{chernozhukov2022locally,ichimura2022influence}. Double machine learning
combines orthogonal scores with cross-fitting \citep{chernozhukov2018double}.
Locally robust IV methods have also been developed for the complier
parameters of \citet{abadie2003semiparametric} with a binary instrument \citep{singhsun2024} and for treatment
effects that vary with observed characteristics
\citep{scheidegger2025inference}. Because $\gamma_0$ is the difference between
the two treatment regressions, I derive one influence function adjustment for each regression and obtain
both Riesz representers in closed form.

The formulas for the representers make the rate requirements for the four regressions explicit.
Each representer is a linear combination of a treatment regression and an
outcome regression.
Substituting the representers into the score yields the exact remainder in
Equation~(\ref{eq:coupled_remainder}), which contains products of outcome and
propensity errors as well as squared propensity errors. For i.i.d.
observations from a fixed distribution, the normalized score
is the efficient influence function. I also allow the number of examiners
and the covariate dimension to increase with the sample through the triangular
array analysis in Section~\ref{ssec:growing}.

The choice of how to estimate the conditional expectations also matters for
the causal interpretation of $\theta_0$. \citet{frandsen2023judging} show that average monotonicity
allows violations of pairwise monotonicity while preserving nonnegative weights
on individual treatment effects. \citet{blandhol2025tsls} show that a weakly
causal linear IV estimand cannot depend on potential outcome levels or place
negative weight on any subgroup. Because
$\mathbb{E}[\gamma_0\mid X]=0$, the term involving the baseline potential
outcome vanishes within each covariate cell. Average monotonicity makes the
remaining weights nonnegative. Under pairwise monotonicity, $\theta_0$ is a
convex combination of conditional pairwise LATEs with explicit weights. The
population difference between conditional expectations and linear projections
can therefore change both the IV ratio and its causal interpretation.

When the number of examiners increases, estimation of the first stage can
overfit, and some examiner categories may be sparse. The broader many-instrument
literature controls the sampling error by regularizing the instrument
covariance matrix or the first-stage fit
\citep{carrasco2012regularization,HANSEN2014290}, estimating optimal instruments
under approximate sparsity \citep{belloni2012sparse}, or replacing many observed
variables with estimated common factors \citep{bai2010instrumental}.
I derive conditions under which estimating all four conditional expectations
preserves valid inference as the number of examiners and the regression
dimensions increase. Corollary~\ref{cor:series} establishes sufficient conditions for
unregularized series regressions. These conditions require $K_n=o(\sqrt n)$
and hence $J_n=o(\sqrt n)$ when the basis contains
unrestricted examiner indicators.

Recent methods develop regularization or jackknife procedures for categorical
instruments, many judges, and clustered samples.
\citet{wiemann2023optimal} exploits the categorical structure of the instrument
while modeling the covariate adjustment as a low-dimensional linear term.
\citet{jochmans2023many} studies many weak judges in a model without observed
covariates, while \citet{chao2023} develop FEJIV, FEFUL, and FELIM for linear
IV models with cluster effects and possibly many included exogenous regressors.
Although they treat covariates differently, all three papers develop inference
within \emph{linear} IV models. The estimator in Section~3 permits nonlinear
covariate adjustment in all four regressions and interactions
between examiner assignment and the covariates in the two long regressions.

The rest of the paper is organized as follows. Section 2 introduces the
examiner IV \mbox{problem} and compares the two instruments.
Section 3 derives the causal interpretation, constructs the estimator, and
establishes its asymptotic distribution. Section 4 presents the Monte Carlo
experiments, Section 5 presents the application to US patent examiners, and Section 6
concludes.

\section{Examiner IV Designs and the Estimation Problem}

To fix ideas, consider the effect of pretrial detention on conviction in a setting where
defendants are assigned to judges \citep{frandsen2023judging}. Let $T_i=1$
indicate detention, and let $Y_i(1)$ and $Y_i(0)$ denote defendant $i$'s
potential conviction outcomes under detention and release. The observed outcome
is
\begin{equation} \label{eq:potoutcome}
Y_i
=
Y_i(0)+\{Y_i(1)-Y_i(0)\}T_i.
\end{equation}

Write $X_i$ for the vector of characteristics that govern assignment, and $Z_i$
for a judge assigned to a defendant $i$. The effect of detention in
\eqref{eq:potoutcome} is $Y_i(1)-Y_i(0)$, which may differ from one defendant
to another. The challenge here for estimating the causal effect of pretrial
detention on conviction is that judges may base detention decisions on
characteristics that affect conviction but are not observed by the researcher.
An OLS regression of conviction on pretrial detention and $X_i$ may therefore
confound the effect of detention with these unobserved characteristics. If judges
are randomly assigned, or plausibly randomly assigned conditional on $X_i$, one
may use the treatment propensity or the identity of these judges as an instrument
for the endogenous detention variable, provided that judge assignment affects
conviction only through the detention decision. The judges must also differ in
their detention rates for at least some values of $X_i$ (instrument relevance).
Under these conditions and either of the monotonicity conditions described below,
the IV ratio in \eqref{eq:estimand} is a weighted average of the effects of
detention across defendants, with nonnegative weights.
Assumption~\ref{ass:identification} provides a formal treatment of these conditions.

Under pairwise monotonicity, judges can be ordered within each value of the
assignment covariates. If judge $j$ has a higher detention propensity than
judge $k$, every defendant detained by $k$ must also be detained by $j$.
Pairwise monotonicity fails if, among defendants with the same $X$, one judge
detains a defendant another would release, while the decisions are
reversed for another defendant
\citep{imbensangrist1994,barualang2016}.

\citet{frandsen2023judging} weaken this requirement with \emph{average
monotonicity}. Their condition requires a nonnegative covariance, across
assigned judges, between a defendant's potential treatment schedule and the
judges' population treatment propensities. Pairwise violations are allowed as
long as the total weight on each defendant's treatment effect is nonnegative.
Both assumptions support a causal interpretation of the same IV ratio. Pairwise
monotonicity also implies the representation in terms of LATEs for pairs of
examiners in Proposition~\ref{prop:identification}. See
\citet{goldsmithpinkham2025leniency} and
\citet{coulibaly2024monotonicity}.

Early applications often measure leniency with the leave-one-out mean
\begin{equation}
\hat{\gamma}^{\mathrm{LOO}}_i
=
\frac{\sum_{i' \neq i} \mathbf{1}\{Z_{i'}=Z_i\}\tilde{T}_{i'}}
{\sum_{i' \neq i} \mathbf{1}\{Z_{i'}=Z_i\}},
\label{eq:loo_iv}
\end{equation}
where $\tilde T_i$ residualizes treatment with respect to the assignment
covariates. Omitting $i$ directly removes $\tilde T_i$, but all dependence on
$T_i$ disappears only if $i$ is also omitted from the residualization
regression. The formula requires $N_z\geq2$, so singleton examiner categories
are excluded. Using the mean as an instrument in two-stage least squares delivers
a version of the jackknife IV estimator in \citet{angrist1999jackknife}.

UJIVE extends this leave-one-out correction to the fitted value of $T$ based on
$X$ alone \citep[Section~6]{kolesarcowles}. Let $L[T\mid X,Z]$ and
$L[T\mid X]$ denote the two population least squares projections. JIVE leaves
observation $i$ out when estimating $L[T_i\mid X_i,Z_i]$, but its estimate of
$L[T_i\mid X_i]$ can still depend on $T_i$. This dependence can produce bias
when the number of covariates is large. By deleting observation $i$ from both
projections, UJIVE removes this source of bias. The population
instrument is
\begin{equation}
\gamma_i^L
=
L[T_i\mid X_i,Z_i]-L[T_i\mid X_i].
\label{eq:diff_linear}
\end{equation}
Under saturation or another correct linear representation,
\begin{equation}
\gamma_i^L
=
\mathbb{E}[T_i\mid X_i,Z_i]-\mathbb{E}[T_i\mid X_i].
\label{eq:diff}
\end{equation}

The right-hand side of (\ref{eq:diff}) is the oracle instrument. Define the two
projection errors by
\[
e_1(X,Z)
=
L[T\mid X,Z]-\mathbb{E}[T\mid X,Z],
\qquad
e_2(X)
=
L[T\mid X]-\mathbb{E}[T\mid X].
\]
Then
\begin{equation}
\gamma^L(X,Z)-\gamma_0(X,Z)
=
e_1(X,Z)-e_2(X).
\label{eq:projection_error_gap}
\end{equation}
The projection and oracle instruments coincide if and only if $e_1=e_2$. This
happens when both regressions are correctly specified or when they share a
common approximation error that depends on $X$. If the two approximation errors
do not cancel, the instrument constructed from the linear projections differs
from the one based on the oracle conditional expectations, and therefore the
estimands based on the two instruments may not coincide.

The difference of conditional expectations has the same interpretation based
on partialling out without a linear model. $\mathbb{E}[T\mid X]$ is the $L^2$
projection of $T$ onto $L^2(X)$, and $\mathbb{E}[T\mid X,Z]$ is the projection
onto $L^2(X,Z)$. Since
$L^2(X)\subseteq L^2(X,Z)$, their difference satisfies
\[
\mathbb{E}[\gamma_0(X,Z)\mid X]=0.
\]
Thus $\gamma_0$ is the part of the treatment propensity that varies with
examiner assignment after conditioning on $X$. Under a correctly specified
linear representation, it is the fitted part attributable to the examiner
regressors after partialling out $X$. Interpreting $\gamma_0$ as variation in
treatment propensities across examiners conditional on $X$ does not require a
linear specification.

UJIVE estimates the projection instrument by
\begin{equation}
\hat{\gamma}_i
=
Z_i'\hat{\beta}_{1/i}
+
X_i'\hat{\beta}_{2/i}
-
X_i'\hat{\alpha}_{/i},
\label{eq:ujive_first_stage}
\end{equation}
where each coefficient estimate omits observation $i$. The IV regression with covariates uses $\hat\gamma_i$ as the excluded
instrument (see Section 6.2 of \citet{kolesarcowles}).

With discrete covariates, saturation forms cells jointly by examiner and
covariates, so the fitted values coincide with the cell means. However, with even a moderate number of discrete covariates, covariate cells may become sparse, which can lead to instability of estimated propensities even when
$\mathbb{E}[\gamma_0^2]$ is positive \citep{bhuller}. To deal with this problem, applied researchers often resort to trimming examiners or cells with small caseloads.  While this may improve stability, it
also removes examiner contrasts and changes their weights, which can change the
parameter being estimated as well as its precision.

A nonsaturated linear specification imposes only a finite collection of
orthogonality restrictions. Let $p(X)$ denote the controls included in both
projections. Their normal equations imply
\[
\mathbb{E}[p(X)\gamma^L(X,Z)]=0.
\]
There is one such restriction for each included control. Conditional centering
requires
\[
\mathbb{E}[\gamma^L(X,Z)\mid X]=0,
\]
or, equivalently,
$\mathbb{E}[h(X)\gamma^L(X,Z)]=0$ for every square-integrable function $h$.
The finite collection of normal equations does not imply this conditional
moment.
\citet{blandhol2025tsls} call the stronger restriction \emph{rich covariates}
and show that it is necessary for their weakly causal interpretation of a
linear IV estimand. Saturation satisfies the restriction because the projection
equals the conditional expectation. The oracle instrument satisfies it without
a finite linear representation:
\[
\begin{aligned}
\mathbb{E}[\gamma_0(X,Z)\mid X]
&=
\mathbb{E}\!\left[\mathbb{E}[T\mid X,Z]\mid X\right]
-\mathbb{E}[T\mid X]\\
&=0.
\end{aligned}
\]
See also \citet{sloczynski2026practical}.

For binary $Z$, the oracle instrument reduces to
$\Delta_T(X)\{Z-\mathbb{E}[Z\mid X]\}$, where $\Delta_T(X)$ is the
conditional first stage. Equation~(\ref{eq:diff}) extends this centering
property to assignment among many examiners.

Nonparametric and regularized methods can approximate the two conditional expectations
without saturation. However, their estimation errors have a first-order effect on the IV
moment even when the regressions are fitted on a separate sample. Sample
splitting excludes each evaluation observation from its own regression fits,
but leaves this population bias unchanged
\citep{chernozhukov2018double,chernozhukov2022locally}. The score developed in Section 3 reduces that bias to products of errors in
the four regressions.

\citet{angrist2022machine} document two failures of machine learning in IV
models. LASSO instrument selection can aggravate many-instrument and pretest
bias, and random forest residuals may remain correlated with the controls,
producing spurious second-stage estimates. For the estimator in Section~3, the
research design determines the examiner assignment variable and the conditioning
set before the regressions are fitted. Machine learning estimates the two treatment regressions that define
$\gamma_0$ and the two outcome regressions used in the influence function adjustments.
Mean square consistency makes the difference between the estimated treatment
regressions converge to $\gamma_0$, for which
$\mathbb{E}[\gamma_0\mid X]=0$. The influence function adjustments remove the
first-order effect of estimation error. In the partially penalized
specification, examiner main effects are left unpenalized, avoiding the
dependence on the omitted reference category noted by
\citet{kolesarmuller2025fragility}.

Valid inference with flexible regressions requires control of their errors
and complexity. \citet{hahnhausman2021} show that cross-fitting alone can leave
many-instrument bias in a nonlinear model with a control variable.
\citet{bonhomme2026neyman} show that first-order orthogonality can also fail to
control bias when nuisance parameters are estimated very imprecisely. For the
score in Section 3, the remainder is a sum of products of propensity and
representer errors. For triangular arrays, these products must be
$o_p(n^{-1/2})$ as the number of examiners increases. For an unrestricted
series first stage, the sufficient conditions in Section 3 require
$J_n=o(\sqrt n)$, together with conditions on support and relevance. The
proportional many-instrument sequences studied by \citet{angrist2022machine} fall
outside these conditions.

\section{Main Results: Causal Interpretation, Estimation and Inference}

This section first derives the weights that $\theta_0$ places on treatment
effects. It then constructs the orthogonal estimator and provides conditions for
inference.

\subsection{The Estimand and Its Causal Interpretation}
\label{sec:estimand}

Let $W=(Y,T,X,Z)$. I study the oracle examiner IV ratio
\begin{equation}
\theta_0
:=
\frac{\operatorname{Cov}(Y,\gamma_0)}{\operatorname{Cov}(T,\gamma_0)}
=
\frac{\mathbb{E}[Y\gamma_0]}{\mathbb{E}[T\gamma_0]},
\label{eq:estimand}
\end{equation}
where
$\gamma_0(X,Z)=\mathbb{E}[T\mid X,Z]-\mathbb{E}[T\mid X]$. The second
equality follows from $\mathbb{E}[\gamma_0]=0$. The estimation results permit
either examiner indicators or observed examiner characteristics in $Z$. For the
causal interpretation below, examiner assignment must have finite support
$\{1,\ldots,J\}$. Define
\[
\pi_j(x)=\Pr(Z=j\mid X=x),\qquad
\mathcal J(x)=\{j:\pi_j(x)>0\}.
\]
For $j\in\mathcal J(x)$, define
\[
p_j(x)=\mathbb{E}[T\mid X=x,Z=j],\qquad
\bar p(x)=\sum_{j\in\mathcal J(x)}\pi_j(x)p_j(x).
\]
The support $\mathcal J(x)$ may vary with $x$, so some examiners have zero
assignment probability at particular covariate values. For unit \(i\), let
\begin{equation}
\omega_i(X_i)
=
\sum_{j\in\mathcal J(X_i)}\pi_j(X_i)
\{p_j(X_i)-\bar p(X_i)\}T_i(j).
\label{eq:average_monotonicity_weight}
\end{equation}
Conditional on $X_i$, $\omega_i(X_i)$ is the covariance across possible
examiner assignments between the unit's potential treatment and the examiner's
population treatment propensity.

The observed treatment and outcome satisfy the usual consistency conditions
$T_i=T_i(Z_i)$ and $Y_i=Y_i(T_i,Z_i)$.

\begin{assumption}[Identification]\label{ass:identification}
\begin{enumerate}[label=(\roman*),leftmargin=*,itemsep=0.25ex,topsep=0.25ex]
\item \emph{Conditional independence:}
\[
Z\perp\{Y(0),Y(1),(T(z))_z\}\mid X.
\]
\item \emph{Exclusion:} Let $Y_i(t,j)$ denote the potential outcome if
treatment were set to $t$ and examiner assignment to $j$. For every
$t\in\{0,1\}$ and $j,k\in\mathcal J(X_i)$,
\[
Y_i(t,j)=Y_i(t,k)\equiv Y_i(t)
\quad\text{almost surely}.
\]
\item \emph{Relevance:} $\mathbb{E}[\gamma_0^2]>0$.
\item \emph{Monotonicity:} either of the following conditions holds:
\begin{enumerate}[label=(\alph*),leftmargin=2em,itemsep=0.15ex,topsep=0.15ex]
\item pairwise monotonicity: for every $j,k\in\mathcal J(X)$,
$p_j(X)\geq p_k(X)$ implies $T_i(j)\geq T_i(k)$ almost surely conditional on
$X$;
\item average monotonicity: $\omega_i(X_i)\geq0$ almost surely.
\end{enumerate}
\end{enumerate}
\end{assumption}

\begin{proposition}[Causal weights for the oracle ratio]
\label{prop:identification}
Suppose Assumption~\ref{ass:identification}(i)--(iii) holds,
\(\mathbb{E}[|Y_i(0)|+|Y_i(1)|]<\infty\), and write
\(\tau_i=Y_i(1)-Y_i(0)\). Then
\begin{align}
\mathbb{E}[T\gamma_0]
&=\mathbb{E}[\gamma_0^2]
=\mathbb{E}[\omega_i(X_i)] \notag\\
&=\mathbb{E}\!\left[
\sum_{\substack{j<k\\j,k\in\mathcal J(X)}}
\pi_j(X)\pi_k(X)\{p_j(X)-p_k(X)\}^2
\right],
\label{eq:identification_denominator}\\
\mathbb{E}[Y\gamma_0]
&=\mathbb{E}[\tau_i\omega_i(X_i)].
\label{eq:identification_numerator_average}
\end{align}
Under average monotonicity,
\[
\theta_0
=\frac{\mathbb{E}[\tau_i\omega_i(X_i)]}
{\mathbb{E}[\omega_i(X_i)]},
\]
so the oracle ratio is a positively weighted average of individual treatment
effects. For an unordered pair with $p_j(x)\ne p_k(x)$, define
\[
\operatorname{LATE}_{\{j,k\}}(x)
=
\frac{\mathbb{E}[\tau_i\{T_i(j)-T_i(k)\}\mid X_i=x]}
{p_j(x)-p_k(x)}.
\]
Interchanging $j$ and $k$ reverses both signs and leaves the ratio unchanged.
Thus $j<k$ below only enumerates unordered pairs. The notation does not order
examiners by leniency. Under pairwise monotonicity,
$\operatorname{LATE}_{\{j,k\}}(x)$ is the average treatment effect among units
whose treatment changes between $j$ and $k$. Set it equal to zero when
$p_j(x)=p_k(x)$, since that pair receives zero weight. The oracle ratio then
satisfies
\begin{equation}
\theta_0
=
\frac{
\mathbb{E}\!\left[
\sum_{\substack{j<k\\j,k\in\mathcal J(X)}}
\pi_j(X)\pi_k(X)\{p_j(X)-p_k(X)\}^2
\operatorname{LATE}_{\{j,k\}}(X)
\right]
}{
\mathbb{E}\!\left[
\sum_{\substack{j<k\\j,k\in\mathcal J(X)}}
\pi_j(X)\pi_k(X)\{p_j(X)-p_k(X)\}^2
\right]
}.
\label{eq:pairwise_late_oracle}
\end{equation}
\end{proposition}

Proposition~\ref{prop:identification} establishes that, under conditional
independence, exclusion, relevance and pairwise monotonicity in
Assumption~\ref{ass:identification}, with integrable potential outcomes, the
oracle ratio is a convex combination of conditional pairwise LATEs.

Average monotonicity in Assumption~\ref{ass:identification}(iv) is the
covariance condition of \citet{frandsen2023judging}. Some individual treatment
contrasts may have the opposite sign from the corresponding differences in
examiner treatment propensities, provided each individual's total weight is
nonnegative. Pairwise monotonicity imposes a common ordering and implies the
representation in (\ref{eq:pairwise_late_oracle}).

Proposition~\ref{prop:identification} also verifies the two conditions for
weak causality in \citet{blandhol2025tsls}. Conditional centering removes the
term involving the untreated potential outcome, and average monotonicity makes
the weights on treatment effects nonnegative. In particular,
$\mathbb{E}[\gamma_0\mid X]=0$ and conditional independence imply
$\mathbb{E}[Y_i(0)\gamma_0\mid X_i]=0$. Equation
(\ref{eq:proof_baseline_zero}) proves the stronger result conditional on both
$X_i$ and the unit's potential treatment responses. The cancellation therefore
holds within each response type. Averaging over response types leaves the
numerator $\mathbb{E}[\tau_i\omega_i(X_i)]$, whose weights are nonnegative under
average monotonicity.

For linear IV, \citet{blandhol2025tsls} use the term \emph{rich covariates}
for the requirement that the linear projection used to residualize the
instrument equal its conditional expectation. \citet{kolesarcowles} maintains
the same equality. For a fixed set of regressors, this equality requires correct
specification. Subtracting $\mathbb{E}[T\mid X]$ from
$\mathbb{E}[T\mid X,Z]$ ensures that $\mathbb{E}[\gamma_0\mid X]=0$, even when
neither conditional expectation is linear in the chosen regressors. I allow
both conditional expectations to be estimated by nonparametric or regularized
methods.

\citet{blandhol2025tsls} distinguish an unconditional average causal response
from a statistically weighted average of conditional LATEs and note that the
partially linear IV estimand estimated by double machine learning may be difficult
to interpret. Under exclusion and average monotonicity,
Proposition~\ref{prop:identification} expresses $\theta_0$ as a weighted average
of individual treatment effects with nonnegative weights. Under pairwise
monotonicity, Equation~(\ref{eq:pairwise_late_oracle}) expresses this average in
terms of conditional LATEs.

To isolate the role of conditional centering, replace the oracle instrument by
an arbitrary square-integrable function $h(X,Z)$ and define
\[
\bar h(X)=\mathbb{E}[h(X,Z)\mid X],
\qquad
\omega_{h,i}(X_i)
=
\sum_{j\in\mathcal J(X_i)}
\pi_j(X_i)h(X_i,j)T_i(j).
\]

\begin{lemma}[Arbitrary examiner instruments]
\label{lem:arbitrary_signal}
Suppose Assumption~\ref{ass:identification}(i)--(ii) holds and the displayed
expectations exist. Then
\begin{align}
\mathbb{E}[Th(X,Z)]
&=\mathbb{E}[\omega_{h,i}(X_i)],
\label{eq:arbitrary_signal_first_stage}\\
\mathbb{E}[Yh(X,Z)]
&=\mathbb{E}[\mathbb{E}[Y_i(0)\mid X_i]\bar h(X_i)]
+\mathbb{E}[\tau_i\omega_{h,i}(X_i)].
\label{eq:arbitrary_signal_outcome}
\end{align}
If $\mathbb{E}[Th]\ne0$, the moment ratio based on $h$ is
\begin{equation}
\frac{\mathbb{E}[Yh]}{\mathbb{E}[Th]}
=
\frac{\mathbb{E}[\mathbb{E}[Y_i(0)\mid X_i]\bar h(X_i)]}
{\mathbb{E}[\omega_{h,i}(X_i)]}
+
\frac{\mathbb{E}[\tau_i\omega_{h,i}(X_i)]}
{\mathbb{E}[\omega_{h,i}(X_i)]}.
\label{eq:arbitrary_signal_ratio}
\end{equation}
\end{lemma}

When $\bar h(X)=0$, the term involving $Y(0)$ in
(\ref{eq:arbitrary_signal_ratio}) vanishes. Conditional centering does not ensure that the weights on treatment effects are
nonnegative. A weakly causal interpretation requires this additional
restriction. For
$h=\gamma_0$,
$\mathbb{E}[h\mid X]=0$ and
$\omega_{h,i}=\omega_i$ in
(\ref{eq:average_monotonicity_weight}).

I next compare first-stage relevance among conditionally centered instruments
with unit variance. Define
\[
\mathcal H_X
=
\{h\in L^2(X,Z):\mathbb{E}[h\mid X]=0,
\ \mathbb{E}[h^2]=1\}.
\]

\begin{lemma}[Maximum first-stage covariance under conditional centering]
\label{lem:maximal_relevance}
If $D=\mathbb{E}[\gamma_0^2]>0$, then
\[
\sup_{h\in\mathcal H_X}|\mathbb{E}[Th]|=\sqrt D.
\]
The only maximizers, up to equality almost surely, are
$h=\gamma_0/\sqrt D$ and $h=-\gamma_0/\sqrt D$. Requiring the covariance to
be positive selects $h=\gamma_0/\sqrt D$.
\end{lemma}

Lemma~\ref{lem:maximal_relevance} shows that $\gamma_0/\sqrt D$ maximizes the
absolute covariance with treatment among instruments with unit variance and
zero conditional expectation given $X$. With heterogeneous treatment effects,
however, the choice of instrument also determines the weights on individual
treatment effects (Lemma~\ref{lem:arbitrary_signal}), so maximizing this
covariance does not determine which weighted average is of economic interest.

\subsection{The Orthogonal Moment Function}

To estimate the population ratio, I construct a moment whose first derivative
with respect to each conditional expectation in $\gamma_0$ is zero.

If $\gamma_0$ were known, the sample oracle estimator would be
\begin{equation}
\hat{\theta}_{\mathrm{or}}
=
\frac{\mathbb{E}_n[Y\gamma_0]}{\mathbb{E}_n[T\gamma_0]}.
\label{eq:oracle_estimator}
\end{equation}
UJIVE replaces the unknown instrument with leave-one-out linear first stages.
When the two propensity functions are estimated nonparametrically or with
regularization, the first-order effect of estimating both components of
$\gamma_0$ must also be corrected. Start from the identifying moment
\begin{equation}
g(W,\gamma,\theta)=(Y-\theta T)\gamma.
\label{eq:identifying_moment}
\end{equation}
At the true parameter and instrument,
\[
\mathbb{E}[(Y-\theta_0T)\gamma_0]
=
\mathbb{E}[Y\gamma_0]-\theta_0\mathbb{E}[T\gamma_0]
=0,
\]
by the definition of the ratio in (\ref{eq:estimand}).

Estimating either conditional expectation changes the identifying moment at first
order. Following the influence function approach of
\citet{newey1994asymptotic} and \citet{chernozhukov2022locally}, I construct
one adjustment for estimation of each conditional expectation and add both
to the identifying moment.
Write the true regression functions as
\[
\gamma_{01}(X,Z)=\mathbb{E}[T\mid X,Z],
\qquad
\gamma_{02}(X)=\mathbb{E}[T\mid X],
\qquad
\gamma_0=\gamma_{01}-\gamma_{02}.
\]
I use $\gamma_1$ and $\gamma_2$ for candidate regression functions and
$\gamma=\gamma_1-\gamma_2$ for their difference.
The projection property of conditional expectations implies the restrictions
used to construct the adjustments. With $\Gamma_1=L^2(X,Z)$ and
$\Gamma_2=L^2(X)$, the residual from each projection is orthogonal to every
function in the corresponding space:
\[
\mathbb{E}\!\left[\delta(X,Z)\{T-\gamma_{01}(X,Z)\}\right]=0
\quad\text{for every }\delta\in\Gamma_1,
\]
and
\[
\mathbb{E}\!\left[\delta(X)\{T-\gamma_{02}(X)\}\right]=0
\quad\text{for every }\delta\in\Gamma_2.
\]

Perturbations of the identifying moment along these two spaces define the bounded
linear functionals
\[
\mathcal{L}_1(\delta)=\mathbb{E}[(Y-\theta_0T)\delta(X,Z)],
\qquad
\mathcal{L}_2(\delta)=-\mathbb{E}[(Y-\theta_0T)\delta(X)].
\]
When $\mathbb{E}[(Y-\theta_0T)^2]<\infty$, the Riesz representation theorem
ensures that these functionals have unique representers:
\[
\alpha_{01}(X,Z)=\mathbb{E}[Y-\theta_0T\mid X,Z],
\qquad
\alpha_{02}(X)=-\mathbb{E}[Y-\theta_0T\mid X].
\]
The two adjustments are
\begin{equation}
\phi(W,\gamma,\alpha,\theta)
=
\alpha_1(X,Z)\{T-\gamma_1(X,Z)\}
+
\alpha_2(X)\{T-\gamma_2(X)\}.
\label{eq3}
\end{equation}

For this parameter, both representers are conditional expectations and have
explicit expressions.\footnote{When such expressions are not available,
automatic Riesz methods can estimate the representers
\citep{chernozhukov2022automatic,chernozhukov2020adversarial}.} Adding
(\ref{eq3}) to the identifying moment defines the orthogonal score
\begin{equation}
\psi(W;\theta,\gamma,\alpha)
=(Y-\theta T)\gamma
+\alpha_1(X,Z)\{T-\gamma_1(X,Z)\}
+\alpha_2(X)\{T-\gamma_2(X)\}.
\label{eq:orthogonal_score}
\end{equation}
At $\theta_0$ and the true nuisance functions, each term has mean zero. The
ratio definition implies
$\mathbb{E}[(Y-\theta_0T)\gamma_0]=0$, and the projection identities imply
$\mathbb{E}[\alpha_{0j}(T-\gamma_{0j})]=0$ for $j=1,2$. These moment
restrictions hold at $\theta_0$ and the true nuisance functions. Using the influence
function arguments in \citet{ichimura2022influence} and
\citet{chernozhukov2022locally}, I show in the proof of
Proposition~\ref{prop:efficiency} that $\psi$ is the influence function for the
identifying moment and that $D^{-1}\psi$ is the efficient influence function
for $\theta_0$.

\begin{proposition}[Efficient influence function]\label{prop:efficiency}
Consider a dominated nonparametric model for the observed law of
$W=(Y,T,X,Z)$ on a fixed measurable support. Suppose the scores of regular
parametric submodels locally equivalent to $F_0$ are dense in the tangent space
$L_0^2(F_0)$. The support may be infinite, and the regression functions along
these submodels are viewed in the fixed spaces $L^2(F_{0,X,Z})$ and
$L^2(F_{0,X})$. Suppose $T\in\{0,1\}$,
$\mathbb{E}[(Y-\theta_0T)^2]<\infty$, and the first stage is
nondegenerate,
$D=\mathbb{E}[T\gamma_0]=\mathbb{E}[\gamma_0^2]>0$. Under i.i.d.\ sampling
from $F_0$, the efficient influence function for
$\theta(F)=\mathbb{E}_F[Y\gamma_F]/\mathbb{E}_F[T\gamma_F]$ is
\[
\varphi_{\mathrm{eff}}(W)
=
D^{-1}\psi(W;\theta_0,\gamma_0,\alpha_0),
\]
where $\psi$ is the score in (\ref{eq:orthogonal_score}) evaluated at the true
nuisances and $\theta_0$. The efficiency bound is
\[
V_{\mathrm{eff}}
=D^{-2}\mathbb{E}\!\left[\psi(W;\theta_0,\gamma_0,\alpha_0)^2\right].
\]
\end{proposition}

Appendix~\ref{app:proofs} derives the pathwise derivative. The efficiency bound applies to $\theta(F)$ as a functional of the observed distribution.
Its causal interpretation also uses Assumption~\ref{ass:identification}. The proposition applies to
i.i.d.\ observations from a fixed distribution with a nondegenerate first
stage. Clustered samples, weak identification, and saturated models with a
growing number of examiners require different arguments.

\subsection{Neyman Orthogonality and Bias Decomposition}

The influence function adjustments remove the first-order effect of estimating the four
regression functions. I first verify Neyman orthogonality and then derive the
remaining bias.

Neyman orthogonality means that each nuisance derivative of the population
score vanishes at the truth. For every admissible direction
$\delta\in\Gamma_1=L^2(X,Z)$,
\begin{align*}
\left.\frac{d}{dt}\mathbb{E}\!\left[
\psi\{W;\theta_0,(\gamma_{01}+t\delta,\gamma_{02}),\alpha_0\}
\right]\right|_{t=0}
&=\mathbb{E}[\{Y-\theta_0T-\alpha_{01}(X,Z)\}\delta(X,Z)]\\
&=0.
\end{align*}
The last equality follows from
$\alpha_{01}(X,Z)=\mathbb{E}[Y-\theta_0T\mid X,Z]$. Varying $\gamma_2$ in a
direction $\delta\in\Gamma_2=L^2(X)$ gives
\[
\left.\frac{d}{dt}\mathbb{E}\!\left[
\psi\{W;\theta_0,(\gamma_{01},\gamma_{02}+t\delta),\alpha_0\}
\right]\right|_{t=0}
=
\mathbb{E}[\{-Y+\theta_0T-\alpha_{02}(X)\}\delta(X)]=0.
\]
The projection residuals also imply
\begin{equation}
\mathbb{E}\left[\phi(W,\gamma_0,\alpha,\theta_0)\right]=0
\quad\text{for every admissible }\alpha\in\mathcal{A}.
\label{eq:first_step_if_mean_zero}
\end{equation}

For fixed $\alpha$, the score in (\ref{eq:orthogonal_score}) is jointly affine
in $(\gamma_1,\gamma_2)$; for fixed $\gamma$, it is affine in $\alpha$.
At the truth, every term that is linear in a single nuisance error has mean
zero. The remaining terms are products of propensity and representer errors.

Let $\mu_0(W)=Y-\theta_0T$ and write
\[
\begin{aligned}
\alpha_{01}(X,Z)&=\mathbb{E}[\mu_0(W)\mid X,Z],
&\alpha_{02}(X)&=-\mathbb{E}[\mu_0(W)\mid X],\\
\gamma_{01}(X,Z)&=\mathbb{E}[T\mid X,Z],
&\gamma_{02}(X)&=\mathbb{E}[T\mid X].
\end{aligned}
\]

\begin{lemma}[Bilinear bias decomposition]\label{lem:mr}
For any square-integrable candidate nuisance functions
$(\alpha_1,\alpha_2,\gamma_1,\gamma_2)$,
\[
\mathbb{E}\left[\psi(W;\theta_0,\gamma,\alpha)\right]
= -\mathbb{E}\left[(\alpha_1-\alpha_{01})(\gamma_1-\gamma_{01})\right]
  -\mathbb{E}\left[(\alpha_2-\alpha_{02})(\gamma_2-\gamma_{02})\right].
\]
Thus the $j$th term is zero whenever either $\alpha_j=\alpha_{0j}$ or
$\gamma_j=\gamma_{0j}$.
\end{lemma}

Lemma~\ref{lem:mr} treats $\alpha$ and $\gamma$ as separate functions. If the representers are estimated separately, the $j$th bias term vanishes
when either $\gamma_j$ or $\alpha_j$ is correct. In the estimator below, the
same four regressions determine both $\alpha$ and $\gamma$, so errors in the
treatment regressions also affect the estimated Riesz representers.

For the estimator used below, let $m_{01}=\mathbb{E}[Y\mid X,Z]$ and
$m_{02}=\mathbb{E}[Y\mid X]$, and write $\delta_{mj}=m_j-m_{0j}$ and
$\delta_{\gamma j}=\gamma_j-\gamma_{0j}$. At $\theta_0$, the formulas in (\ref{eq:alpha_closed}) imply
\[
\delta_{\alpha1}=\delta_{m1}-\theta_0\delta_{\gamma1},
\qquad
\delta_{\alpha2}=-\delta_{m2}+\theta_0\delta_{\gamma2},
\]
and substituting these shared errors into Lemma~\ref{lem:mr} gives
\begin{align}
\mathbb{E}[\psi(W;\theta_0,\gamma,\alpha)]
={}&-\mathbb{E}[\delta_{m1}\delta_{\gamma1}]
+\theta_0\mathbb{E}[\delta_{\gamma1}^2] \notag\\
&+\mathbb{E}[\delta_{m2}\delta_{\gamma2}]
-\theta_0\mathbb{E}[\delta_{\gamma2}^2].
\label{eq:coupled_remainder}
\end{align}
If both outcome regressions are correct, the bias reduces to
$\theta_0\{\mathbb{E}[\delta_{\gamma1}^2]-\mathbb{E}[\delta_{\gamma2}^2]\}$,
which is generally nonzero. All four terms are products of regression errors. Assumption~\ref{ass2} makes
their sum negligible after multiplication by $\sqrt n$.

\subsection{The Estimator}\label{mom_est}

I estimate the four regressions by cross-fitting. Partition the sample into folds
$I_l$, $l=1,\ldots,L$. For observations in $I_l$, fit the regressions using only
$I_l^c$:
\[
\hat m_{1l}^Y(x,z)\approx \mathbb{E}[Y\mid X=x,Z=z],\quad
\hat m_{2l}^Y(x)\approx \mathbb{E}[Y\mid X=x],
\]
\[
\hat\gamma_{1l}(x,z)\approx \mathbb{E}[T\mid X=x,Z=z],\quad
\hat\gamma_{2l}(x)\approx \mathbb{E}[T\mid X=x].
\]
The cross-fitted instrument is
$\hat\gamma_l(x,z)=\hat\gamma_{1l}(x,z)-\hat\gamma_{2l}(x)$. In the population,
the treatment propensities lie between zero and one, but fitted values can
fall outside this range. One can impose these bounds through constrained
estimation or an appropriate link function. In the simulations and empirical
application, I instead truncate the fitted values before evaluating the score.

At a candidate value $\theta$, the population representers are
\begin{equation} \label{eq:alpha_closed}
\alpha_{01}(X,Z;\theta)=\mathbb{E}[Y\mid X,Z]-\theta\mathbb{E}[T\mid X,Z],
\qquad
\alpha_{02}(X;\theta)=-\mathbb{E}[Y\mid X]+\theta\mathbb{E}[T\mid X].
\end{equation}
I estimate these functions by
\begin{equation} \label{eq:alpha_feasible_closed}
\hat\alpha_{1l}(x,z;\theta)=\hat m_{1l}^Y(x,z)-\theta\hat\gamma_{1l}(x,z),
\qquad
\hat\alpha_{2l}(x;\theta)=-\hat m_{2l}^Y(x)+\theta\hat\gamma_{2l}(x).
\end{equation}
The maps are affine in $\theta$, so the estimating equation does not require a
preliminary estimate of $\theta_0$.

After substituting the expressions in (\ref{eq:alpha_feasible_closed}), the
score can be written in residualized form.
For candidate outcome regressions $m_1(X,Z)$ and $m_2(X)$, let
$V(X,Z)=\gamma_1(X,Z)-\gamma_2(X)$ and
$\eta=(m_1,m_2,\gamma_1,\gamma_2)$. Then
\begin{align}
\psi(W;\theta,\eta)
={}&
\{Y-m_2(X)-\theta[T-\gamma_2(X)]\}V(X,Z) \notag\\
&+\{T-\gamma_1(X,Z)\}
\{m_1(X,Z)-m_2(X)-\theta V(X,Z)\}.
\label{eq:residualized_score}
\end{align}

The cross-fitted sample moment is
\begin{align} \label{debiasedm}
\hat{\psi}(\theta)
&=
\frac{1}{n} \sum_{l=1}^L \sum_{i \in I_{l}}(Y_i-\theta T_i)\hat\gamma_l(X_i,Z_i) \notag\\
&\quad +
\frac{1}{n} \sum_{l=1}^L \sum_{i \in I_{l}} \hat{\alpha}_{1l}(X_i,Z_i;\theta)\left(T_i-\hat{\gamma}_{1l}(X_i,Z_i)\right) \notag\\
&\quad +
\frac{1}{n} \sum_{l=1}^L \sum_{i \in I_{l}} \hat{\alpha}_{2l}(X_i;\theta)\left(T_i-\hat{\gamma}_{2l}(X_i)\right).
\end{align}
The estimator solves $\hat\psi(\theta)=0$. Because the representer maps in
(\ref{eq:alpha_feasible_closed}) are affine in $\theta$, the sample moment is
affine as well:
\begin{equation} \label{eq:closed_form_score}
\hat\psi(\theta)=\widehat A+\theta\widehat Q,
\end{equation}
where
\begin{align}
\widehat A
&=
\frac{1}{n}\sum_{l=1}^L\sum_{i\in I_l}
\left[
Y_i\hat\gamma_l(X_i,Z_i)
+\hat m_{1l}^Y(X_i,Z_i)\{T_i-\hat\gamma_{1l}(X_i,Z_i)\}
-\hat m_{2l}^Y(X_i)\{T_i-\hat\gamma_{2l}(X_i)\}
\right],
\label{eq:Ahat_closed}\\
\widehat Q
&=
\frac{1}{n}\sum_{l=1}^L\sum_{i\in I_l}
\left[
-2T_i\hat\gamma_l(X_i,Z_i)
+\hat\gamma_{1l}(X_i,Z_i)^2
-\hat\gamma_{2l}(X_i)^2
\right].
\label{eq:Qhat_closed}
\end{align}
Equivalently, with $\widehat D=-\widehat Q$,
\begin{equation} \label{eq:closed_form_theta}
\hat\theta=-\frac{\widehat A}{\widehat Q}
=
\frac{\widehat A}{\widehat D},
\qquad
\widehat D
=
\frac{1}{n}\sum_{l=1}^L\sum_{i\in I_l}
\left[
2T_i\hat\gamma_l(X_i,Z_i)
-\hat\gamma_{1l}(X_i,Z_i)^2
+\hat\gamma_{2l}(X_i)^2
\right].
\end{equation}

The numerator $\widehat A$ estimates $\mathbb{E}[Y\gamma_0]$ with the influence
function adjustments included. The denominator $\widehat D$ estimates
$\mathbb{E}[T\gamma_0]=\mathbb{E}[\gamma_0^2]$ with the same correction. Their
ratio is $\hat\theta$, and $\widehat D$ measures instrument strength.

At the true distribution,
\[
\gamma_{02}(x)
=
\int \gamma_{01}(x,z)\,P_{Z\mid X}(dz\mid x),
\]
where $P_{Z\mid X}(A\mid x)=\Pr(Z\in A\mid X=x)$. With discrete examiner
assignment, the integral is
\[
\gamma_{02}(x)
=
\sum_{j\in\mathcal J(x)}\pi_j(x)\gamma_{01}(x,j),
\qquad
\pi_j(x)=\Pr(Z=j\mid X=x).
\]
One could use this identity to estimate $\gamma_2$ by integrating the fitted
$\gamma_1$ against an estimate of $P_{Z\mid X}$. The estimated propensities
would then satisfy the averaging identity with respect to the estimated
assignment distribution. Error in the short regression would come from both
the fitted long regression and the estimated assignment probabilities.

Alternatively, fitting $\gamma_2$ directly requires fewer conditioning
variables than fitting $\gamma_1$ and no estimate of the assignment
probabilities, $\pi_j(x)=\Pr(Z=j\mid X=x)$. Each regression can then use its own method and tuning
parameter. The theory covers either approach when the estimates satisfy
Assumptions~\ref{ass1} and~\ref{ass2}.

Each score contribution uses regression estimates fitted on the other folds.
Cross-fitting avoids Donsker conditions and keeps an observation's own outcome
and treatment out of the regressions used in its score
\citep{chernozhukov2018double,chernozhukov2022debiased,chernozhukov2022locally}.
With many examiners, cross-fitting alone does not ensure that the regression
errors satisfy the convergence and product conditions in
Assumptions~\ref{ass1} and~\ref{ass2}.

When cases handled by the same examiner are correlated, splitting cases
across folds can induce dependence between the training and evaluation samples, since
both samples may contain cases from that examiner. In
Appendix~\ref{app:cluster_inference}, I provide additional conditions for
inference with dependence within clusters, including conditions on the errors
from estimating the regressions.

\subsection{Asymptotic Theory}
\label{sec:asymptotic_theory}
For a fixed distribution, the regression estimates must converge in $L^2$,
and the products of their errors must be $o_p(n^{-1/2})$. Assumption~\ref{ass1}
controls the random variation introduced by replacing the population
regressions with estimates. Assumption~\ref{ass2} controls the population bias in
Lemma~\ref{lem:mr}. I also assume that the first stage is bounded away from
zero.

\begin{assumption}[Mean square consistency]\label{ass1}
For each fold $l$ and each $j\in\{1,2\}$,
\[
\|\hat\gamma_{jl}-\gamma_{0j}\| \overset{p}{\to}0,
\qquad
\|\hat\alpha_{jl}(\theta_0)-\alpha_{0j}(\theta_0)\| \overset{p}{\to}0,
\]
where $\|a\|=(\mathbb{E}[a(W)^2])^{1/2}$ is the $L^2$ norm. For a
pair $a=(a_1,a_2)$, write
$\|a\|=\{\|a_1\|^2+\|a_2\|^2\}^{1/2}$.
For the two treatment regressions, let
$\hat{\boldsymbol\gamma}_l=(\hat\gamma_{1l},\hat\gamma_{2l})$ and
$\boldsymbol\gamma_0=(\gamma_{01},\gamma_{02})$. Define
$\hat{\boldsymbol\alpha}_l$ and $\boldsymbol\alpha_0$ analogously.
\end{assumption}

\begin{assumption}[Product condition]\label{ass2}
For each fold $l$, the nuisance estimates satisfy
\[
\|\hat{\boldsymbol\alpha}_l(\theta_0)-\boldsymbol\alpha_0(\theta_0)\|\,
\|\hat{\boldsymbol\gamma}_l-\boldsymbol\gamma_0\|
=o_p(n^{-1/2}).
\]
\end{assumption}

Assumption~\ref{ass2} holds under either of the following conditions:
(i) both factors are $o_p(n^{-1/4})$; or (ii) one factor converges more slowly,
but the other converges sufficiently quickly that their product is still
$o_p(n^{-1/2})$.

Notice that Assumption~\ref{ass1} requires convergence in $L^2$ but does not
specify how quickly the errors vanish. Assumption~\ref{ass2}, on the other
hand, provides the rate that ensures the bilinear remainder converges to zero
in probability after multiplication by $\sqrt n$. Both conditions pertain to
convergence at $\theta_0$. Since the feasible score is affine in $\theta$,
these conditions obviate the need for a uniform convergence argument over a
compact parameter set.

\begin{assumption}[Score bounds and continuity]\label{ass3}
The true representers $\alpha_{0j}(\theta_0)$ and the cross-fitted estimates
$\hat\alpha_{jl}(\theta_0)$ are uniformly bounded with probability approaching
one. The fitted treatment regressions used in
(\ref{eq:alpha_feasible_closed}) are also uniformly bounded with probability
approaching one. Moreover,
\[
\mathbb{E}[\psi(W;\theta_0,\gamma_0,\alpha_0)^2]<\infty.
\]
Along each cross-fitted nuisance sequence, the score evaluated at $\theta_0$ is
mean square continuous in $(\gamma,\alpha)$ at
$(\gamma_0,\alpha_0)$.
\end{assumption}

\begin{assumptionalternative}[Fourth moment alternative]\label{ass3prime}
The outcome residual $Y-\theta_0T$, the true representers, and the cross-fitted
representers have uniformly bounded fourth moments. Uniformly over folds and
$j\in\{1,2\}$,
\[
\|\hat\gamma_{jl}-\gamma_{0j}\|_4=o_p(1),
\qquad
\|\hat\alpha_{jl}(\theta_0)-\alpha_{0j}(\theta_0)\|_4=o_p(1),
\]
and
\[
\max_{l,j}
\|\hat\alpha_{jl}(\theta_0)-\alpha_{0j}(\theta_0)\|_4
\|\hat\gamma_{jl}-\gamma_{0j}\|_4
=o_p(n^{-1/2}).
\]
The oracle score has a finite second moment.
\end{assumptionalternative}

There are two ways to control the sample score. Assumption~\ref{ass3} uses
uniform boundedness and continuity in $L^2$. Under the fourth moment
bounds in Assumption~\ref{ass3prime}, a Cauchy--Schwarz inequality controls
the bilinear remainder as well as empirical terms involving a single
nuisance error. Clipping fitted propensities to $[0,1]$ when $T$ is binary
makes the propensity errors uniformly bounded. Mean square
consistency then also implies the $L^4$ consistency required of those
propensity estimates.

\begin{theorem}[$\sqrt n$ expansion of the debiased sample moment]\label{thm1}
Suppose the observations are independent and identically distributed and the
fold partition is deterministic or generated independently of the observations.
Let $L$ be fixed, with fold proportions $n_l/n$ bounded away from zero and one.
If Assumptions~\ref{ass1} and~\ref{ass2} hold together with either
Assumption~\ref{ass3} or Assumption~\ref{ass3prime}, then
\[
\sqrt{n}\,\hat\psi(\theta_0)
=
\frac{1}{\sqrt n}\sum_{i=1}^n \psi(W_i;\theta_0,\gamma_0,\alpha_0)+o_p(1).
\]
\end{theorem}

Lemma~\ref{lem:mr} expresses the conditional expectation of the score error as
a sum of products of regression errors, and
Assumption~\ref{ass2} makes it $o_p(n^{-1/2})$. Cross-fitting, convergence in
$L^2$, and either Assumption~\ref{ass3} or Assumption~\ref{ass3prime} control
the remaining sample variation without Donsker restrictions.
The proof is in Appendix~\ref{app:proofs}.

For inference, the sample Jacobian is the coefficient $\widehat Q$ in
(\ref{eq:Qhat_closed}):
\[
\widehat Q
=
\frac{\partial \hat\psi(\theta)}{\partial \theta}
=
\frac{1}{n}\sum_{l=1}^L\sum_{i\in I_l}
\left[
-2T_i\hat\gamma_l(X_i,Z_i)
+\hat\gamma_{1l}(X_i,Z_i)^2
-\hat\gamma_{2l}(X_i)^2
\right].
\]
The population Jacobian is
\[
Q=\mathbb{E}\left[-2T\{\gamma_{01}(X,Z)-\gamma_{02}(X)\}+\gamma_{01}(X,Z)^2-\gamma_{02}(X)^2\right]
=-\mathbb{E}\left[\left(\gamma_{01}(X,Z)-\gamma_{02}(X)\right)^2\right].
\]
For the second equality, write $\gamma_0=\gamma_{01}-\gamma_{02}$. Then
\[
\mathbb{E}[T\gamma_0]
=\mathbb{E}[\gamma_{01}\gamma_0]
=\mathbb{E}[\gamma_0^2],
\]
where the first step uses $\mathbb{E}[T\mid X,Z]=\gamma_{01}$ and the second uses
$\mathbb{E}[\gamma_0\mid X]=0$.

\begin{assumption}[Strong relevance]\label{ass4}
For some constant $c>0$,
\[
D=\mathbb{E}[\gamma_0^2]\ge c.
\]
\end{assumption}

The condition $D\ge c>0$ requires nontrivial residual variation in examiner
propensities. Under the nuisance conditions above, the sample Jacobian converges
to $-D$.

\begin{lemma}[Consistency of the sample Jacobian]\label{lem:Qhat}
Under Assumption~\ref{ass1}, either Assumption~\ref{ass3} or
Assumption~\ref{ass3prime}, and the sampling conditions of
Theorem~\ref{thm1}, $\widehat Q\overset{p}{\to}Q=-D$. Under
Assumption~\ref{ass4}, it follows that $|\widehat Q|\ge c/2$ with probability
approaching one.
\end{lemma}

Appendix~\ref{app:proofs} proves Lemma~\ref{lem:Qhat} by conditioning on each
fold's training sample and controlling the quadratic Jacobian terms in $L^1$.

Let $l(i)$ denote the fold containing observation $i$, and define the cross-fitted score contribution
\[
\begin{aligned}
\hat\psi_i(\theta)
={}&(Y_i-\theta T_i)\hat\gamma_{l(i)}(X_i,Z_i)\\
&+\hat\alpha_{1,l(i)}(X_i,Z_i;\theta)
\{T_i-\hat\gamma_{1,l(i)}(X_i,Z_i)\}\\
&+\hat\alpha_{2,l(i)}(X_i;\theta)
\{T_i-\hat\gamma_{2,l(i)}(X_i)\}.
\end{aligned}
\]
In this expression, $\hat\gamma_l=\hat\gamma_{1l}-\hat\gamma_{2l}$.
For independent observations, estimate the variance by
\[
\widehat\Omega_{\mathrm{iid}}
=\frac{1}{n}\sum_{i=1}^n\left(\hat\psi_i(\hat\theta)-\frac{1}{n}\sum_{k=1}^n\hat\psi_k(\hat\theta)\right)^2,
\qquad
\widehat V_{\mathrm{iid}}=\widehat Q^{-2}\widehat\Omega_{\mathrm{iid}}.
\]
Thus \(\widehat V_{\mathrm{iid}}\) estimates the asymptotic variance of
\(\sqrt n(\hat\theta-\theta_0)\), and the reported sample standard error is
\(\sqrt{\widehat V_{\mathrm{iid}}/n}\).

\begin{corollary}[Asymptotic normality]\label{coro}
Under Assumptions~\ref{ass1}, \ref{ass2}, and \ref{ass4}, together with
either Assumption~\ref{ass3} or Assumption~\ref{ass3prime}, and the i.i.d.\
central limit theorem,
\[
\sqrt n(\hat\theta-\theta_0)\overset{d}{\to}N(0,V_{\mathrm{iid}}),
\qquad
V_{\mathrm{iid}}=Q^{-2}\mathbb{E}[\psi(W;\theta_0,\gamma_0,\alpha_0)^2],
\]
and $\widehat V_{\mathrm{iid}}\overset{p}{\to}V_{\mathrm{iid}}$.
\end{corollary}

In empirical examiner designs,
observations are often dependent within examiner. Let $C_g$ denote the cases handled by examiner $g$, for
$g=1,\ldots,G$. The clustered variance estimator is
\[
\widehat\Omega_{\mathrm{cl}}
=\frac{1}{n}\frac{G}{G-1}\sum_{g=1}^G
\left(\sum_{i\in C_g}\hat\psi_i(\hat\theta)\right)^2,
\qquad
\widehat V_{\mathrm{cl}}=\widehat Q^{-2}\widehat\Omega_{\mathrm{cl}}.
\]
With independent cases and a fixed number of examiners, the i.i.d.\ theorem applies
and $\widehat V_{\mathrm{iid}}$ is the relevant variance estimator. The clustered variance estimator is consistent when the number of independent
examiner clusters grows, the sums of scores within clusters satisfy a central
limit theorem, and the regression errors are negligible after aggregation
within clusters (Appendix~\ref{app:cluster_inference}).

With a fixed number of examiners, arbitrary dependence among cases handled by
the same examiner does not support conventional cluster asymptotics. The
appropriate variance estimator therefore depends on the sampling and assignment
process \citep{goldsmithpinkham2025leniency}.

The denominator of the IV ratio measures instrument strength. Applications
should therefore report both
\[
-\widehat Q=\widehat D
\qquad\text{and}\qquad
n^{-1}\sum_i\hat\gamma_{l(i)}(X_i,Z_i)^2.
\]
At the true nuisance functions, both diagnostics are unbiased and consistent
for $D=\mathbb{E}[\gamma_0^2]$.

When the estimated first stage is close to zero, small sampling errors in the
denominator can produce large changes in the IV ratio and make the Wald
approximation unreliable. I therefore also invert the score test in
Appendix~\ref{app:score_inversion}, which does not divide by the estimated
Jacobian. The accepted parameter values can form a bounded set, an unbounded
set, or the whole real line. I report these sets as a sensitivity analysis
because their coverage has not been established uniformly along sequences
with $D_n\to0$.

\subsection{Growing Dimension}\label{ssec:growing}

The number of examiners, the covariate basis, and the estimand may all vary with
$n$. The convergence bounds and the expansion must then hold uniformly along
the sequence
\citep{chernozhukov2018double,chernozhukov2022locally}.

In this subsection, $\mathbb{E}_n$ denotes population expectation under the
$n$-th design. Let $W_{ni}=(Y_{ni},T_{ni},X_{ni},Z_{ni})$, $i=1,\ldots,n$, denote the $n$-th
row of the array. Under the $n$-th design, define
$\gamma_{1n}(X,Z)=\mathbb{E}_n[T\mid X,Z]$,
$\gamma_{2n}(X)=\mathbb{E}_n[T\mid X]$,
$m_{1n}^Y(X,Z)=\mathbb{E}_n[Y\mid X,Z]$, and
$m_{2n}^Y(X)=\mathbb{E}_n[Y\mid X]$. The estimand in row $n$ is
\[
\theta_n
=
\frac{\mathbb{E}_n[Y\gamma_n]}{\mathbb{E}_n[T\gamma_n]},
\qquad
\gamma_n(X,Z)=\gamma_{1n}(X,Z)-\gamma_{2n}(X),
\]
I write $D_n=\mathbb{E}_n[\gamma_n^2]$ for instrument strength.

\begin{assumption}[Triangular array conditions]\label{ass:tri}
Within each row, the observations are i.i.d.\ from the $n$-th design, and the
fold partition is deterministic or generated independently of the observations
in that row. The number of folds $L$ is fixed, and fold proportions are bounded
away from zero and one.
The sequence $\theta_n$ is bounded, $\sup_n|\theta_n|<\infty$. Uniformly along
the array, the parameter and nuisance functions satisfy either the
conditions on bounds and continuity in Assumption~\ref{ass3} or the
fourth moment conditions in Assumption~\ref{ass3prime}. For each fold
$l$ and each $j\in\{1,2\}$:
\begin{enumerate}
    \item \textbf{Mean square consistency:} $\|\hat\gamma_{jl}-\gamma_{jn}\|_n+\|\hat m_{jl}^Y-m_{jn}^Y\|_n\xrightarrow{p}0$, where $\|\cdot\|_n$ is the $L^2$ norm under the $n$-th design.
    \item \textbf{Product condition:} with $r_{g,n}=\max_{l\le L}\max_j\|\hat\gamma_{jl}-\gamma_{jn}\|_n$ and $r_{m,n}=\max_{l\le L}\max_j\|\hat m_{jl}^Y-m_{jn}^Y\|_n$,
    \[
    r_{g,n}\,(r_{g,n}+r_{m,n})=o_p(n^{-1/2}).
    \]
    \item \textbf{Uniform relevance:} $\liminf_{n\to\infty}D_n>0$.
    \item \textbf{Lindeberg CLT:} let
    \[
    \xi_{ni}=\psi_n(W_{ni};\theta_n,\gamma_n,\alpha_n),
    \qquad
    \Omega_n=\mathrm{Var}_n(\xi_{ni}).
    \]
    Then $\Omega_n\to\Omega_\infty\in(0,\infty)$ and, for every
    $\varepsilon>0$,
    \[
    \frac{1}{n\Omega_n}\sum_{i=1}^n
    \mathbb{E}_n\!\left[\xi_{ni}^2\mathbf{1}\!\left\{
    |\xi_{ni}|>\varepsilon\sqrt{n\Omega_n}
    \right\}\right]
    \longrightarrow 0.
    \]
\end{enumerate}
\end{assumption}

Parts (i) and (ii) impose the same convergence requirements as in the
result for a fixed distribution, now uniformly along the array. The first controls
sample variation in the score; the second controls the remainder in
(\ref{eq:coupled_remainder}). Part (iii) rules out a weak first stage, and part
(iv) imposes a Lindeberg condition on the oracle score in each row.

For independent observations, suppose that, for some \(\delta>0\),
\[
\sup_n\mathbb{E}_n\!\left[
\left|\frac{\xi_{n1}}{\sqrt{\Omega_n}}\right|^{2+\delta}
\right]<\infty.
\]
Together with \(\Omega_n\to\Omega_\infty\in(0,\infty)\), this bound implies the
displayed Lindeberg condition \citep[see][Section~27]{billingsley1995probability}.

\begin{theorem}[Triangular array asymptotic normality]\label{thm:tri}
Under Assumption~\ref{ass:tri},
\[
\sqrt n\bigl(\hat\theta-\theta_n\bigr)
=
D_n^{-1}\,n^{-1/2}\sum_{i=1}^n\psi_n(W_{ni};\theta_n,\gamma_n,\alpha_n)+o_p(1),
\]
and consequently
\[
\frac{\sqrt n(\hat\theta-\theta_n)}{\sqrt{\Omega_n/D_n^2}}\Rightarrow N(0,1),
\]
where $\Omega_n=\mathrm{Var}_n[\psi_n(W;\theta_n,\gamma_n,\alpha_n)]$.
\end{theorem}

Feasible studentization requires uniform versions of the plug-in variance
conditions in Corollary~\ref{coro}. In particular, suppose
\[
\frac1n\sum_{i=1}^n
\left\{\hat\psi_i(\hat\theta)
-\psi_n(W_{ni};\theta_n,\gamma_n,\alpha_n)\right\}^2=o_p(1)
\]
and suppose the sample second moment of the oracle score consistently estimates
$\Omega_n$. Then
\(\widehat\Omega_{\mathrm{iid}}-\Omega_n=o_p(1)\) and
\(\widehat Q+D_n=o_p(1)\). Hence
\[
\frac{\sqrt n(\hat\theta-\theta_n)}{\sqrt{\widehat V_{\mathrm{iid}}}}
\Rightarrow N(0,1).
\]
For clustered data, analogous uniform conditions must hold for the cluster
score sums and the estimated score errors.

Theorem~\ref{thm:tri} assumes convergence rates for the four regressions.
Corollary~\ref{cor:series} derives such rates for series estimators.

\begin{corollary}[Series first stages under increasing dimension]
\label{cor:series}
Suppose each of the four regression functions is estimated by a series regression.
For regression $r\in\{g_1,g_2,m_1,m_2\}$, let $b_{r,n}(V_r)$ be its normalized
basis and let $K_{r,n}\leq K_n$ be its dimension. Let $B_{r,l}$ contain the
basis observations in training fold $I_l^c$. Assume the following conditions
hold uniformly over regressions and folds:
\begin{enumerate}[label=(\roman*),leftmargin=*,itemsep=0.3ex,topsep=0.3ex]
\item The eigenvalues of
$\mathbb{E}_n[b_{r,n}(V_r)b_{r,n}(V_r)']$ lie in $[c,C]$, and the sample Gram
matrices satisfy
\[
P_n\!\left(
c\leq
\lambda_{\min}\!\left\{\frac{B_{r,l}'B_{r,l}}{|I_l^c|}\right\}
\leq
\lambda_{\max}\!\left\{\frac{B_{r,l}'B_{r,l}}{|I_l^c|}\right\}
\leq C
\ \text{for every }r,l
\right)\to1.
\]
\item The maximum leverage satisfies
\begin{equation}
\max_{r,l}\max_{i\in I_l^c}
 b_{r,n}(V_{ri})'(B_{r,l}'B_{r,l})^{-1}b_{r,n}(V_{ri})
=o_p(1).
\label{eq:series_leverage}
\end{equation}
\item The regression disturbances have uniformly bounded conditional second
moments.
\item The approximation errors are bounded by $b_{K_n}$ in $L^2$.
\item Examiner shares satisfy
\[
\frac{c}{J_n}\leq P_n(Z=j)\leq\frac{C}{J_n},
\qquad
\frac{n\pi_{\min,n}}{\log J_n}\to\infty,
\quad
\pi_{\min,n}=\min_j P_n(Z=j).
\]
\item If a basis contains interactions between examiner indicators and
covariates, the Gram and leverage conditions hold for the complete interacted
basis.
\end{enumerate}
Then
\[
r_{g,n},r_{m,n}
=O_p\!\left(\sqrt{\frac{K_n}{n}}+b_{K_n}\right),
\]
and Assumption~\ref{ass:tri}(ii) holds when
\[
K_n=o(\sqrt n),
\qquad
b_{K_n}=o(n^{-1/4}).
\]
\end{corollary}

With unrestricted examiner indicators, \(K_n\geq J_n\), so these sufficient
conditions for unregularized series require \(J_n=o(\sqrt n)\).

The rate calculation in Appendix~\ref{app:proofs} uses the Gram matrix
and leverage conditions for the complete basis to control the product
remainder under $K_n=o(\sqrt n)$. For normalized examiner indicators,
condition (v) ensures uniform cell occupancy and vanishing leverage.
Adding interactions with covariates, or other terms that link observations
across cells, requires checking the Gram matrix and leverage for the
enlarged basis as well. A dimension bound alone is insufficient.

There may be many examiner coefficients to estimate. The baseline sample
in \citet{bhuller}, for example, contains about 33{,}500 cases and 500 judges,
with $J/\sqrt n\approx2.7$. Although a finite-sample ratio cannot establish
whether an asymptotic sequence satisfies the corollary, this example shows
why the dimension restriction may be demanding in practice. Pooling
examiner effects or imposing other structure could improve the convergence
rate, subject to an analysis of both the estimation error and the
approximation error it introduces.

One option is partial penalization, which leaves examiner main effects
unrestricted and imposes sparsity on the remaining coefficients. To describe the rate, split the regressors
for each $r\in\{g_1,g_2,m_1,m_2\}$ into $q_{r,n}$ unpenalized terms in
$U_{r,n}$ and $d_{r,n}$ penalized terms in $H_{r,n}$, at most $s_{r,n}$ of
which are relevant. The $J_n$ examiner main effects can be included in
$U_{r,n}$ for the long regressions $g_1$ and $m_1$. For $g_2$ and $m_2$,
which condition on $X$ alone, this block contains only functions of $X$.
Write
\[
q_n=\max_r q_{r,n},\qquad d_n=\max_r d_{r,n},\qquad
s_n=\max_r s_{r,n},
\]
and let $M_{U_{r,n}}$ project off the span of $U_{r,n}$.

Suppose $H_{r,n}'M_{U_{r,n}}H_{r,n}/n$ satisfies a restricted eigenvalue
bound on the usual LASSO cones of order $s_n$, the maximum leverage of
$U_{r,n}$ vanishes, and the $L^2$ approximation error is bounded by $a_n$,
uniformly over regressions and folds. For the long regressions, also suppose
examiner shares lie between $c/J_n$ and $C/J_n$. Choose penalties of order
$\sqrt{\log(d_n)/n}$ that dominate the relevant empirical scores uniformly
over folds. Under the preceding restricted eigenvalue, leverage,
approximation, and score conditions, partially penalized LASSO or Post-LASSO has
the prediction rate
\[
r_{g,n},r_{m,n}
=O_p\!\left(\sqrt{\frac{q_n+s_n\log d_n}{n}}+a_n\right).
\]
Assumption~\ref{ass:tri}(ii) therefore holds when
\[
q_n+s_n\log d_n=o(\sqrt n),
\qquad
a_n=o(n^{-1/4}).
\]
If the long regressions contain all examiner indicators, then $q_n\geq J_n$.
With the examiner effects unpenalized, changing the omitted category leaves
the fitted values unchanged. Estimating all of these coefficients still contributes
$\sqrt{J_n/n}$ to the rate. Partial pooling or shrinkage may reduce that term,
with any gain assessed against the additional approximation error.

The calculation for partially penalized ridge in Appendix~\ref{app:primitive}
does not impose approximate sparsity. Its rate consequently depends on all
$d_n$ penalized terms, replacing the $s_n\log d_n$ dependence above.

Dependence among cases handled by the same examiner calls for a central limit
theorem for the cluster sums
$\sum_{i\in C_g}\psi_n(W_{ni};\theta_n,\gamma_n,\alpha_n)$, together with
negligible error from the four fitted regressions after aggregation within
clusters. The proof for independent observations does not apply when cases
handled by the same examiner are correlated, and assigning those cases to
different folds can induce dependence between the training and evaluation
samples. If examiner clusters are independent, assigning whole examiners to
folds avoids dependence between the samples, but then a regression with
unrestricted examiner indicators has no training observations from which to
estimate the coefficients for examiners in the evaluation fold.

A regression on observed examiner types or characteristics can be evaluated
for examiners absent from the training sample. Suppose $Z$ contains such
characteristics of fixed dimension, examiner identity defines independent
clusters, and whole clusters are assigned to folds. Primitive sufficient
conditions are $G_n\to\infty$, uniformly bounded cluster size, and the mean
square and product conditions for regressions trained on the remaining
clusters. A Cauchy--Schwarz inequality converts the individual $L^2$ bounds
to bounds on cluster score sums, giving consistency of
$\widehat V_{\mathrm{cl}}$ under arbitrary dependence within a cluster.
The regressions in this argument use observed examiner characteristics
and do not include unrestricted examiner indicators.

For an unrestricted saturated estimator with $J_n/n\to\kappa>0$, cells
defined by examiner and covariates can remain sparse enough that nuisance
estimation error has a first-order effect on the IV estimator. The regression convergence conditions
then fail even if $D_n$ stays bounded away from zero. Uniform inference
requires methods for many instruments, many covariates, or fixed effects
estimated by leave-out procedures
\citep{cattaneo2018two,cattaneo2018alternative,mikusheva2022inference,chao2023,kline2020leaveout}.

Accurate estimation of the treatment regressions does not by itself ensure a
strong first stage. When examiner propensities become similar conditional
on $X$, $D_n=\mathbb{E}_n[\gamma_n^2]$ can converge to zero even if the
regressions are estimated consistently. Conversely, sparse combinations of
examiner assignment and covariates can make the regression estimates
unstable while $D_n$ remains bounded away from zero.
Theorem~\ref{thm:tri} therefore requires regression consistency as well as
instrument strength bounded away from zero.

\section{Monte Carlo Evidence}\label{sec:montecarlo}

Design 1 has an additive treatment propensity for which both the linear UJIVE
first stage and the sieve regressions are correctly specified.
In Design 2, examiner effects vary with covariates, separating the IV estimands
based on conditional expectations and linear projections. In Design 3, the
fitted sieve omits part of a nonlinear treatment propensity. These designs
distinguish error from estimating the four regressions, differences between
the two population instruments, and error from approximating the conditional
expectations.

The comparison includes the orthogonal estimator, a sieve plug-in estimator,
and linear UJIVE\@. The plug-in uses the same cross-fitted instrument and omits the influence
function adjustment:
\[
\widehat\theta_{\mathrm{plug}}
=
\frac{\sum_{l=1}^L\sum_{i\in I_l}
Y_i\widehat\gamma_l(X_i,Z_i)}
{\sum_{l=1}^L\sum_{i\in I_l}
T_i\widehat\gamma_l(X_i,Z_i)}.
\]
The flexible sieve contains all four regression functions in Designs 1 and 2.
Within a replication, every estimator uses the same simulated sample.

\paragraph{Design 1: correctly specified additive first stage.} The sample has
$n=1{,}800$ observations assigned with equal probability to $J=18$ examiners.
The coordinates of $X=(X_1,\ldots,X_6)$ are independent uniform draws on
$[-1,1]$. Examiner leniency parameters $\ell_j$ are equally spaced on
$[-1.25,1.25]$. Treatment satisfies
\[
T\mid X,Z=j\sim\operatorname{Bernoulli}\{p(X,j)\},\qquad
p(X,j)=0.50+0.05X_1+0.30\ell_j.
\]
Outcomes satisfy
\[
Y=\mu(X)+(1+0.30X_1)T+0.20\varepsilon,\qquad \varepsilon\sim N(0,1),
\]
where
\[
\mu(X)=0.6+0.50\sin(0.8X_1)-0.30X_2+0.35X_3X_4-0.30X_5^2+0.25\,1\{X_2>0\}+0.20X_6^2.
\]
The sieve is correctly specified for both the treatment and outcome regressions.
Because the propensity is additively separable in $X$ and $\ell_j$, the oracle
instrument is $\gamma_0=0.30\ell_j$ and is independent of $X$. Therefore
\[
\theta_0
=\frac{\mathbb{E}[(1+0.30X_1)\gamma_0^2]}{\mathbb{E}[\gamma_0^2]}
=\mathbb{E}[1+0.30X_1]
=1,
\]
where the last equality uses $\mathbb{E}[X_1]=0$.

\paragraph{Design 2: interactions between examiners and covariates.}
The design has \(n=2{,}400\) observations and \(J=8\) equally likely examiners,
with the same covariate and \(\ell_j\) distributions as in Design 1.
Let \(D_2=1\{X_2>0\}\), and set
\[
p(X,j)=0.50+0.08X_1+\ell_j(0.16+0.16D_2),
\qquad
Y=\mu(X)+(1+0.60D_2)T+0.20\varepsilon.
\]
To span all four regression functions, I include examiner indicators and their
interactions with \(D_2\) in the \((X,Z)\) regressions, and include
\(X_1D_2\) in the \(X\) regressions. Independent examiner assignment and
symmetry of the leniency grid then imply the oracle estimand
\[
\theta_0
=\frac{\mathbb{E}[(1+0.60D_2)(0.16+0.16D_2)^2]}
{\mathbb{E}[(0.16+0.16D_2)^2]}
=1.48.
\]
Omitting the interactions between examiner indicators and \(D_2\), as the
additive linear UJIVE first stage does, changes the population ratio to
\[
\theta^L
=\frac{\mathbb{E}[(1+0.60D_2)(0.16+0.16D_2)]}
{\mathbb{E}[0.16+0.16D_2]}
=1.40.
\]
The sieve estimator and additive linear UJIVE thus estimate different ratios:
the former is based on the conditional expectations, the latter on their two
linear projections.

\paragraph{Design 3: nonlinear first stage outside the sieve space.} I return
to $n=1{,}800$ observations and $J=18$ examiners, with independent standard
normal coordinates of $X$ and leniency parameters $\ell_j$ equally spaced on
$[-1.5625,1.5625]$. Treatment satisfies
\[
\begin{aligned}
T\mid X,Z=j&\sim\operatorname{Bernoulli}\!\left(\Lambda[\eta(X,j)]\right),\\
\eta(X,j)&=\ell_j\{1+0.95X_1-0.75\,1\{X_2>0\}\}+1.10\sin(0.8X_1)\\
&\quad-0.25X_2+0.50X_3X_4-0.40X_5^2+0.25X_6,
\end{aligned}
\]
where $\Lambda(u)=1/(1+e^{-u})$. Outcomes satisfy
\[
Y=\mu_s(X)+\{1+0.40X_1+0.25\,1\{X_2>0\}\}T+0.25\varepsilon,
\]
with $\mu_s(X)=0.5+0.35X_1-0.20X_2+0.25X_3X_4-0.25X_5+0.15X_6^2$.
The logistic propensity lies outside the linear span of the fitted sieve,
although the basis includes interactions between examiner indicators and
covariates. Numerical integration using a low-discrepancy sequence yields
$\theta_0=1.36871$. Because the fixed sieve is misspecified, I do not report
Wald coverage for the feasible estimators.

\paragraph{Estimation details.} The treatment and outcome regressions
use the same basis for a given conditioning set. Let
\(D_2=1\{X_2>0\}\) and \(E_j=1\{Z=j\}\), with examiner 1 omitted. In Designs 1
and 3, the basis for regressions conditional on \(X\) is
\[
\begin{aligned}
b_X={}&(1,X_1,\ldots,X_6,X_1^2,X_2^2,X_5^2,X_6^2,
X_1X_2,X_1X_3,X_3X_4,\\
&\sin(0.8X_1),D_2)',
\end{aligned}
\]
and the basis for regressions conditional on \((X,Z)\) is
\[
b_{XZ}=(b_X',E_2,\ldots,E_J,E_2X_1,\ldots,E_JX_1,
E_2D_2,\ldots,E_JD_2)'.
\]
Design 2 uses
\[
b_X=(1,X_1,X_2,X_3X_4,X_5^2,X_6^2,\sin(0.8X_1),D_2,X_1D_2)'
\]
and appends \((E_2,\ldots,E_J,E_2D_2,\ldots,E_JD_2)\) to form \(b_{XZ}\).
These bases span all four regression functions in Designs 1 and 2. Design 3 places
the propensity outside the linear span. Treatment predictions are clipped to
\([10^{-6},1-10^{-6}]\), and the orthogonal estimate solves the affine estimating
equation in (\ref{eq:closed_form_score}).

I compute UJIVE using the leave-one-out construction with covariates in
\citet{kolesarcowles}. Its long first-stage matrix $B^L$ contains the included
covariates and examiner indicators, while the short matrix $B^S$ contains the
covariates alone. For each $r\in\{L,S\}$, let $\hat e_{ri}$ be the residual
from fitting treatment on the full sample and let $h_{rii}$ be the $i$th
diagonal element of that regression's hat matrix. The leave-one-out
fitted treatment is computed by the identity for linear regression
\[
\widehat T_{r,-i}
=T_i-\frac{\hat e_{ri}}{1-h_{rii}}.
\]
The excluded instrument is
$\widehat\gamma_i^{\mathrm{UJIVE}}=\widehat T_{L,-i}-\widehat T_{S,-i}$,
and the reported estimate is
\[
\widehat\theta_{\mathrm{UJIVE}}
=
\frac{\sum_i Y_i\widehat\gamma_i^{\mathrm{UJIVE}}}
{\sum_i T_i\widehat\gamma_i^{\mathrm{UJIVE}}}.
\]
For numerical stability, I add $10^{-6}I$ to each first-stage Gram
matrix.\footnote{In 1{,}000 replications of Design 1, omitting this adjustment
changed UJIVE by at most $1.1\times10^{-8}$ and left every reported table
entry unchanged.}

In Design 2, I fit four ridge regressions, penalizing the covariates and their
interactions with examiner indicators but not the intercept or examiner main
effects. Within each outer training sample $\mathcal T_l$, I center and scale
the penalized regressors to unit variance. With $P_l$ the diagonal selector
for penalized coefficients, the ridge regression for response $R$ solves
\[
\widehat\beta_l(\lambda)
=\mathop{\arg\min}_{\beta}
\left\{
\frac{1}{|\mathcal T_l|}
\sum_{i\in\mathcal T_l}(R_i-b_i'\beta)^2
+\lambda\|P_l\beta\|_2^2
\right\}.
\]
Appendix~\ref{app:primitive} uses the same normalization.

I form five independently randomized folds in Design 2, following
Theorem~\ref{thm1}. Within each outer training sample, exact leave-one-out
cross-validation selects separate treatment and outcome penalties from
\(\{10^{-8},10^{-6},10^{-4},10^{-3},10^{-2},10^{-1}\}\). I use the selected
treatment penalty in both treatment regressions and the selected outcome
penalty in both outcome regressions. Neither choice uses observations from
the outer evaluation fold. Appendix Tables~\ref{tab:mc_tuning}
and~\ref{tab:mc_prediction} compare this partially penalized specification
with ridge regressions that penalize every coefficient.

Designs 1 and 3 use fixed nuisance specifications with a penalty close to
zero: $10^{-6}$ is added to the unnormalized Gram matrix. I use five random
folds in Design 1 and three in Design 3, with 1{,}000 replications in each
design. Evaluating the score at the four true regressions removes both
estimation and approximation error, permitting a comparison of its sampling
variance with the efficiency bound.

Table~\ref{tab:mc} reports i.i.d.\ coverage and average standard errors for
the feasible orthogonal estimator in the two correctly specified designs,
and for the oracle score in all three. UJIVE is included for comparison of
point estimates. Its variance under heterogeneous effects and many
instruments is not consistently estimated by the conventional sandwich
formula, which treats the fitted instrument as fixed
\citep{evdokimovkolesar2018}. UJIVE standard errors and coverage are
therefore omitted.
\begin{table}[H]
\centering
\small
\setlength{\tabcolsep}{5.1pt}
\caption{Monte Carlo results}
\label{tab:mc}
\begin{tabular}{llrrrrrr}
\toprule
Design & Estimator & Mean & Bias & RMSE & SD & Coverage & Avg.\ SE \\
\midrule
Additive first stage & Orthogonal & 1.000 & 0.000 & 0.034 & 0.034 & 0.953 & 0.033 \\
 & Oracle score & 1.001 & 0.001 & 0.026 & 0.026 & 0.957 & 0.026 \\
 & Sieve plug-in & 0.999 & $-$0.001 & 0.033 & 0.033 & --- & --- \\
 & Linear UJIVE & 1.001 & 0.001 & 0.029 & 0.029 & --- & --- \\
\addlinespace
Examiner interactions & Orthogonal & 1.490 & 0.010 & 0.033 & 0.031 & 0.943 & 0.033 \\
 & Oracle score & 1.480 & 0.000 & 0.029 & 0.029 & 0.960 & 0.030 \\
 & Sieve plug-in & 1.478 & $-$0.002 & 0.031 & 0.031 & --- & --- \\
 & Linear UJIVE & 1.401 & $-$0.079 & 0.086 & 0.034 & --- & --- \\
\addlinespace
Misspecified sieve & Orthogonal & 1.441 & 0.073 & 0.111 & 0.084 & --- & --- \\
 & Oracle score & 1.369 & 0.000 & 0.056 & 0.056 & 0.951 & 0.055 \\
 & Sieve plug-in & 1.253 & $-$0.115 & 0.151 & 0.098 & --- & --- \\
 & Linear UJIVE & 1.586 & 0.217 & 0.249 & 0.121 & --- & --- \\
\bottomrule
\end{tabular}
\begin{minipage}{0.96\textwidth}
\footnotesize\emph{Notes:} The oracle estimands are 1, 1.48, and 1.36871.
Each design uses 1{,}000 replications. Coverage and average standard errors are
based on the i.i.d.\ variance implied by the influence function for that row. The
infeasible ``Oracle score'' estimates solve the sample estimating equation
using the four true regression functions. UJIVE is reported only for point
estimation because its standard errors and coverage require a separate
many-instrument variance procedure, not implemented in these simulations.
\end{minipage}
\end{table}

The additive design favors the parsimonious UJIVE first stage, which already
contains the examiner component of the treatment propensity. All three feasible
estimators have negligible bias, but UJIVE is more precise: its RMSE is 0.029,
compared with 0.033 for the sieve plug-in and 0.034 for the orthogonal
estimator. Coverage for the orthogonal estimator with five folds is 0.953
(Monte Carlo standard error 0.007). The average estimated standard error, 0.033, is close
to the empirical standard deviation, 0.034. Evaluating the score at the true
regressions reduces RMSE to 0.026, with coverage of 0.957.

In the additive design, the efficiency bound can be calculated in closed form.
Let
$g=0.30\ell_Z$, $h=0.30X_1$, $p=p(X,Z)$, and
$\bar p=\mathbb{E}[p(X,Z)\mid X]$. At $\theta_0=1$, the oracle score simplifies
to
\[
\psi=g\{h(2T-p-\bar p)+0.20\varepsilon\}.
\]
Integrating this expression gives $D=0.05239$ and
$V_{\mathrm{eff}}=1.17137$, or a standard error of 0.02551 at $n=1{,}800$.
The oracle score's Monte Carlo standard deviation of 0.026 is close to this
value. The orthogonal estimator and UJIVE are more variable at this sample size, with
Monte Carlo variance equal to 1.81 times $V_{\mathrm{eff}}/n$ for the
orthogonal estimator and 1.30 times the bound for UJIVE\@. Thus UJIVE has the
smaller finite-sample RMSE despite the orthogonal estimator's asymptotic
efficiency.

In Design 2, linear UJIVE is centered at the value implied by its population
projections: its Monte Carlo mean is 1.401, compared with the oracle estimand
of 1.48. The difference of 0.079 amounts to 5.3 percent of the oracle estimand
and 2.3 times UJIVE's Monte Carlo standard deviation. Since the sieve spans
all four regression functions, it permits a separate assessment of nuisance
estimation error. Relative to the oracle ratio, the orthogonal estimator has
bias 0.010 and RMSE 0.033, and the sieve plug-in has bias $-0.002$ and RMSE
0.031. Appendix Tables~\ref{tab:mc_tuning} and~\ref{tab:mc_prediction} report
the tuning results, out-of-fold prediction errors, and sensitivity to fixed
penalties.

Increasing the sample size brings the feasible estimator closer to its oracle
counterpart (Appendix Table~\ref{tab:mc_n_sequence}). In the interaction
design, the increase from $n=1{,}200$ to $n=9{,}600$ reduces feasible bias
from 0.029 to 0.002 and the ratio of feasible to oracle variance from 1.37 to
1.04.

Misspecification of the sieve leaves bias even after the orthogonal
correction. In Design 3, the correction lowers RMSE from 0.151 for the
plug-in estimator to 0.111, compared with 0.249 for UJIVE, but the remaining
bias is 0.073. The fitted regressions only approximate the true conditional
expectations, and orthogonality does not eliminate this approximation error. With
the true regressions, the oracle score has zero bias, RMSE 0.056, and coverage
0.951. Appendix Tables~\ref{tab:mc_j}--\ref{tab:mc_linear} examine the
number of examiners, fold counts, weak identification, tuning, and correct
linear specification.

\section{Empirical Application: Patent Grants and Subsequent Patent Applications}
\label{sec:patent_application}

I use US patent applications to compare estimates of the effect of a grant on
subsequent filing by the same inventors. Keeping the sample, treatment, and
outcome fixed, I compare instruments constructed from linear projections and
flexible estimates of the conditional expectations.

A patent may raise the return to further invention or facilitate financing,
licensing, and commercialization. \citet{farremensa2020patent} use examiner
assignment to estimate the effect of a startup's first patent on firm growth and
subsequent innovation. \citet{goldsmithpinkham2025leniency} reexamine that application using UJIVE and
related diagnostics. I compare the instrument defined by conditional expectations with
the instruments obtained from linear UJIVE specifications.

\subsection{Institutional Setting and Data}

The USPTO places an application in an art unit based on subject matter and then
assigns it to an examiner within that unit. I use Art Unit 2872, which covers
eye examination, technologies for correcting vision, and optical systems and
elements \citep{uspto2022artunits}. I chose the unit for its sample size, examiner support, first-stage relevance,
and covariate coverage, before constructing the outcome. Restricting the
sample to this unit still leaves variation in finer technology and inventor
characteristics, as well as differences in treatment propensities across
examiners. The estimate pertains to this art unit and its assignment process.

I restrict the sample to applications filed during 2001--2017, first docketed
before October 1, 2020, and with a recorded final disposition by January 1,
2020. The docketing restriction excludes applications assigned under the
automated system introduced in October 2020. \citet{pairolero2026routing}
show that this system increased the technical similarity between new
assignments and examiners' prior work and effectively ended the earlier
quasi-random assignment regime. The disposition cutoff permits a complete
three-year filing window.

I link the 2022 Patent Examination Research Dataset to PatentsView pregrant
publication records \citep{graham2018patex,uspto2026patentsview} and apply the
sample restrictions in Appendix~\ref{app:patent_details}. The sample contains
12,919 applications handled by 77 examiners and naming 26,677 inventors,
each of whom appears on only one focal application. The grant rate is 0.706.

Examiner assignment is measured using the examiner recorded by PatEx at the
time of the data extract. Reassignment during prosecution may therefore
cause the recorded examiner to differ from the examiner who first received
the application. If reassignment depends on application quality after
conditioning on the observed characteristics, the recorded assignment would
violate the conditional independence assumption.

Treatment $T_i$ indicates a patent grant on the focal application. I follow
the inventors for three years after disposition and set $Y_i=1$ if at least
one files another application within that period that is subsequently
published. This measure of subsequent filing includes continuations,
divisionals, and other filings in the same continuity family, as well as
new inventions. Its sample mean is 0.711. For granted applications, follow-up
begins on the patent issue date; for other applications, it begins on the
recorded final status date. An examiner who affects disposition timing
therefore also affects the start of the outcome window.

Although the focal applications have no inventors in common, their inventors
may subsequently file together. I link focal applications whose inventors
share a later filing and use the graph's 11,029 connected components as
clusters for variance estimation. Applications in the same component are
assigned to the same cross-fitting fold. Appendix~\ref{app:patent_details}
reports the component sizes.

\subsection{Assignment and Estimation}

Examiners specialized in particular technologies even before automated
assignment. I therefore condition on detailed application and applicant
characteristics, rather than art unit and filing year alone. The predetermined
covariates $X_i$ include docket and filing dates, USPC class and subclass,
title terms, foreign priority, continuation status, applicant characteristics,
inventor location, and the number of inventors. Conditional on $X_i$, I assume
that examiner assignment is independent of potential grant decisions and
potential outcomes.

The exclusion restriction requires examiner assignment to affect subsequent
filing only through the grant decision. Examiners may, however, influence
prosecution duration, claim scope, or the advice applicants receive during
examination. Any effect of these aspects of examination on subsequent filing
other than through the grant decision would violate exclusion. Under
exclusion, conditional assignment, relevance, and average monotonicity,
Proposition~\ref{prop:identification} expresses the estimand as a positively
weighted average effect of a patent grant among applications whose grant
status responds to examiner assignment.

The detailed technology classifications make saturation impractical. The
sample contains 77 USPC classes and 1,296 combinations of class and subclass.
After also conditioning on docket year, half of the observations belong to
cells with fewer than three examiners. A saturated regression would divide
these cells further by examiner assignment, leaving many cells with few
observations and little or no variation for estimating treatment propensities.

An unsaturated linear specification avoids estimating a separate propensity
in every cell by restricting how examiner propensities vary with technology
and the other application characteristics in $X$. If these restrictions
produce different approximation errors in the two linear projections,
Equation~(\ref{eq:projection_error_gap}) implies that the population UJIVE
instrument differs from $\gamma_0$.

For each conditional expectation, I fit a series regression with a ridge penalty.
In the regressions on $(X,Z)$, examiner main effects are unpenalized and
examiner assignment is interacted with docket cohort, technology, application
characteristics, and title terms. The regressions on $X$ contain functions of
the covariates and titles alone.

Estimation uses grouped five-fold cross-fitting, repeated over eleven
partitions. Within each training sample, a validation split selects a penalty
for each regression, leaving the evaluation fold unused for tuning. I combine estimates and
variances using the median rule of \citet{chernozhukov2018double}. The primary
variance estimate clusters by connected component, and examiner clustering is a
sensitivity calculation. I include UJIVE estimates for comparison but omit
their standard errors because a conventional sandwich formula does not account
for estimation of the leave-one-out instrument \citep{evdokimovkolesar2018}.
Appendix~\ref{app:patent_details} details the complete regression specifications,
tuning procedure, variance formula, and inference discussion.

\subsection{Results}

Table~\ref{tab:patent_application} reports two linear UJIVE estimates and one
estimate based on the flexible ridge regressions.

\begin{table}[H]
\centering
\scriptsize
\setlength{\tabcolsep}{3.0pt}
\caption{Patent grants and subsequent patent applications}
\label{tab:patent_application}
\begin{tabular}{@{}llrrrrr@{}}
\toprule
Estimator & Specification & Estimate & Clustered SE & $\widehat D$
& $\mathbb E_n[\widehat\gamma^2]$ & MSE($T$) \\
\midrule
Linear UJIVE & Docket year & 0.088 & --- & 0.0259 & 0.0271 & 0.1804 \\
Linear UJIVE & Additive controls & 0.129 & --- & 0.0245 & 0.0257 & 0.1791 \\
Orthogonal & Flexible & 0.042 & 0.036 & 0.0199 & 0.0246 & 0.1774 \\
\bottomrule
\end{tabular}
\begin{minipage}{0.98\textwidth}
\footnotesize\emph{Notes:} The outcome equals one if an inventor on the focal
application files, within three years after disposition, another application
that is later published. There are 12,919 focal applications and 77 examiners,
with no inventor appearing on more than one focal application. UJIVE uses
leave-one-out linear first stages and is reported for comparison of point
estimates. Appendix~\ref{app:patent_details} describes the specifications and
the reason for omitting UJIVE standard errors. The orthogonal estimate is the
median over 11 grouped five-fold partitions. Its standard error incorporates
variation across partitions and clusters by connected components of
applications linked through a common later filing. For UJIVE, $\widehat D$
equals $\mathbb E_n[T_i\widehat\gamma_i^{\mathrm{UJIVE}}]$.
For the orthogonal estimator, it is the corrected denominator in
Equation~(\ref{eq:closed_form_theta}). MSE($T$) is calculated from the fitted
treatment regression on $(X,Z)$. Treatment MSE is also reported as the
median over the eleven partitions.
\end{minipage}
\end{table}

Adding the assignment covariates linearly raises the UJIVE estimate from
0.088, with docket year indicators alone, to 0.129. The orthogonal estimate
using flexible ridge regressions is 0.042. Yet their regressions of treatment on
both the covariates and examiner assignment predict almost equally well: the
mean squared errors on the same folds are
0.1804, 0.1791, and 0.1774, respectively, a difference of at most 1.7 percent.
The point estimates differ by as much as 0.087.

The UJIVE comparison changes the controls included in the linear projections.
The flexible regressions for $\mathbb E[T\mid X,Z]$ and
$\mathbb E[T\mid X]$ additionally allow examiner propensities to vary with
docket cohort, technology, application characteristics, and title terms. Similar
treatment prediction errors can thus accompany different fitted instruments
when examiner propensities vary with application characteristics. A similar comparison appears in Monte Carlo
Design 2, where the projection and oracle instruments are known to differ.

Estimation and regularization also affect the denominator: for the flexible
specification, its corrected value is 0.0199, about 19 percent below the
sample mean of $\widehat\gamma_i^2$, which is 0.0246.

The flexible estimate is imprecise. Its standard error is 0.036 when clustered
by connected component, giving a 95 percent confidence interval of
$[-0.029,0.113]$, and rises to 0.046 under examiner clustering.

\section{Conclusion}

\begingroup
\widowpenalty=10000
In examiner IV designs, the covariates needed to make examiners' treatment
propensities conditionally ignorable, and hence valid as instruments under
exclusion and relevance, can make a saturated first stage impractical. To
solve this challenge, I have provided a locally robust estimator that allows
nonparametric and regularized covariate
adjustment while accounting for estimation of both conditional expectations
defining the instrument. The influence function adjustments have closed-form
expressions and can be estimated using four regressions: namely, the conditional
expectations of treatment and outcome given both covariates and examiner
assignment, and given covariates only.

Validity of inference depends on the accuracy of these regressions.
With cross-fitting, the remaining bias consists of products of regression
errors. Under the regularity and rate conditions in Section~3, the estimator
is asymptotically normal and attains the semiparametric efficiency bound for
i.i.d.\ observations from a fixed distribution. I also establish conditions
for inference when the number of examiners and regressors increases with
sample size. These conditions require sufficiently accurate regression
estimates and instrument strength bounded away from zero.

The choice of regression specification also matters for the causal parameter.
Under conditional assignment, exclusion, relevance, and average monotonicity,
the ratio based on conditional expectations is a positively weighted average
of treatment effects. Replacing those expectations with linear projections
can change that ratio even when estimation error is small, an observation
borne out by simulation evidence. In the application to US patent examiners, the treatment
prediction errors differ by at most 1.7 percent, yet the IV estimates range
from 0.042 to 0.129. Similar predictive performance can therefore conceal
economically important differences between fitted instruments.

\endgroup

% Bibstyle aea.bst version 2009.05.20

\clearpage
\appendix

\section{Additional Monte Carlo Evidence}\label{app:mc}

The additional experiments vary sample size, the number of examiners, the number of
cross-fitting folds, ridge tuning, and instrument strength. The final design is
a correctly specified linear benchmark.
Throughout the simulations, treatment is assigned using an independent uniform draw, and
$\varepsilon$ is independent of that draw and of $(X,Z)$.

\begin{table}[H]
\centering
\scriptsize
\setlength{\tabcolsep}{3.0pt}
\caption{Sample size sequences}
\label{tab:mc_n_sequence}
\begin{tabular}{lrrrrrrrrr}
\toprule
Design & $n$ & Reps. & Bias & $\sqrt n$ Bias & SD & Avg.\ SE & Coverage &
Coverage, oracle score & Var.\ ratio \\
\midrule
Additive & 1{,}200 & 1{,}000 & $-$0.0039 & $-$0.135 & 0.0511 & 0.0474 & 0.945 & 0.958 & 2.44 \\
 & 2{,}400 & 1{,}000 & $-$0.0009 & $-$0.044 & 0.0265 & 0.0268 & 0.958 & 0.946 & 1.40 \\
 & 4{,}800 & 1{,}000 & $-$0.0003 & $-$0.020 & 0.0175 & 0.0171 & 0.946 & 0.947 & 1.19 \\
 & 9{,}600 & 1{,}000 & $-$0.0005 & $-$0.052 & 0.0113 & 0.0116 & 0.958 & 0.958 & 1.10 \\
\addlinespace
Examiner interactions & 1{,}200 & 500 & 0.0295 & 1.020 & 0.0520 & 0.0526 & 0.928 & 0.936 & 1.37 \\
 & 2{,}400 & 1{,}000 & 0.0098 & 0.482 & 0.0311 & 0.0325 & 0.943 & 0.960 & 1.13 \\
 & 4{,}800 & 500 & 0.0045 & 0.309 & 0.0207 & 0.0218 & 0.948 & 0.966 & 1.10 \\
 & 9{,}600 & 500 & 0.0024 & 0.236 & 0.0145 & 0.0150 & 0.956 & 0.956 & 1.04 \\
\bottomrule
\end{tabular}
\begin{minipage}{0.98\textwidth}
\footnotesize\emph{Notes:} The table reports the feasible orthogonal estimator.
The second coverage column uses the infeasible estimator obtained by solving
the sample estimating equation with the four true regression functions. The
variance ratio is the empirical variance of the feasible estimator divided by
the empirical variance of this infeasible estimator. The interaction design uses
the partially penalized specification described in the main text. Its
$n=2{,}400$ row uses the 1{,}000-replication run reported in
Table~\ref{tab:mc}; the other interaction rows use 500 replications.
\end{minipage}
\end{table}

\begin{table}[H]
\centering
\small
\caption{Sensitivity to the number of examiners in the additive design}
\label{tab:mc_j}
\begin{tabular}{rrlrrrr}
\toprule
$J$ & Reps. & Estimator & Bias & RMSE & SD & Coverage \\
\midrule
8 & 1{,}000 & Orthogonal & 0.001 & 0.027 & 0.027 & 0.944 \\
8 & 1{,}000 & Sieve plug-in & $-$0.001 & 0.028 & 0.028 & --- \\
8 & 1{,}000 & Linear UJIVE & 0.001 & 0.027 & 0.027 & --- \\
\addlinespace
18 & 1{,}000 & Orthogonal & $-$0.002 & 0.038 & 0.038 & 0.938 \\
18 & 1{,}000 & Sieve plug-in & $-$0.004 & 0.038 & 0.038 & --- \\
18 & 1{,}000 & Linear UJIVE & 0.000 & 0.029 & 0.029 & --- \\
\addlinespace
50 & 1{,}000 & Orthogonal & $-$0.046 & 0.126 & 0.117 & 0.928 \\
50 & 1{,}000 & Sieve plug-in & $-$0.013 & 0.061 & 0.060 & --- \\
50 & 1{,}000 & Linear UJIVE & $-$0.002 & 0.032 & 0.032 & --- \\
\bottomrule
\end{tabular}
\begin{minipage}{0.94\textwidth}
\footnotesize\emph{Notes:} The true value is 1. Coverage is i.i.d.\ Wald
coverage for the orthogonal estimator. All rows use the same basis, penalty,
and three folds stratified by examiner. The \(J=50\) design has comparatively
many examiner terms. Corollary~\ref{cor:series} imposes a dimension
restriction on an asymptotic sequence, not a finite-sample cutoff.
\end{minipage}
\end{table}

Bias and RMSE rise sharply at $J=50$, when the sieve is large relative to the
sample. The deterioration accords with the dimension restriction in
Corollary~\ref{cor:series}, although this comparison cannot identify a
finite-sample threshold for $J/\sqrt n$. UJIVE remains
precise because its additive linear first stage is correctly specified.

\begin{table}[H]
\centering
\small
\caption{Sensitivity to the number of cross-fitting folds}
\label{tab:mc_folds}
\begin{tabular}{rrrrrrr}
\toprule
Folds & Reps. & Bias & RMSE & SD & Coverage & Avg.\ SE \\
\midrule
2 & 1{,}000 & $-$0.003 & 0.051 & 0.051 & 0.897 & 0.040 \\
3 & 1{,}000 & $-$0.001 & 0.039 & 0.039 & 0.928 & 0.035 \\
5 & 1{,}000 & 0.000 & 0.034 & 0.034 & 0.953 & 0.033 \\
\bottomrule
\end{tabular}
\begin{minipage}{0.92\textwidth}
\footnotesize\emph{Notes:} Correctly specified additive design, orthogonal
estimator, and i.i.d.\ variance based on the influence function. All runs use random folds.
The true value is 1.
\end{minipage}
\end{table}

With five folds, each regression is trained on more observations than with two. In this
comparison, dispersion falls and coverage moves closer to 0.95, with little
change in the point estimates. The experiment does not establish which
number of folds is optimal.

\begin{table}[H]
\centering
\small
\caption{Tuning sensitivity in Design 2}
\label{tab:mc_tuning}
\begin{tabular}{llrrrr}
\toprule
Tuning & Estimator & Bias & RMSE & SD & Coverage \\
\midrule
Partially penalized ridge (primary) & Orthogonal & 0.010 & 0.033 & 0.031 & 0.943 \\
 & Sieve plug-in & $-$0.002 & 0.031 & 0.031 & --- \\
All coefficients, common & Orthogonal & 0.007 & 0.032 & 0.031 & 0.939 \\
 & Sieve plug-in & $-$0.001 & 0.029 & 0.029 & --- \\
All coefficients, separate & Orthogonal & 0.008 & 0.032 & 0.031 & 0.940 \\
 & Sieve plug-in & 0.017 & 0.034 & 0.030 & --- \\
All coefficients, fixed $10^{-6}$ & Orthogonal & 0.009 & 0.034 & 0.033 & 0.928 \\
 & Sieve plug-in & 0.000 & 0.032 & 0.032 & --- \\
All coefficients, fixed $\lambda=50$ & Orthogonal & 0.018 & 0.034 & 0.029 & 0.842 \\
 & Sieve plug-in & 0.045 & 0.053 & 0.028 & --- \\
\bottomrule
\end{tabular}
\begin{minipage}{0.92\textwidth}
\footnotesize\emph{Notes:} Design 2 with 1{,}000 replications and five
independently randomized folds. The common specification selects one penalty for the
two treatment regressions and one for the two outcome regressions,
applying each penalty to both conditioning sets. Partially penalized ridge uses the normalized
objective in the main text. The other rows reproduce the original
unnormalized specification, which penalizes every coefficient. The specification
with separate penalties tunes all four regressions.
The specification with fixed penalty $\lambda=50$ is not selected by
cross-validation.
\end{minipage}
\end{table}

\begin{table}[H]
\centering
\small
\renewcommand{\arraystretch}{1.08}
\caption{Ridge penalties and out-of-fold prediction errors in Design 2}
\label{tab:mc_prediction}
\begin{tabular}{lrrrr}
\toprule
Tuning & Mean $\lambda_{g_1}$ & Mean $\lambda_{g_2}$ &
Mean $\lambda_{m_1}$ & Mean $\lambda_{m_2}$ \\
\midrule
Partially penalized ridge (normalized) & 0.0200 & 0.0200 & 0.0073 & 0.0073 \\
All coefficients, common & 0.991 & 0.991 & 0.841 & 0.841 \\
All coefficients, separate & 0.991 & 8.377 & 0.841 & 3.800 \\
All coefficients, fixed $10^{-6}$ & $10^{-6}$ & $10^{-6}$ & $10^{-6}$ & $10^{-6}$ \\
All coefficients, fixed $50$ & 50.000 & 50.000 & 50.000 & 50.000 \\
\bottomrule
\end{tabular}
\medskip

\begin{tabular}{lrrrr}
\toprule
& \multicolumn{2}{c}{Out-of-fold RMSE for $T$}
& \multicolumn{2}{c}{Out-of-fold RMSE for $Y$} \\
\cmidrule(lr){2-3}\cmidrule(lr){4-5}
Tuning & $(X,Z)$ & $X$ & $(X,Z)$ & $X$ \\
\midrule
Partially penalized ridge (primary) & 0.456 & 0.499 & 0.624 & 0.695 \\
All coefficients, common & 0.456 & 0.499 & 0.624 & 0.695 \\
All coefficients, separate & 0.456 & 0.499 & 0.624 & 0.696 \\
All coefficients, fixed $10^{-6}$ & 0.456 & 0.499 & 0.624 & 0.696 \\
All coefficients, fixed $50$ & 0.459 & 0.499 & 0.632 & 0.697 \\
\bottomrule
\end{tabular}
\begin{minipage}{0.92\textwidth}
\footnotesize\emph{Notes:} Partially penalized ridge uses the normalized grid
$\{10^{-8},10^{-6},10^{-4},10^{-3},10^{-2},10^{-1}\}$. Specifications that
penalize every coefficient use the unnormalized grid
$\{10^{-6},10^{-4},10^{-2},10^{-1},1,10\}$. RMSE compares fitted values
with realized $T$ or $Y$ in the outer evaluation fold, including variation
of each response around its conditional expectation as well as regression error.
It is not the $L^2$ error against the population regression functions.
\end{minipage}
\end{table}

Under a common penalty for both conditioning sets, penalizing every
coefficient changes little relative to partially penalized ridge. RMSE is similar and
coverage is 0.939, compared with 0.943 under partial penalization.
Separate tuning of the regressions has a larger effect. It selects much
larger penalties for the $X$ regressions, and plug-in bias rises from
$-0.001$ to 0.017. Using one penalty for the $(X,Z)$ and $X$ regressions
of each response, as in the primary specification, limits this difference
in shrinkage. In particular, it applies the same penalty to the two
treatment regressions defining the instrument.

With every coefficient subject to the fixed penalty $\lambda=50$,
shrinkage is substantial. The orthogonal correction lowers bias from
0.045 to 0.018 and RMSE from 0.053 to 0.034, but coverage is only 0.842.
The correction removes a first-order effect of estimation error. At this
degree of shrinkage, the remaining error is still too large for an accurate
Wald approximation.

\paragraph{Designs with weak first stages.} I reduce the coefficient on
$\ell_j$ in Design 1 from 0.30 to 0.06. This weakens identification while
leaving the sieve correctly specified, so there is no approximation error.
A second experiment uses the nonlinear Design 3, compressing the examiner
grid by a factor of 0.35 from $[-1.5625,1.5625]$ to
$[-0.546875,0.546875]$.

Both experiments use $n=1{,}800$, $J=18$, the Design 1 sieve, and a
penalty of $10^{-6}$. I form five folds within each examiner and clip
treatment predictions to $[10^{-6},1-10^{-6}]$. In each replication,
inverting the 5 percent score test over the real line amounts to solving
a quadratic acceptance inequality. The reported frequencies distinguish
bounded intervals, one half-line, two half-lines, and the entire real line.
\begin{table}[H]
\centering
\scriptsize
\caption{Diagnostics for weak first stages}
\label{tab:mc_weak}
\begin{tabular}{lrrrrr}
\toprule
Design & True value & Median & 5th pct. & 95th pct. & Med.\ abs.\ err. \\
\midrule
Additive, weak & 1.000 & 1.001 & 0.771 & 1.229 & 0.077 \\
Nonlinear, weak & 1.434 & 1.167 & $-$3.840 & 7.037 & 0.939 \\
\bottomrule
\end{tabular}
\medskip

\begin{tabular}{lrrrrrr}
\toprule
Design & Set coverage & Bounded & One half-line & Two half-lines & All $\mathbb R$ & Med.\ bounded length \\
\midrule
Additive, weak & 0.950 & 0.782 & 0.000 & 0.064 & 0.154 & 0.542 \\
Nonlinear, weak & 0.755 & 0.062 & 0.000 & 0.301 & 0.637 & 3.348 \\
\bottomrule
\end{tabular}
\begin{minipage}{0.96\textwidth}
\footnotesize\emph{Notes:} Each row uses 1{,}000 replications of the
orthogonal estimator and five folds stratified by examiner. Set coverage is
the fraction of replications in which the set obtained by inverting the
5 percent score test contains the true estimand. Median length refers to
bounded intervals only. Wald coverage is omitted: a noisy denominator can
produce very wide intervals and hence high coverage with little information
about the ratio.
\end{minipage}
\end{table}

The corrected denominator \(\widehat D\) averages $-0.0071$ in the weak
additive design and $-0.0003$ in the weak nonlinear design. Both means are
negative, whereas the means of the squared fitted instruments
are 0.0112 and 0.0126. Population relevance in the weak additive design is
itself small: \(D=0.06^2\mathbb{E}[\ell_Z^2]=0.00210\).

The negative mean of \(\widehat D\) reflects second-order estimation bias.
Sampling variation produces further dispersion around that mean and
negative denominators in individual replications. For a fitted propensity pair trained on an
independent sample \(\mathcal T_l\), iterated expectations give
\[
\begin{aligned}
&\mathbb{E}\!\left[
2T\{\widehat\gamma_{1l}(X,Z)-\widehat\gamma_{2l}(X)\}
-\widehat\gamma_{1l}(X,Z)^2+\widehat\gamma_{2l}(X)^2
\,\middle|\,\mathcal T_l\right]-D \\
&\qquad=
-\|\widehat\gamma_{1l}-\gamma_{01}\|^2
+\|\widehat\gamma_{2l}-\gamma_{02}\|^2.
\end{aligned}
\]
In these designs, the $(X,Z)$ regression is harder to estimate than the
$X$ regression. Its squared error has a negative sign in the bias
expression and can be large enough to make the mean corrected denominator
negative.

In the weak additive design, coverage under inversion is 0.950. Yet the
set is unbounded in 0.218 of replications, leaving limited information about
the ratio despite correct sieve specification and nominal coverage.
The finite-sample coverage does not establish uniform validity.
In the weak nonlinear design, the fixed sieve is misspecified and coverage
falls to 0.755. Inversion avoids division by the noisy denominator, but
approximation error still affects the moment being inverted. Uniform
validity with weak identification therefore also requires control of
nuisance estimation along the relevant sequences.

\paragraph{Correctly specified linear benchmark.} I also consider a linear
design with $n=3{,}200$ observations and $J=15$ examiners. Leniency is
equally spaced on $[-1.25,1.25]$, the covariates are independent with
$X_1,X_2,X_3\sim U[-1,1]$, and treatment and outcomes satisfy
\[
\begin{aligned}
p(X,j)&=0.50+0.20\ell_j+0.10X_1-0.08X_2+0.05X_3,\\
Y&=0.5+0.35X_1-0.20X_2+0.15X_3+(1+0.30X_1)T+0.30\varepsilon,\\
&\hspace{4cm}\varepsilon\sim N(0,1).
\end{aligned}
\]
The propensity lies in $[0.02,0.98]$ throughout the support. Both the
linear UJIVE first stage and the sieve regressions are correctly specified,
and symmetry implies $\theta_0=1$.

\begin{table}[H]
\centering
\small
\caption{Linear benchmark at $n=3{,}200$ and $J=15$}
\label{tab:mc_linear}
\begin{tabular}{lrrrrrr}
\toprule
Estimator & Mean & Bias & RMSE & SD & Coverage & Avg.\ SE \\
\midrule
Orthogonal & 1.002 & 0.002 & 0.054 & 0.054 & 0.941 & 0.053 \\
Sieve plug-in & 0.999 & $-$0.001 & 0.046 & 0.046 & --- & --- \\
Linear UJIVE & 1.000 & 0.000 & 0.038 & 0.038 & --- & --- \\
\bottomrule
\end{tabular}
\begin{minipage}{0.92\textwidth}
\footnotesize\emph{Notes:} The table uses 1{,}000 replications. Coverage
and average standard errors are reported for the orthogonal estimator.
UJIVE is included for comparison of point estimates.
\end{minipage}
\end{table}

All three estimators are well centered in this correctly specified design,
and UJIVE is the most precise: its RMSE is 0.038, compared with 0.046 for
the sieve plug-in and 0.054 for the orthogonal estimator. The latter has
an average standard error of 0.053, close to its Monte Carlo standard
deviation of 0.054, and coverage of 0.941.

\clearpage

\section{Additional Inference Results}\label{app:additional_inference}

\subsection{Clustered Inference}\label{app:cluster_inference}

For clustered inference, define
\[
\psi_i^0=\psi(W_i;\theta_0,\gamma_0,\alpha_0),
\qquad
S_g^0=\sum_{i\in C_g}\psi_i^0,
\qquad
\hat S_g(\theta)=\sum_{i\in C_g}\hat\psi_i(\theta).
\]
Suppose the following conditions hold:
\begin{enumerate}[label=(\roman*),leftmargin=*,itemsep=0.3ex,topsep=0.3ex]
\item The clusters are independent, $G\to\infty$, and
\[
n^{-1/2}\sum_{g=1}^G S_g^0
\overset{d}{\longrightarrow}N(0,\Omega_{\mathrm{cl}}),
\qquad
n^{-1}\sum_{g=1}^G(S_g^0)^2
\overset{p}{\longrightarrow}\Omega_{\mathrm{cl}}.
\]
\item The cross-fitted score error satisfies
\[
n^{-1/2}\sum_{g=1}^G\sum_{i\in C_g}
\{\hat\psi_i(\theta_0)-\psi_i^0\}=o_p(1).
\]
\item The sample Jacobian satisfies $\widehat Q\overset{p}{\to}Q=-D$.
\item The estimated cluster score second moment satisfies
\[
\frac1n\sum_{g=1}^G
\{\hat S_g(\hat\theta)^2-(S_g^0)^2\}=o_p(1).
\]
\end{enumerate}
The affine estimating equation implies
\[
\sqrt n(\hat\theta-\theta_0)
\overset{d}{\longrightarrow}
N(0,Q^{-2}\Omega_{\mathrm{cl}}),
\qquad
\widehat V_{\mathrm{cl}}
\overset{p}{\longrightarrow}Q^{-2}\Omega_{\mathrm{cl}}.
\]
The cluster central limit theorem and the two conditions on estimated
cluster scores are additional to the assumptions for independent
observations. They hold in the case with observed examiner characteristics and bounded cluster size developed in
Section~\ref{ssec:growing}, which also explains why unrestricted examiner
indicators create a further difficulty.

\subsection{Score Inversion}\label{app:score_inversion}

Anderson--Rubin or Stock--Wright inversion tests candidate parameter values
without dividing by the estimated Jacobian \citep{andrews2019weak}. It can
therefore be used when that estimate is close to zero. At each candidate $\theta$,
evaluate the affine representer maps in
(\ref{eq:alpha_feasible_closed}) at that value, form $\hat\psi_i(\theta)$, and
compute
\[
S_n(\theta)=n^{-1/2}\sum_i\hat\psi_i(\theta),
\qquad
\widehat\Omega_{\mathrm{iid}}(\theta)
=\frac1n\sum_i\bigl(\hat\psi_i(\theta)-\overline{\hat\psi}(\theta)\bigr)^2,
\quad
\overline{\hat\psi}(\theta)=\frac1n\sum_i\hat\psi_i(\theta).
\]
For clustered data, $\widehat\Omega_{\mathrm{cl}}(\theta)$ is formed from the
cluster score sums. The confidence set consists of the values satisfying
\[
S_n(\theta)^2
\leq\chi^2_{1,1-\alpha}\widehat\Omega(\theta).
\]

Because the score is affine, the inversion problem reduces to a
quadratic inequality in $\theta$, as in \citet{fieller1954some}. With
$\widehat D$ well separated from zero, the accepted values usually form
a bounded interval. Under a weak first stage, two half-lines or the entire
real line may satisfy the inequality, so the set admits arbitrarily large
ratios.

Uniform validity as $D_n=\mathbb{E}_n[\gamma_n^2]\to0$ requires more than
this algebraic construction. It requires a null central limit theorem along
sequences with weak identification, nuisance error controlled uniformly over candidate values,
and uniform consistency of $\widehat\Omega(\theta)$. I use inversion as
a sensitivity analysis in the absence of these results. The set remains
computable when the estimated Jacobian is close to zero.

\section{US Patent Applications: Data and Estimation Details}
\label{app:patent_details}

\subsection{Data and Sample Construction}

The 2022 Patent Examination Research Dataset records application and filing
dates, examiner identity, art unit, USPC class and subclass, grant status,
continuations, foreign priority, applicant characteristics, and inventor
information \citep{graham2018patex}. Linking these records to the publication
identifiers and disambiguated inventor identities in the PatentsView pregrant
files \citep{uspto2026patentsview} identifies 7,734,016 public applications. I then
restrict the data to Art Unit 2872 and apply the sample criteria below.

I downloaded the PatentsView files on August 4, 2026. Their publication table
runs through December 25, 2025, covering publications after the latest
eligible disposition date, January 1, 2020, and the end of the three-year
filing window, January 1, 2023. Qualifying filings count toward the outcome
only if they appear as publications in these pregrant records.

An application is eligible if it has reached disposition, has a pregrant
publication with recorded inventor identities, and has a complete outcome
window. Obtaining identities from pregrant records avoids having to recover
them from the grant itself. There are 35,454 eligible applications with linked
inventors. I order these by filing date and application number, admitting an
application only if none of its inventors appears on an application already
included. Requiring at least 25 focal cases per examiner then leaves 12,919
applications, 77 examiners, and 26,677 inventors, without repeated inventors
across focal applications. Selection does not use the realized outcome.

Selecting focal applications with no inventors in common does not rule out
shared later filings: inventors from different focal applications may
subsequently file together. I form a graph with focal applications as nodes,
joining two nodes whenever their inventors appear on the same later
application. The graph has 11,029 connected
components. Most are singletons: 10,366 components, or 94.0 percent, contain
one focal application and account for 80.2 percent of the observations. The 99th
percentile contains four applications, and the two largest contain 210 and 202.

\subsection{Regression Specifications and Sample Splitting}

The first linear UJIVE specification includes indicators for the year in which
the application was first docketed. The second adds docket month, filing year and month, grouped USPC class, indicators
for foreign priority and continuation, organizational applicant status,
inventor location, and a quadratic in the number of inventors.

For response $R$ and dictionary $b_r$, a ridge fit on training sample
$\mathcal T$ solves
\[
\widehat\beta_{r,\lambda}
=
\arg\min_{\beta}
\left\{
\frac{1}{|\mathcal T|}
\sum_{i\in\mathcal T}(R_i-b_{ri}'\beta)^2
+\lambda\beta'P_r\beta
\right\},
\]
where $b_r$ denotes the dictionary after rescaling by the second moments in the
training sample, and $P_r$ is diagonal with entries equal to zero or one. For
regressions on $X$, only the intercept has a zero entry. In the $(X,Z)$
regressions, zero entries leave the intercept and 76 included examiner
indicators unpenalized, and unit entries penalize all covariate, title, and
interaction columns. The interactions are formed from examiner identifiers, independently of
which examiner category I omit from the main effects. Changing that category
thus changes only the parameterization of the unpenalized main effects,
leaving fitted values unchanged.

For functions of $X$, the dictionary has 3,819 columns. It includes the
covariates described in Section~\ref{sec:patent_application}, interactions
between docket year and USPC class, and 2,048 hashed title terms. For functions
of $(X,Z)$, I add examiner indicators and their interactions with docket year,
docket month, USPC class, and application characteristics. I use a separate signed hash with 2,048 columns for interactions between
examiner indicators and title terms. The dictionary for regressions
on covariates and examiner assignment has 8,629 columns.

After dropping a short list of stop words, I take lowercase alphanumeric
words and adjacent word pairs from each title. Each term contributes to the
column selected by CRC32 with the sign assigned by Adler32. I add
contributions that share a column and normalize the title vector to unit
Euclidean norm. To construct the interactions, I hash the examiner identifier
jointly with the title term. These deterministic maps use neither a random
seed nor an estimated vocabulary and are fixed before sample splitting.
Column rescaling is specific to each ridge fit: I use second moments from
that fit's training sample and cap the multipliers at 100.

I assign whole connected components to five folds, balancing examiner
frequencies across folds. For each of $\gamma_1$, $\gamma_2$, $m_1$, and
$m_2$, I select a penalty from a 17-point grid between 0.1 and 100 using a
further training and validation split within each outer training sample.
The outer evaluation fold is therefore unused for penalty selection.
Treatment and outcome predictions are truncated to $[0.001,0.999]$ before
I evaluate the score. For the
two UJIVE specifications in Table~\ref{tab:patent_application}, I assess
treatment prediction by fitting least squares on each training sample.

I repeat grouped five-fold cross-fitting for eleven partitions. Let
$\widehat\theta_s$ denote the estimate in partition $s$, and let
$\widehat{\operatorname{Var}}_s(\widehat\theta_s)$ denote its estimated
finite-sample variance, equal to the square of the standard error clustered by
component. Following the rule for repeated sample splitting in
\citet{chernozhukov2018double}, I report
\[
\widetilde\theta=\operatorname{median}_s\widehat\theta_s,\qquad
\widetilde{\operatorname{Var}}
=\operatorname{median}_s
\left\{\widehat{\operatorname{Var}}_s(\widehat\theta_s)
+(\widehat\theta_s-\widetilde\theta)^2\right\}.
\]
The reported standard error,
$\sqrt{\widetilde{\operatorname{Var}}}$, combines sampling uncertainty within
a partition with variation across partitions.

\subsection{Asymptotic Conditions}

Theorem~\ref{thm1} requires the score formed from the estimated regressions to
be sufficiently close to the population score. Clustered inference additionally requires a central limit theorem for sums
of scores within clusters. Grouping related outcomes and balancing examiner
frequencies makes the application's partition depend on observed data, so
Theorem~\ref{thm1} does not cover it. The ridge regression with 8,629
regressors also falls outside the unregularized series sequence in
Corollary~\ref{cor:series}. Both the partition and the regression
specification therefore require further arguments before the asymptotic
results can be applied directly.

Across eleven partitions, the estimates lie between 0.023 and 0.062 and the
corrected denominators between 0.0196 and 0.0204. Stability across partitions and prediction errors in evaluation folds help
assess the fitted regressions, but neither establishes the product condition
in Assumption~\ref{ass2}.

I use connected components for the primary variance estimate and examine
clustering by examiner as a sensitivity calculation. With 77 examiner
clusters and unrestricted examiner indicators in the first stage, the latter
calculation lies outside the sufficient conditions for clustered inference
in Section~\ref{ssec:growing}.
For UJIVE, a conventional sandwich formula treats the leave-one-out
instrument as known. Since valid inference with many instruments requires
a different variance procedure \citep{evdokimovkolesar2018}, I report
its point estimates without standard errors.

\section{Proofs for Section 3}\label{app:proofs}

I prove the identification, efficiency, and asymptotic results of Section 3
for a fixed distribution, then extend the asymptotic argument as the number
of examiners or covariates grows.

\subsection{Proof of Proposition~\ref{prop:identification}}

Fix a value \(x\) in the support of \(X\). Conditional independence implies
\[
p_j(x)
=\mathbb{E}[T_i(j)\mid X_i=x]
\]
for every $j\in\mathcal J(x)$. Examiners outside $\mathcal J(x)$ have zero
conditional assignment probability and are excluded from all sums below.
When \(Z_i=j\), the oracle instrument is
\(\gamma_0(x,j)=p_j(x)-\bar p(x)\). Therefore, by iterated
expectations,
\begin{align}
\mathbb{E}[T_i\gamma_0(X_i,Z_i)\mid X_i=x]
&=\sum_{j\in\mathcal J(x)}\pi_j(x)p_j(x)\{p_j(x)-\bar p(x)\} \notag\\
&=\sum_{j\in\mathcal J(x)}\pi_j(x)\{p_j(x)-\bar p(x)\}^2.
\label{eq:proof_ident_tgamma}
\end{align}
The last equality uses
\(\sum_j\pi_j(x)\{p_j(x)-\bar p(x)\}=0\). Applying iterated expectations to
the other two quantities gives
\begin{align}
\mathbb{E}[\gamma_0(X_i,Z_i)^2\mid X_i=x]
&=\sum_{j\in\mathcal J(x)}\pi_j(x)\{p_j(x)-\bar p(x)\}^2, \\
\mathbb{E}[\omega_i(X_i)\mid X_i=x]
&=\sum_{j\in\mathcal J(x)}\pi_j(x)\{p_j(x)-\bar p(x)\}
  \mathbb{E}[T_i(j)\mid X_i=x] \notag\\
&=\sum_{j\in\mathcal J(x)}\pi_j(x)\{p_j(x)-\bar p(x)\}^2.
\label{eq:proof_ident_omega}
\end{align}
Thus the three conditional expectations in
(\ref{eq:proof_ident_tgamma})--(\ref{eq:proof_ident_omega}) coincide.

For completeness, the weighted variance identity is
\begin{align}
\sum_{j\in\mathcal J(x)}\pi_j(x)\{p_j(x)-\bar p(x)\}^2
&=\sum_{j\in\mathcal J(x)}\pi_j(x)p_j(x)^2-\bar p(x)^2 \notag\\
&=\frac12\sum_{j\in\mathcal J(x)}\sum_{k\in\mathcal J(x)}
  \pi_j(x)\pi_k(x)\{p_j(x)-p_k(x)\}^2 \notag\\
&=\sum_{\substack{j<k\\j,k\in\mathcal J(x)}}\pi_j(x)\pi_k(x)
  \{p_j(x)-p_k(x)\}^2.
\label{eq:proof_weighted_variance}
\end{align}
Integrating over \(X\) proves (\ref{eq:identification_denominator}).

For the numerator, exclusion and consistency imply
\(Y_i=Y_i(0)+\tau_i T_i(Z_i)\). Conditional independence continues to hold
after conditioning on the potential treatment schedule. Hence
\begin{align}
&\mathbb{E}\!\left[
Y_i(0)\gamma_0(X_i,Z_i)
\,\middle|\,
X_i=x,\{T_i(j)\}_{j\in\mathcal J(x)}
\right] \notag\\
&\qquad=
\mathbb{E}\!\left[
Y_i(0)
\,\middle|\,
X_i=x,\{T_i(j)\}_{j\in\mathcal J(x)}
\right]
\sum_{j\in\mathcal J(x)}
\pi_j(x)\{p_j(x)-\bar p(x)\}
=0.
\label{eq:proof_baseline_zero}
\end{align}
Integrating over the potential treatment schedule gives
$\mathbb{E}[Y_i(0)\gamma_0(X_i,Z_i)\mid X_i=x]=0$.
For the component involving treatment effects, condition first on all potential variables
and $X_i=x$, and then average over the independent examiner draw:
\begin{align}
&\mathbb{E}\!\left[
\tau_i T_i(Z_i)\gamma_0(X_i,Z_i)\mid X_i=x
\right] \notag\\
&\qquad=
\mathbb{E}\!\left[
 \tau_i\sum_{j\in\mathcal J(x)}\pi_j(x)
 \{p_j(x)-\bar p(x)\}T_i(j)
\,\bigg|\,X_i=x
\right] \notag\\
&\qquad=
\mathbb{E}[\tau_i\omega_i(X_i)\mid X_i=x].
\label{eq:proof_treatment_component}
\end{align}
Equations (\ref{eq:proof_baseline_zero}) and
(\ref{eq:proof_treatment_component}), followed by integration over \(X\),
prove (\ref{eq:identification_numerator_average}). If average monotonicity
holds, \(\omega_i(X_i)\geq0\) almost surely. Relevance and
(\ref{eq:identification_denominator}) imply
\(\mathbb{E}[\omega_i(X_i)]>0\), so normalizing these weights gives the
positively weighted average in Proposition~\ref{prop:identification}.

It remains to establish the pairwise representation. Define
\[
q_j(x)=\mathbb{E}[\tau_i T_i(j)\mid X_i=x].
\]
Applying the weighted covariance identity to \(p_j(x)\) and \(q_j(x)\)
yields
\begin{align}
&\sum_{j\in\mathcal J(x)}\pi_j(x)\{p_j(x)-\bar p(x)\}q_j(x) \notag\\
&\qquad=
\sum_{\substack{j<k\\j,k\in\mathcal J(x)}}\pi_j(x)\pi_k(x)
\{p_j(x)-p_k(x)\}\{q_j(x)-q_k(x)\}.
\label{eq:proof_pairwise_covariance}
\end{align}
Whenever \(p_j(x)\ne p_k(x)\), the definition in the proposition implies
\[
q_j(x)-q_k(x)
=\{p_j(x)-p_k(x)\}\operatorname{LATE}_{\{j,k\}}(x).
\]
The ratio $\operatorname{LATE}_{\{j,k\}}(x)$ is unchanged when $j$ and $k$
are interchanged, because its numerator and denominator both change sign.
I can therefore orient the pair from lower to higher propensity; under
pairwise monotonicity, the potential treatment difference is then an
indicator for switching treatment, so
$\operatorname{LATE}_{\{j,k\}}(x)$ averages the treatment effects of
switchers conditional on the covariates. If \(p_j(x)=p_k(x)\), that term in
(\ref{eq:proof_pairwise_covariance}) is zero. Substitution into
(\ref{eq:proof_pairwise_covariance}) and integration over \(X\) give the
numerator of (\ref{eq:pairwise_late_oracle});
(\ref{eq:proof_weighted_variance}) is its denominator. Every pairwise weight
is nonnegative, and relevance makes their total strictly positive. This proves
the representation with convex weights. \qed

\subsection{Proof of Lemma~\ref{lem:arbitrary_signal}}

Fix a value $x$ in the support of $X$. Conditional on the potential variables
and $X_i=x$, conditional independence makes the realized examiner an independent
draw with probabilities $\{\pi_j(x):j\in\mathcal J(x)\}$. Consistency of the
potential treatments then gives
\begin{align}
&\mathbb{E}\!\left[
T_i h(X_i,Z_i)
\,\middle|\,
X_i=x,(T_i(j))_{j\in\mathcal J(x)}
\right] \notag\\
&\qquad=
\sum_{j\in\mathcal J(x)}
\pi_j(x)h(x,j)T_i(j).
\label{eq:proof_arbitrary_first_stage_conditional}
\end{align}
Taking expectations over the potential treatments conditional on $X_i=x$ and
then integrating over $X_i$ proves
\[
\mathbb{E}[T_i h(X_i,Z_i)]
=\mathbb{E}[\omega_{h,i}(X_i)],
\]
which is (\ref{eq:arbitrary_signal_first_stage}).

For the outcome moment, exclusion and consistency imply
$Y_i=Y_i(0)+\tau_i T_i(Z_i)$. Separating the baseline outcome from the treatment
effect, conditional independence implies
\begin{align}
&\mathbb{E}[Y_i(0)h(X_i,Z_i)\mid X_i=x] \notag\\
&\qquad=
\mathbb{E}[Y_i(0)\mid X_i=x]
\sum_{j\in\mathcal J(x)}\pi_j(x)h(x,j) \notag\\
&\qquad=
\mathbb{E}[Y_i(0)\mid X_i=x]\bar h(x).
\label{eq:proof_arbitrary_baseline}
\end{align}
For the term involving treatment effects, condition first on the potential variables and
then average over the assignment draw:
\begin{align}
&\mathbb{E}[\tau_i T_i(Z_i)h(X_i,Z_i)\mid X_i=x] \notag\\
&\qquad=
\mathbb{E}\!\left[
\tau_i
\sum_{j\in\mathcal J(x)}
\pi_j(x)h(x,j)T_i(j)
\,\middle|\,X_i=x
\right] \notag\\
&\qquad=
\mathbb{E}[\tau_i\omega_{h,i}(X_i)\mid X_i=x].
\label{eq:proof_arbitrary_effect}
\end{align}
Adding (\ref{eq:proof_arbitrary_baseline}) and
(\ref{eq:proof_arbitrary_effect}) and integrating over $X_i$ proves
(\ref{eq:arbitrary_signal_outcome}). When the first-stage moment is nonzero,
division by (\ref{eq:arbitrary_signal_first_stage}) establishes
(\ref{eq:arbitrary_signal_ratio}). \qed

\subsection{Proof of Lemma~\ref{lem:maximal_relevance}}

Let $h\in\mathcal H_X$. Iterated expectations and the definition of
$\gamma_{01}$ give
\[
\mathbb{E}[Th]
=
\mathbb{E}[\mathbb{E}[T\mid X,Z]h(X,Z)]
=
\mathbb{E}[\gamma_{01}(X,Z)h(X,Z)].
\]
Because $\gamma_{02}(X)$ is measurable with respect to $X$ and
$\mathbb{E}[h\mid X]=0$,
\[
\mathbb{E}[\gamma_{02}(X)h(X,Z)]
=
\mathbb{E}[\gamma_{02}(X)\mathbb{E}[h(X,Z)\mid X]]
=0.
\]
Subtracting this zero term yields
\begin{equation}
\mathbb{E}[Th]
=
\mathbb{E}[\{\gamma_{01}(X,Z)-\gamma_{02}(X)\}h(X,Z)]
=
\mathbb{E}[\gamma_0h].
\label{eq:proof_maximal_covariance}
\end{equation}
Cauchy--Schwarz and the unit variance normalization imply
\[
|\mathbb{E}[Th]|
\leq
\{\mathbb{E}[\gamma_0^2]\}^{1/2}
\{\mathbb{E}[h^2]\}^{1/2}
=\sqrt D.
\]
The bound is attainable because
$h_+(X,Z)=\gamma_0(X,Z)/\sqrt D$ satisfies
$\mathbb{E}[h_+\mid X]=0$ and $\mathbb{E}[h_+^2]=1$, and its negative also belongs
to $\mathcal H_X$. For equality in the Cauchy--Schwarz bound, $h$ must be
proportional to $\gamma_0$ almost surely, say $h=c\gamma_0$ for a constant
$c$. Unit variance then leaves two choices, $c=1/\sqrt D$ and
$c=-1/\sqrt D$. By (\ref{eq:proof_maximal_covariance}), the covariances are
$\mathbb{E}[Th]=\sqrt D$ and $\mathbb{E}[Th]=-\sqrt D$, respectively.
The requirement of positive covariance selects the first, $h_+$. \qed

\subsection{Proof of Proposition~\ref{prop:efficiency}}
First consider regular parametric submodels
$\{F_t:t\in(-\epsilon,\epsilon)\}$ that are locally equivalent to $F_0$ and
have bounded scores $s(W)$. Such scores form a dense subset of
$L_0^2(F_0)$, and the derivative obtained below is a bounded linear functional,
so the calculation extends by continuity to the full tangent space. Write
\[
\mu_0(W)=Y-\theta_0T,
\qquad
M(F,\theta)=\mathbb{E}_F[(Y-\theta T)\gamma_F(X,Z)],
\]
where
$\gamma_F(X,Z)=\mathbb{E}_F[T\mid X,Z]-\mathbb{E}_F[T\mid X]$.
The parameter $\theta(F)$ is defined by $M(F,\theta(F))=0$. Differentiating this
identity at $t=0$ gives
\begin{equation}\label{eq:eff_path_identity}
\begin{aligned}
\dot\theta
&=-\bigl[\partial_\theta M(F_0,\theta_0)\bigr]^{-1}\dot M,\\
\dot M
&=\frac{d}{dt}M(F_t,\theta_0)\Big|_{t=0}.
\end{aligned}
\end{equation}
The derivative with respect to the parameter is
\[
\partial_\theta M(F_0,\theta_0)
=-\mathbb{E}[T\gamma_0]
=-D.
\]
The last equality uses
$D=\mathbb{E}[T\gamma_0]=\mathbb{E}[\gamma_0^2]$, which follows from the two
identities for conditional expectations defining $\gamma_0$.

It remains to calculate the pathwise derivative $\dot M$. Holding $\gamma_F$
fixed at $\gamma_0$, differentiation of the expectation contributes
\begin{equation}\label{eq:eff_fixed_gamma}
\mathbb{E}[\mu_0(W)\gamma_0(X,Z)s(W)].
\end{equation}
The remaining contribution comes from the dependence of $\gamma_F$ on $F$.

For $V\in\{(X,Z),X\}$, let
$m_t(V)=\mathbb{E}_{F_t}[T\mid V]$. Local equivalence allows every $m_t$ to be
regarded as an element of the fixed Hilbert space $L^2(F_{0,V})$. Moreover,
$T\in\{0,1\}$ implies
$\operatorname*{ess\,sup}_v\mathbb{E}_{F_t}[T^2\mid V=v]\leq 1$, so the
conditional second moment required for $L^2$ pathwise differentiability of a
regression function is automatically bounded. The pathwise differentiability
result for regression functions in
\citet{luedtkechung2024one} therefore implies
\[
\|m_t-m_0-t\dot m\|_{L^2(F_{0,V})}=o(t).
\]
To identify the derivative, write $r_t=dF_t/dF_0$. Then
\[
m_t(V)
=
\frac{\mathbb{E}_0[Tr_t\mid V]}{\mathbb{E}_0[r_t\mid V]},
\]
and differentiation at $t=0$, where $r_0=1$ and $\dot r_0=s$, gives
\[
\dot m(V)
=\mathbb{E}_0[(T-m_0(V))s(W)\mid V].
\]
Conditional Jensen's inequality also gives
\[
\|\dot m\|_{L^2(F_{0,V})}^2
\leq
\mathbb{E}_0[(T-m_0(V))^2s(W)^2]
\leq
\mathbb{E}_0[s(W)^2],
\]
so the derivative is a bounded linear map from $L^2_0(F_0)$ into
$L^2(F_{0,V})$. The derivative concerns the regression function as an element of
$L^2$, rather than its value at a fixed $v$. The support of $V$ may therefore be
continuous or infinite. Applying the derivative
with $V=(X,Z)$ and $V=X$ yields
\[
\dot\gamma_1(X,Z)
=\mathbb{E}[(T-\gamma_{01}(X,Z))s(W)\mid X,Z],
\]
\[
\dot\gamma_2(X)
=\mathbb{E}[(T-\gamma_{02}(X))s(W)\mid X].
\]
Because $\gamma_F=\gamma_{1F}-\gamma_{2F}$, the derivative through the generated
instrument is
\begin{align}
\mathbb{E}[\mu_0\{\dot\gamma_1(X,Z)-\dot\gamma_2(X)\}]
&=\mathbb{E}[\alpha_{01}(X,Z)\dot\gamma_1(X,Z)]
  +\mathbb{E}[\alpha_{02}(X)\dot\gamma_2(X)] \notag\\
&=\mathbb{E}\!\left[
\left\{
\alpha_{01}(X,Z)(T-\gamma_{01}(X,Z))
+\alpha_{02}(X)(T-\gamma_{02}(X))
\right\}s(W)
\right].
\label{eq:eff_generated_gamma}
\end{align}
The first equality uses
$\alpha_{01}(X,Z)=\mathbb{E}[\mu_0\mid X,Z]$ and
$\alpha_{02}(X)=-\mathbb{E}[\mu_0\mid X]$. The negative sign in
$\alpha_{02}$ offsets the subtraction of $\dot\gamma_2$.

Adding (\ref{eq:eff_fixed_gamma}) and (\ref{eq:eff_generated_gamma}) gives
\[
\dot M
=\mathbb{E}[\psi(W;\theta_0,\gamma_0,\alpha_0)s(W)].
\]
Substitution into (\ref{eq:eff_path_identity}) therefore yields
\[
\dot\theta
=\mathbb{E}\!\left[
D^{-1}\psi(W;\theta_0,\gamma_0,\alpha_0)s(W)
\right]
\]
for every score in the dense class just described, and hence for every element
of the tangent space by continuity. Each component of the oracle score has
mean zero. In addition, binary treatment implies
$|\gamma_0|\leq 1$ and $|T-\gamma_{0j}|\leq 1$ for $j=1,2$, while conditional
Jensen's inequality and the maintained moment condition give
\[
\|\alpha_{01}\|_2\leq\|\mu_0\|_2,
\qquad
\|\alpha_{02}\|_2\leq\|\mu_0\|_2.
\]
Consequently, $D^{-1}\psi(W;\theta_0,\gamma_0,\alpha_0)$ is mean zero and
square-integrable. It is therefore an influence function for $\theta(F)$. In the
nonparametric model, the tangent space is the full space of square-integrable
functions with mean zero, so the influence function is the unique canonical
gradient and hence the efficient influence function. Its variance is
\[
V_{\mathrm{eff}}
=D^{-2}\mathbb{E}[\psi(W;\theta_0,\gamma_0,\alpha_0)^2],
\]
which is the efficiency bound
\citep{newey1994asymptotic,bickel1993efficient,van_der_Vaart_1998}. \qed

\subsection{Proof of Lemma~\ref{lem:mr}}
Let
\[
\delta_{\gamma 1}=\gamma_1-\gamma_{01},\qquad
\delta_{\gamma 2}=\gamma_2-\gamma_{02},\qquad
\delta_{\alpha 1}=\alpha_1-\alpha_{01},\qquad
\delta_{\alpha 2}=\alpha_2-\alpha_{02}.
\]
Also write $\mu_0(W)=Y-\theta_0T$. The score evaluated at generic nuisance functions is
\[
\psi(W;\theta_0,\gamma,\alpha)
=
\mu_0(\gamma_1-\gamma_2)
+\alpha_1(T-\gamma_1)
+\alpha_2(T-\gamma_2),
\]
while the oracle score uses $(\gamma_0,\alpha_0)$. Subtracting the oracle score and collecting terms gives
\begin{align*}
&\psi(W;\theta_0,\gamma,\alpha)-\psi(W;\theta_0,\gamma_0,\alpha_0)\\
&\quad = (\mu_0-\alpha_{01})\delta_{\gamma 1}+(-\mu_0-\alpha_{02})\delta_{\gamma 2}
+\delta_{\alpha 1}(T-\gamma_{01})+\delta_{\alpha 2}(T-\gamma_{02})
-\delta_{\alpha 1}\delta_{\gamma 1}-\delta_{\alpha 2}\delta_{\gamma 2}.
\end{align*}
The first four terms are linear in a single nuisance error, and each has
expectation zero by iterated expectations. For the first term,
\[
\mathbb{E}\{(\mu_0-\alpha_{01})\delta_{\gamma 1}\}
=
\mathbb{E}\!\left[
\delta_{\gamma 1}(X,Z)
\mathbb{E}\{\mu_0-\alpha_{01}(X,Z)\mid X,Z\}
\right]
=0,
\]
because $\alpha_{01}(X,Z)=\mathbb{E}[\mu_0\mid X,Z]$. Similarly,
\[
\mathbb{E}\{(-\mu_0-\alpha_{02})\delta_{\gamma 2}\}
=
\mathbb{E}\!\left[
\delta_{\gamma 2}(X)
\mathbb{E}\{-\mu_0-\alpha_{02}(X)\mid X\}
\right]
=0,
\]
because $\alpha_{02}(X)=-\mathbb{E}[\mu_0\mid X]$. The third and fourth terms vanish because
\[
\mathbb{E}\{T-\gamma_{01}(X,Z)\mid X,Z\}=0,
\qquad
\mathbb{E}\{T-\gamma_{02}(X)\mid X\}=0.
\]
The oracle score also satisfies $\mathbb{E}[\psi(W;\theta_0,\gamma_0,\alpha_0)]=0$, so only the two bilinear terms remain:
\[
\mathbb{E}\left[\psi(W;\theta_0,\gamma,\alpha)\right]
=-\mathbb{E}[\delta_{\alpha 1}\delta_{\gamma 1}]-\mathbb{E}[\delta_{\alpha 2}\delta_{\gamma 2}],
\]
which proves the result. \qed

\subsection{Proof of Theorem~\ref{thm1}}
The proof applies the cross-fitted locally robust expansion in Lemma 8 of
\citet{chernozhukov2022locally} to (\ref{eq:orthogonal_score}). I distinguish
the terms controlled by orthogonality from those controlled by the product condition.
If the fold partition is randomized, condition first on its realization.
Independence of the partition and the observations then leaves the evaluation
observations i.i.d.\ within each fold.

Let $l(i)$ denote the fold containing observation $i$. For each fold $l$, define the nuisance errors
\[
\delta_{\gamma 1,l}=\hat\gamma_{1l}-\gamma_{01},\qquad
\delta_{\gamma 2,l}=\hat\gamma_{2l}-\gamma_{02},
\]
and
\[
\delta_{\alpha 1,l}=\hat\alpha_{1l}(\theta_0)-\alpha_{01}(\theta_0),
\qquad
\delta_{\alpha 2,l}=\hat\alpha_{2l}(\theta_0)-\alpha_{02}(\theta_0).
\]
The dependence of these functions on $(X,Z)$ or $X$ is suppressed when no
confusion can arise. Assumption~\ref{ass1} implies that the $L^2$ norms of all
four errors converge to zero at $\theta_0$.

For $i\in I_l$, the algebra in Lemma~\ref{lem:mr} gives the following
decomposition for the score of observation $i$:
\[
\begin{aligned}
&\psi(W_i;\theta_0,\hat\gamma_l,\hat\alpha_l(\theta_0))
-\psi(W_i;\theta_0,\gamma_0,\alpha_0)\\
&\quad =
(\mu_{0i}-\alpha_{01,i})\delta_{\gamma 1,l,i}
{}+
(-\mu_{0i}-\alpha_{02,i})\delta_{\gamma 2,l,i}\\
&\qquad
+\delta_{\alpha 1,l,i}(T_i-\gamma_{01,i})
+\delta_{\alpha 2,l,i}(T_i-\gamma_{02,i})\\
&\qquad
-\delta_{\alpha 1,l,i}\delta_{\gamma 1,l,i}
-\delta_{\alpha 2,l,i}\delta_{\gamma 2,l,i},
\end{aligned}
\]
where $\mu_{0i}=Y_i-\theta_0T_i$. The identity separates each observation's score error into four terms
linear in a single nuisance error and two bilinear interactions.

All nuisance functions used to evaluate the score in $I_l$ are estimated
outside $I_l$ and are therefore fixed once the training sample is given.
Conditional on that training sample, the first linear term has mean zero:
\[
\mathbb{E}\!\left[
(\mu_0-\alpha_{01})\delta_{\gamma 1,l}(X,Z)
\mid \text{training sample}
\right]=0,
\]
because $\mathbb{E}[\mu_0-\alpha_{01}(X,Z)\mid X,Z]=0$. The same argument,
using iterated expectations, shows that the other three linear terms also have mean zero:
\[
\mathbb{E}[(-\mu_0-\alpha_{02})\delta_{\gamma 2,l}(X)\mid \text{training sample}]=0,
\]
\[
\mathbb{E}[\delta_{\alpha 1,l}(X,Z)(T-\gamma_{01})\mid \text{training sample}]=0,
\qquad
\mathbb{E}[\delta_{\alpha 2,l}(X)(T-\gamma_{02})\mid \text{training sample}]=0.
\]
Define the two bilinear interaction terms by
\[
B_l(W)=
-\delta_{\alpha 1,l}(X,Z)\delta_{\gamma 1,l}(X,Z)
-\delta_{\alpha 2,l}(X)\delta_{\gamma 2,l}(X).
\]
Since the four linear terms have zero conditional expectation, the bias from nuisance
estimation is $\mathbb{E}[B_l(W)\mid\text{training sample}]$.
By Cauchy--Schwarz,
\[
|\mathbb{E}[B_l(W)\mid\text{training sample}]|
\le
\|\delta_{\alpha 1,l}\|_2\|\delta_{\gamma 1,l}\|_2
+\|\delta_{\alpha 2,l}\|_2\|\delta_{\gamma 2,l}\|_2
\le
\|\hat{\boldsymbol\alpha}_l(\theta_0)-\boldsymbol\alpha_0(\theta_0)\|
\|\hat{\boldsymbol\gamma}_l-\boldsymbol\gamma_0\|.
\]
Under Assumption~\ref{ass2}, the product on the right is $o_p(n^{-1/2})$.
The conditional expectation of the score error is consequently negligible after
multiplication by $\sqrt n$. We must also bound the empirical fluctuation
around that conditional expectation.
Let
\[
\Delta_l(W)
=
\psi(W;\theta_0,\hat\gamma_l,\hat\alpha_l(\theta_0))
-\psi(W;\theta_0,\gamma_0,\alpha_0).
\]
Assumption~\ref{ass1}, together with either
Assumption~\ref{ass3} or Assumption~\ref{ass3prime}, implies, conditionally on the
training sample,
\[
\mathbb{E}\!\left[\Delta_l(W)^2\mid\text{training sample}\right]=o_p(1).
\]
Under Assumption~\ref{ass3}, this follows from mean square continuity along a
consistent nuisance sequence. Under
Assumption~\ref{ass3prime}, the same conclusion follows term by term. For
example,
\[
\mathbb{E}\!\left[
(\mu_0-\alpha_{01})^2\delta_{\gamma 1,l}^2
\mid\text{training sample}\right]
\le
\|\mu_0-\alpha_{01}\|_4^2\|\delta_{\gamma 1,l}\|_4^2=o_p(1),
\]
while binary treatment implies
\[
\mathbb{E}\!\left[
\delta_{\alpha 1,l}^2(T-\gamma_{01})^2
\mid\text{training sample}\right]
\le \|\delta_{\alpha 1,l}\|_2^2=o_p(1).
\]
The same inequalities bound the two linear terms involving the second
nuisance component. For each bilinear term, Cauchy--Schwarz bounds its
second moment by the product of the squared fourth moment norms of its
two errors.

The observations in $I_l$ are i.i.d.\ given the training
sample. Therefore,
\[
\operatorname{Var}\!\left[
\mathbb{E}_{n,l}\Delta_l(W)
\mid\text{training sample}\right]
\le
\frac{1}{n_l}
\mathbb{E}\!\left[\Delta_l(W)^2\mid\text{training sample}\right]
=o_p(n_l^{-1}).
\]
By conditional Chebyshev's inequality, the centered average over $I_l$ is of
order $o_p(n_l^{-1/2})$, or $o_p(n^{-1/2})$ because fold proportions are
bounded away from zero. The product bound implies the same order for the
conditional expectation of the uncentered average, so restoring this
expectation leaves the average score error $o_p(n^{-1/2})$. Averaging
over the fixed number of folds preserves this order. Combining the oracle score with the preceding error terms
gives
\[
\sqrt n\,\hat\psi(\theta_0)=\frac{1}{\sqrt n}\sum_{i=1}^n\psi(W_i;\theta_0,\gamma_0,\alpha_0)+o_p(1).
\]
This is the claimed expansion. \qed

\subsection{Proof of Lemma~\ref{lem:Qhat}}
For any pair $\gamma=(\gamma_1,\gamma_2)$, define
\[
q(W;\gamma)
=-2T(\gamma_1-\gamma_2)+\gamma_1^2-\gamma_2^2.
\]
Then
\[
\widehat Q
=\sum_{l=1}^L\frac{n_l}{n}\,\mathbb{E}_{n,l}
[q(W;\hat\gamma_l)],
\qquad
\mathbb{E}_{n,l}f
=\frac{1}{n_l}\sum_{i\in I_l}f(W_i).
\]
Fix a fold $l$ and condition on its training sample. The fitted functions
$\hat\gamma_{1l}$ and $\hat\gamma_{2l}$ are then fixed, while the observations in
$I_l$ are i.i.d.\ draws from the population. Write
$\delta_{1l}=\hat\gamma_{1l}-\gamma_{01}$ and
$\delta_{2l}=\hat\gamma_{2l}-\gamma_{02}$. Direct expansion gives
\begin{align}
q(W;\hat\gamma_l)-q(W;\gamma_0)
={}&2(\gamma_{01}-T)\delta_{1l}
+2(T-\gamma_{02})\delta_{2l} \notag\\
&+\delta_{1l}^2-\delta_{2l}^2.
\label{eq:qhat_difference}
\end{align}
The two linear terms have conditional expectation zero because
\[
\mathbb{E}[T-\gamma_{01}(X,Z)\mid X,Z]=0,
\qquad
\mathbb{E}[T-\gamma_{02}(X)\mid X]=0.
\]
Consequently,
\[
\mathbb{E}[q(W;\hat\gamma_l)-q(W;\gamma_0)
\mid\text{training sample}]
=\|\delta_{1l}\|_2^2-\|\delta_{2l}\|_2^2=o_p(1)
\]
by Assumption~\ref{ass1}.

Since the true propensities are bounded, mean square consistency under
Assumption~\ref{ass1} implies that the fitted propensities are $O_p(1)$
in $L^2$. I use this norm bound and boundedness of $T$ to control the
empirical average over $I_l$. Equation
(\ref{eq:qhat_difference}) and Cauchy--Schwarz therefore imply
\[
\mathbb{E}\!\left[
|q(W;\hat\gamma_l)-q(W;\gamma_0)|
\mid\text{training sample}
\right]
=o_p(1).
\]
Conditional Markov's inequality therefore implies
\[
\mathbb{E}_{n,l}
|q(W;\hat\gamma_l)-q(W;\gamma_0)|=o_p(1).
\]
Moreover, $q(W;\gamma_0)$ is integrable, so the ordinary law of large numbers
implies
\[
\mathbb{E}_{n,l}q(W;\gamma_0)
\overset{p}{\longrightarrow}
\mathbb{E}[q(W;\gamma_0)].
\]
Combining these two displays yields
\[
\mathbb{E}_{n,l}q(W;\hat\gamma_l)
\overset{p}{\longrightarrow}
\mathbb{E}[q(W;\gamma_0)]
\]
for every fold. Since $L$ is fixed and the fold proportions are bounded away from
zero, averaging across folds preserves the limit.

Finally, iterated expectations give
$\mathbb{E}[T\gamma_{01}]=\mathbb{E}[\gamma_{01}^2]$,
$\mathbb{E}[T\gamma_{02}]=\mathbb{E}[\gamma_{02}^2]$, and
$\mathbb{E}[\gamma_{01}\gamma_{02}]=\mathbb{E}[\gamma_{02}^2]$. Hence
\begin{align*}
\mathbb{E}[q(W;\gamma_0)]
&=-\mathbb{E}[\gamma_{01}^2-\gamma_{02}^2]\\
&=-\mathbb{E}[(\gamma_{01}-\gamma_{02})^2]
=-D=Q.
\end{align*}
Thus $\widehat Q\overset{p}{\to}-D$. Under Assumption~\ref{ass4}, $D\ge c$,
so $|\widehat Q|\ge c/2$ with probability approaching one. \qed

\subsection{Proof of Corollary~\ref{coro}}
The corollary uses the expansion in Theorem~\ref{thm1} and the fact that the
sample moment is affine in $\theta$ once the cross-fitted estimates of the
conditional expectations are held fixed. Holding these estimates fixed,
\[
\hat\psi(\theta)
=
\hat A+\theta \widehat Q,
\]
where
\[
\widehat Q
=
\frac{1}{n}\sum_{l=1}^L\sum_{i\in I_l}
\left[
-2T_i\hat\gamma_l(X_i,Z_i)
+\hat\gamma_{1l}(X_i,Z_i)^2
-\hat\gamma_{2l}(X_i)^2
\right].
\]
Since $\hat\theta$ solves the estimating equation,
\[
0=\hat\psi(\hat\theta)=\hat\psi(\theta_0)+\widehat Q(\hat\theta-\theta_0).
\]
Rearranging yields the identity
\[
\sqrt n(\hat\theta-\theta_0)=-\widehat Q^{-1}\sqrt n\,\hat\psi(\theta_0).
\]
By Theorem~\ref{thm1},
\[
\sqrt n\,\hat\psi(\theta_0)
=
\frac{1}{\sqrt n}\sum_{i=1}^n
\psi(W_i;\theta_0,\gamma_0,\alpha_0)+o_p(1).
\]
Under the central limit theorem for independent observations,
\[
\frac{1}{\sqrt n}\sum_{i=1}^n
\psi(W_i;\theta_0,\gamma_0,\alpha_0)
\overset{d}{\to}
N(0,\Omega_{\mathrm{iid}}),
\qquad
\Omega_{\mathrm{iid}}=\mathbb{E}[\psi(W;\theta_0,\gamma_0,\alpha_0)^2],
\]
because the oracle score has mean zero. Lemma~\ref{lem:Qhat} establishes $\widehat Q\overset{p}{\to}Q=-D\neq0$ under Assumption~\ref{ass4}, so Slutsky's theorem implies
\[
\sqrt n(\hat\theta-\theta_0)
\overset{d}{\to}
N(0,Q^{-2}\Omega_{\mathrm{iid}}).
\]

It remains to justify the plug-in variance estimator. The preceding
identity, Theorem~\ref{thm1}, and Lemma~\ref{lem:Qhat} first imply
\[
\hat\theta-\theta_0=O_p(n^{-1/2})=o_p(1).
\]
Define the oracle score, the fitted score at the true parameter, and their
difference by
\[
\psi_i^0=\psi(W_i;\theta_0,\gamma_0,\alpha_0),
\qquad
\tilde\psi_i=\hat\psi_i(\theta_0),
\qquad
\Delta_i=\tilde\psi_i-\psi_i^0.
\]
Fix a fold $l$ and condition on its training sample. Assumption~\ref{ass1},
together with either Assumption~\ref{ass3} or
Assumption~\ref{ass3prime}, implies
\[
\mathbb{E}[\Delta_i^2\mid\text{training sample for fold }l]=o_p(1),
\qquad i\in I_l.
\]
The average $n_l^{-1}\sum_{i\in I_l}\Delta_i^2$ has conditional expectation
equal to this population mean square error.
Conditional Markov's inequality therefore yields
\[
\frac1{n_l}\sum_{i\in I_l}\Delta_i^2=o_p(1).
\]
Because the number of folds is fixed and their proportions are bounded away
from zero, combining the bounds for each fold gives
\begin{equation}
\frac1n\sum_{i=1}^n\Delta_i^2=o_p(1).
\label{eq:variance_delta_l2}
\end{equation}

For each observation, the fitted score is affine in $\theta$. In particular,
\[
\hat\psi_i(\hat\theta)
=
\tilde\psi_i+(\hat\theta-\theta_0)\hat q_i,
\]
where, writing $l=l(i)$,
\[
\hat q_i
=
-2T_i\hat\gamma_l(X_i,Z_i)
+\hat\gamma_{1l}(X_i,Z_i)^2
-\hat\gamma_{2l}(X_i)^2.
\]
Under Assumption~\ref{ass3}, bounded fitted propensities imply
$n^{-1}\sum_i\hat q_i^2=O_p(1)$. Under Assumption~\ref{ass3prime}, the same
conclusion follows from their uniformly bounded fourth moments. Hence
\begin{align}
&\frac1n\sum_{i=1}^n
\{\hat\psi_i(\hat\theta)-\psi_i^0\}^2 \notag\\
&\qquad\leq
\frac2n\sum_{i=1}^n\Delta_i^2
+2(\hat\theta-\theta_0)^2
\frac1n\sum_{i=1}^n\hat q_i^2
=o_p(1).
\label{eq:variance_score_l2}
\end{align}

Let $e_i=\hat\psi_i(\hat\theta)-\psi_i^0$. The ordinary law of large
numbers implies
\[
\frac1n\sum_{i=1}^n(\psi_i^0)^2
\overset{p}{\longrightarrow}\Omega_{\mathrm{iid}},
\qquad
\frac1n\sum_{i=1}^n\psi_i^0\overset{p}{\longrightarrow}0.
\]
By (\ref{eq:variance_score_l2}) and Cauchy--Schwarz,
\[
\left|\frac1n\sum_{i=1}^n e_i\right|
\leq
\left(\frac1n\sum_{i=1}^n e_i^2\right)^{1/2}
=o_p(1),
\]
and
\begin{align*}
&\left|
\frac1n\sum_{i=1}^n\hat\psi_i(\hat\theta)^2
-\frac1n\sum_{i=1}^n(\psi_i^0)^2
\right|\\
&\quad\leq
2\left(\frac1n\sum_{i=1}^n(\psi_i^0)^2\right)^{1/2}
\left(\frac1n\sum_{i=1}^n e_i^2\right)^{1/2}
+\frac1n\sum_{i=1}^n e_i^2
=o_p(1).
\end{align*}
Thus the fitted score mean converges to zero and the fitted second moment
converges to the second moment of the oracle score. Subtracting the squared
sample mean proves
$\widehat\Omega_{\mathrm{iid}}\overset{p}{\to}\Omega_{\mathrm{iid}}$. Finally,
$\widehat Q\overset{p}{\to}Q$ by Lemma~\ref{lem:Qhat}, so
$\widehat V_{\mathrm{iid}}\overset{p}{\to}
Q^{-2}\Omega_{\mathrm{iid}}=V_{\mathrm{iid}}$. \qed

\subsection{Proof of Theorem~\ref{thm:tri}}

The proof follows the argument for Theorem~\ref{thm1}, with all population
quantities evaluated in row $n$ and a Lindeberg theorem replacing the i.i.d.\
central limit theorem. For an independently randomized
partition, condition first on its realization. Let $l(i)$ denote the fold
containing
observation $i$, and write $\|\cdot\|_n$ for the $L^2$ norm under the $n$-th
design. All expectations and probabilities below are taken under the
$n$-th design, and the bounds hold uniformly over the fixed set of folds. For
each fold $l$, define
\[
\delta_{\gamma j,n,l}=\hat\gamma_{jl}-\gamma_{jn},
\qquad
\delta_{m j,n,l}=\hat m_{jl}^Y-m_{jn}^Y,
\qquad j=1,2,
\]
and let $\delta_{\alpha j,n,l}=\hat\alpha_{jl}(\theta_n)-\alpha_{jn}(\theta_n)$ denote the representer errors evaluated at $\theta_n$. Assumption~\ref{ass:tri} implies $|\theta_n|=O(1)$.

The identities $\alpha_{1n}(\theta_n)=m_{1n}^Y-\theta_n\gamma_{1n}$ and $\alpha_{2n}(\theta_n)=-m_{2n}^Y+\theta_n\gamma_{2n}$ imply that the representer errors decompose as
\[
\delta_{\alpha 1,n,l}=\delta_{m 1,n,l}-\theta_n\delta_{\gamma 1,n,l},
\qquad
\delta_{\alpha 2,n,l}=-\delta_{m 2,n,l}+\theta_n\delta_{\gamma 2,n,l}.
\]
Hence $\|\delta_{\alpha j,n,l}\|_n\lesssim r_{m,n}+r_{g,n}$ uniformly in $j,l$, and Assumption~\ref{ass:tri}(i) yields $\|\delta_{\alpha j,n,l}\|_n\xrightarrow{p}0$.

\paragraph{Score decomposition.}
Write $\mu_{0n,i}=Y_{ni}-\theta_n T_{ni}$, $\alpha_{0j,n,i}=\alpha_{jn}(\theta_n)$ evaluated at observation $i$, and $\gamma_{0j,n,i}=\gamma_{jn}$ evaluated at observation $i$. For $i\in I_l$, the algebra of Lemma~\ref{lem:mr} applied at $\theta_n$ under the $n$-th design gives
\[
\begin{aligned}
&\psi_n(W_{ni};\theta_n,\hat\gamma_l,\hat\alpha_l(\theta_n))
-\psi_n(W_{ni};\theta_n,\gamma_n,\alpha_n(\theta_n))\\
&\quad=
(\mu_{0n,i}-\alpha_{01,n,i})\delta_{\gamma 1,n,l,i}
+(-\mu_{0n,i}-\alpha_{02,n,i})\delta_{\gamma 2,n,l,i}\\
&\qquad+\delta_{\alpha 1,n,l,i}(T_{ni}-\gamma_{01,n,i})
+\delta_{\alpha 2,n,l,i}(T_{ni}-\gamma_{02,n,i})\\
&\qquad-\delta_{\alpha 1,n,l,i}\delta_{\gamma 1,n,l,i}
-\delta_{\alpha 2,n,l,i}\delta_{\gamma 2,n,l,i}.
\end{aligned}
\]
The first four terms are linear in one nuisance error at a time, and the last two are bilinear interaction terms.

\paragraph{Conditional expectation of the score error.}
For each fold $l$, the nuisance estimates
$(\hat\gamma_l,\hat\alpha_l(\theta_n))$ are trained on observations outside
$I_l$, so they are nonrandom when conditioning on the training sample for fold
$l$. For the first linear term,
\[
\mathbb{E}_n\!\left[(\mu_{0n}-\alpha_{01,n})\delta_{\gamma 1,n,l}\,\bigl|\,\text{training}\right]
=
\mathbb{E}_n\!\left[\delta_{\gamma 1,n,l}(X,Z)\,\mathbb{E}_n[\mu_{0n}-\alpha_{01,n}(X,Z)\mid X,Z]\right]
=0,
\]
because $\alpha_{01,n}(X,Z)=\mathbb{E}_n[\mu_{0n}\mid X,Z]$. By iterated expectations, the remaining three linear terms also have zero
conditional expectation, since
$\alpha_{02,n}(X)=-\mathbb{E}_n[\mu_{0n}\mid X]$,
$\gamma_{01,n}(X,Z)=\mathbb{E}_n[T\mid X,Z]$, and
$\gamma_{02,n}(X)=\mathbb{E}_n[T\mid X]$.

For each $j\in\{1,2\}$, the conditional expectation of the bilinear term satisfies
\[
\left|\mathbb{E}_n\!\left[\delta_{\alpha j,n,l}\delta_{\gamma j,n,l}\,\bigl|\,\text{training}\right]\right|
\le
\|\delta_{\alpha j,n,l}\|_n\,\|\delta_{\gamma j,n,l}\|_n
=O_p\!\bigl(r_{g,n}(r_{g,n}+r_{m,n})\bigr)
=o_p(n^{-1/2})
\]
by Cauchy--Schwarz and Assumption~\ref{ass:tri}(ii). Hence the conditional
expectation of the complete score error is negligible at the $\sqrt n$ scale.

\paragraph{Empirical fluctuation.}
Let
\[
\Delta_{n,l}(W)
=
\psi_n(W;\theta_n,\hat\gamma_l,\hat\alpha_l(\theta_n))
-\psi_n(W;\theta_n,\gamma_n,\alpha_n(\theta_n)).
\]
The uniform mean square continuity condition and
Assumption~\ref{ass:tri}(i) imply
\[
\max_{l\leq L}
\mathbb{E}_n\!\left[
\Delta_{n,l}(W)^2\mid\text{training}
\right]
=o_p(1).
\]
The mean square condition controls the four linear terms and the centered bilinear terms
jointly, without requiring the outcome residual to be uniformly bounded. Under
the uniform fourth moment alternative, the conclusion also follows directly
from
\[
\begin{aligned}
\mathbb{E}_n\!\left[
(\mu_{0n}-\alpha_{01,n})^2\delta_{\gamma 1,n,l}^2
\mid\text{training}\right]
&\leq
\|\mu_{0n}-\alpha_{01,n}\|_{n,4}^2
\|\delta_{\gamma 1,n,l}\|_{n,4}^2,\\
\mathbb{E}_n\!\left[
\delta_{\alpha 1,n,l}^2(T-\gamma_{01,n})^2
\mid\text{training}\right]
&\leq
\|\delta_{\alpha 1,n,l}\|_n^2,
\end{aligned}
\]
with analogous bounds for the second component and a Cauchy--Schwarz bound that
uses fourth moments for each bilinear term.

Conditional on the training sample, the observations in $I_l$ are i.i.d. Hence
\[
\operatorname{Var}_n\!\left[
\mathbb{E}_{n,l}\Delta_{n,l}(W)\mid\text{training}
\right]
\leq
\frac{1}{n_l}
\mathbb{E}_n\!\left[
\Delta_{n,l}(W)^2\mid\text{training}
\right]
=o_p(n_l^{-1}).
\]
Conditional Chebyshev's inequality makes the centered average over $I_l$
$o_p(n^{-1/2})$, uniformly over the fixed number of folds. The conditional expectation
of the uncentered average score error equals the sum of the expectations of
the two bilinear terms and is
$o_p(n^{-1/2})$ by Assumption~\ref{ass:tri}(ii). Each displayed component, and hence their sum, is negligible at the $\sqrt n$
scale.
\paragraph{Asymptotically linear expansion.}
Combining the four linear and two bilinear terms with the oracle score,
\begin{equation}\label{eq:exp_tri}
\sqrt n\,\hat\psi(\theta_n)
=
\frac{1}{\sqrt n}\sum_{i=1}^n\psi_n(W_{ni};\theta_n,\gamma_n,\alpha_n)+o_p(1).
\end{equation}

\paragraph{Affine identity and Jacobian.}
For the estimator in closed form, $\hat\psi(\theta)=\hat A+\theta\widehat Q$, where
\[
\widehat Q
=
\frac{1}{n}\sum_{l=1}^L\sum_{i\in I_l}
\left[
-2T_{ni}\hat\gamma_l(X_{ni},Z_{ni})
+\hat\gamma_{1l}(X_{ni},Z_{ni})^2
-\hat\gamma_{2l}(X_{ni})^2
\right].
\]
In row $n$, the population Jacobian is
\[
Q_n^\partial=\mathbb{E}_n\!\left[-2T\gamma_n+\gamma_{1n}^2-\gamma_{2n}^2\right]=-\mathbb{E}_n[\gamma_n^2]=-D_n,
\]
where the second equality follows from iterated expectations and $\mathbb{E}_n[\gamma_n\mid X]=0$. By Assumption~\ref{ass:tri}(i), $\widehat Q-Q_n^\partial=o_p(1)$, so $\widehat Q\xrightarrow{p}-D_n$ along the array, and by Assumption~\ref{ass:tri}(iii), $|\widehat Q|\ge c/2$ with probability approaching one.

Since $\hat\theta$ solves the estimating equation, $0=\hat\psi(\hat\theta)=\hat\psi(\theta_n)+\widehat Q(\hat\theta-\theta_n)$, and rearranging gives the identity
\[
\sqrt n(\hat\theta-\theta_n)=-\widehat Q^{-1}\sqrt n\,\hat\psi(\theta_n)
=
D_n^{-1}\sqrt n\,\hat\psi(\theta_n)+o_p(1).
\]

\paragraph{Central limit theorem for the triangular array.}
By Assumption~\ref{ass:tri}(iv), the oracle score satisfies a Lindeberg CLT under the $n$-th design:
\[
\frac{1}{\sqrt n}\sum_{i=1}^n\psi_n(W_{ni};\theta_n,\gamma_n,\alpha_n)/\sqrt{\Omega_n}
\Rightarrow N(0,1),
\]
where $\Omega_n=\mathrm{Var}_n[\psi_n(W;\theta_n,\gamma_n,\alpha_n)]$, with $\Omega_n\to\Omega_\infty\in(0,\infty)$. Combining with the expansion in (\ref{eq:exp_tri}),
\[
\frac{\sqrt n(\hat\theta-\theta_n)}{\sqrt{\Omega_n/D_n^2}}
=
\frac{D_n^{-1}\sqrt n\,\hat\psi(\theta_n)+o_p(1)}{\sqrt{\Omega_n/D_n^2}}
=
\frac{n^{-1/2}\sum_{i=1}^n\psi_n(W_{ni};\theta_n,\gamma_n,\alpha_n)}{\sqrt{\Omega_n}}+o_p(1)
\Rightarrow N(0,1),
\]
using $\liminf D_n>0$ by Assumption~\ref{ass:tri}(iii) to control the studentization. \qed

\subsection{Proof of Corollary~\ref{cor:series}}
Under the conditions of Corollary~\ref{cor:series}, the standard series
prediction bound applies uniformly across the fixed number of
cross-fitting folds. Thus, for every regression and every fold
$l$,
\[
\|\hat r_l-r_{0n}\|_n
=
O_p\!\left(\sqrt{\frac{K_n}{n}}+b_{K_n}\right).
\]
Because the four regressions obey the same bound, the nuisance rates in
Assumption~\ref{ass:tri} satisfy
\[
r_{g,n}
=
O_p(a_n),
\qquad
r_{m,n}
=
O_p(a_n),
\qquad
a_n=\sqrt{\frac{K_n}{n}}+b_{K_n}.
\]
The rate conditions in the corollary imply $K_n/n=o(n^{-1/2})$ and
$b_{K_n}^2=o(n^{-1/2})$. The cross term has the same order or smaller because
\[
2b_{K_n}\sqrt{\frac{K_n}{n}}
\le
b_{K_n}^2+\frac{K_n}{n}.
\]
Consequently,
\[
a_n^2
=
\frac{K_n}{n}
+2b_{K_n}\sqrt{\frac{K_n}{n}}
+b_{K_n}^2
=o(n^{-1/2}),
\]
and therefore
\[
r_{g,n}\{r_{g,n}+r_{m,n}\}
=O_p(a_n^2)
=o_p(n^{-1/2}).
\]
Hence Assumption~\ref{ass:tri}(ii) holds. The same conditions also imply
$a_n=o(1)$, which ensures the mean square consistency required by
Assumption~\ref{ass:tri}(i).

Finally, an unrestricted series containing one indicator for each examiner has
at least $J_n$ basis functions. Thus $K_n\ge J_n$, and
$K_n=o(\sqrt n)$ implies $J_n=o(\sqrt n)$ for the estimator analyzed in the
corollary. Pooling, shrinkage, and grouping define different estimators, whose
rates require separate arguments. \qed

\section{Primitive Conditions for Assumptions~\ref{ass1} and~\ref{ass2}}
\label{app:primitive}

This section derives sufficient conditions for Assumptions~\ref{ass1} and
\ref{ass2} in terms of the four regression functions. The first
proposition reduces the assumptions to bounds on the four regression errors.

Let
\[
m_{01}(X,Z)=\mathbb{E}[Y\mid X,Z], \qquad m_{02}(X)=\mathbb{E}[Y\mid X],
\]
\[
g_{01}(X,Z)=\mathbb{E}[T\mid X,Z], \qquad g_{02}(X)=\mathbb{E}[T\mid X].
\]
By Equations~(\ref{eq:alpha_closed}) and~(\ref{eq:alpha_feasible_closed}),
\[
\alpha_{01}(X,Z;\theta_0)=m_{01}(X,Z)-\theta_0 g_{01}(X,Z),
\qquad
\alpha_{02}(X;\theta_0)=-m_{02}(X)+\theta_0 g_{02}(X).
\]
For each fold $l$, the nuisance estimates are
\[
\hat{\alpha}_{1l}(x,z;\theta)=\hat{m}_{1l}(x,z)-\theta \hat{g}_{1l}(x,z),
\qquad
\hat{\alpha}_{2l}(x;\theta)=-\hat{m}_{2l}(x)+\theta \hat{g}_{2l}(x).
\]
Here $\hat{m}_{1l}$ and $\hat{m}_{2l}$ estimate the two outcome regressions,
and $\hat{g}_{1l}$ and $\hat{g}_{2l}$ estimate the two treatment regressions.

\begin{proposition}[Reduction to rates for conditional expectations]
\label{prop:primitive_reduction}
Suppose
\[
\|\hat{m}_{1l}-m_{01}\|+\|\hat{m}_{2l}-m_{02}\|+\|\hat{g}_{1l}-g_{01}\|+\|\hat{g}_{2l}-g_{02}\|\xrightarrow{p}0.
\]
Then Assumption~\ref{ass1} holds for the representer estimates above.
Moreover,
\[
\left\|\hat{\alpha}_l(\theta_0)-\alpha_0(\theta_0)\right\|
\lesssim
\sum_{j=1}^2 \left(
\|\hat{m}_{jl}-m_{0j}\|+\|\hat{g}_{jl}-g_{0j}\|
\right),
\]
where $m_{0j}$ and $g_{0j}$ denote the true conditional expectations. Assumption~\ref{ass2} therefore follows if
\[
\left[
\sum_{j=1}^2
\left(
\|\hat{m}_{jl}-m_{0j}\|
+
\|\hat{g}_{jl}-g_{0j}\|
\right)
\right]
\left(
\|\hat{g}_{1l}-g_{01}\|+\|\hat{g}_{2l}-g_{02}\|
\right)
=o_p(n^{-1/2}).
\]
In particular, if all four regressions converge at rate $o_p(n^{-1/4})$ in
$L^2$, then Assumptions~\ref{ass1} and~\ref{ass2} are satisfied.
\end{proposition}

\begin{proof}
At the true parameter, the first representer error is
\[
\hat{\alpha}_{1l}(\theta_0)-\alpha_{01}(\theta_0)
=
(\hat{m}_{1l}-m_{01})-\theta_0(\hat{g}_{1l}-g_{01}).
\]
The triangle inequality therefore gives
\[
\|\hat{\alpha}_{1l}(\theta_0)-\alpha_{01}(\theta_0)\|
\le
\|\hat{m}_{1l}-m_{01}\|
{}+|\theta_0|
\|\hat{g}_{1l}-g_{01}\|
\xrightarrow{p}0.
\]
For the second representer,
\[
\hat{\alpha}_{2l}(\theta_0)-\alpha_{02}(\theta_0)
=
-\{\hat{m}_{2l}-m_{02}\}
+\theta_0\{\hat{g}_{2l}-g_{02}\},
\]
and hence
\[
\|\hat{\alpha}_{2l}(\theta_0)-\alpha_{02}(\theta_0)\|
\le
\|\hat{m}_{2l}-m_{02}\|
{}+
|\theta_0|\|\hat{g}_{2l}-g_{02}\|.
\]
The maintained convergence of the four estimates of conditional expectations implies
Assumption~\ref{ass1}. Summing the two bounds proves the displayed inequality for
$\|\hat{\alpha}_l(\theta_0)-\alpha_0(\theta_0)\|$. In addition,
\[
\|\hat{\boldsymbol\gamma}_l-\boldsymbol\gamma_0\|
\le
\|\hat{g}_{1l}-g_{01}\|+\|\hat{g}_{2l}-g_{02}\|
\]
by another application of the triangle inequality. Multiplying the two bounds
establishes the displayed sufficient condition for Assumption~\ref{ass2}. If each of
the four regression errors is $o_p(n^{-1/4})$, the first factor and the second
factor are both $o_p(n^{-1/4})$, so their product is $o_p(n^{-1/2})$.
\end{proof}

\paragraph{Ridge sieves.}
Consider first ridge regression with every coefficient penalized. For each of the
four conditional expectations
$r_{0j}\in\{m_{01},m_{02},g_{01},g_{02}\}$, suppose there is an approximation
$r_{K_j}=b_{K_j}'\beta_{K_j}$ in a $K_j$-dimensional sieve space with error
$a_{n,j}=\|r_{K_j}-r_{0j}\|$. Also suppose the basis functions have uniformly
bounded second moments and the Gram matrix of $b_{K_j}$ has eigenvalues bounded
away from zero and infinity. Under these standard conditions, ridge least
squares satisfies
\[
\|\hat{r}_{jl}-r_{0j}\|
=
O_p\!\left(a_{n,j}+\sqrt{K_j/n}+\lambda_{n,j}\|\beta_{K_j}\|_2\right)
\]
for each regression (see \citet{CHEN20075549}). Here $\lambda_{n,j}$ is the
penalty in the normalized criterion
\[
\frac{1}{n_l^{\mathrm{tr}}}
\sum_{i\in I_l^c}\{R_i-b_{K_j}(V_i)'\beta\}^2
+\lambda_{n,j}\|\beta\|_2^2,
\qquad n_l^{\mathrm{tr}}=|I_l^c|.
\]
If the objective instead uses the unnormalized residual sum of squares plus
$\lambda_{\mathrm{sim},j}\|\beta\|_2^2$, then
$\lambda_{n,j}=\lambda_{\mathrm{sim},j}/n_l^{\mathrm{tr}}$.
To apply Proposition~\ref{prop:primitive_reduction}, require all four
regression errors to vanish and the sum of those errors, multiplied by the
sum of the two treatment regression errors, to be $o_p(n^{-1/2})$.
These conditions imply Assumptions~\ref{ass1} and~\ref{ass2}. With
balanced rates and negligible ridge bias, a convenient sufficient condition is
\[
a_{n,j}+\sqrt{K_j/n}=o_p(n^{-1/4})
\qquad
\text{for each } r_{0j}\in\{m_{01},m_{02},g_{01},g_{02}\}.
\]

I derive the bound for partially penalized ridge in the following
proposition by first residualizing the penalized regressors with respect
to the unpenalized block.

\begin{proposition}[Partially penalized ridge]
\label{prop:partial_ridge_rate}
For one of the four regressions, write
\[
R=r_0(V)+\varepsilon,
\qquad
\mathbb E[\varepsilon\mid V]=0,
\qquad
\mathbb E[\varepsilon^2\mid V]\leq C.
\]
On a training sample of $N\asymp n$ independent observations, let $U$ contain $q_n$
unpenalized regressors and let $H$ contain $d_n$ penalized regressors. Suppose
there are coefficient vectors $(\delta_n,\beta_n)$ such that
\[
r_{K_n}(v)=u(v)'\delta_n+h(v)'\beta_n,
\qquad
\|r_{K_n}-r_0\|=a_n.
\]
Suppose that $U$ has full column rank, and let $P_U=U(U'U)^{-1}U'$ and
$M_U=I-P_U$. Suppose that, with probability approaching one,
\[
c\leq\lambda_{\min}(H'M_UH/N),
\qquad
\lambda_{\max}(H'M_UH/N)\leq C.
\]
Assume also that the population norm is bounded by the sample norm on the sieve
space: with probability approaching one,
\[
\|u'\delta+h'\beta\|
\leq C\|U\delta+H\beta\|_N
\]
for every $(\delta,\beta)$, where
$\|v\|_N^2=N^{-1}\sum_{i=1}^Nv_i^2$. Finally, suppose that
$q_n+d_n=o(N)$. If $(\widehat\delta,\widehat\beta)$ minimizes
\[
\frac{1}{N}\|R-U\delta-H\beta\|_2^2+\lambda_n\|\beta\|_2^2,
\]
then
\[
\|u'\widehat\delta+h'\widehat\beta-r_0\|
=
O_p\!\left(
a_n+\sqrt{\frac{q_n+d_n}{n}}
+\lambda_n\|\beta_n\|_2
\right).
\]
The same bound holds for a data-dependent penalty, with its contribution
$\lambda_n\|\beta_n\|_2$ included in the rate.
\end{proposition}

\begin{proof}
Let $a_i=r_0(V_i)-r_{K_n}(V_i)$ on the training sample, and collect these
approximation errors in the vector $a$. Write
$\Delta_\beta=\widehat\beta-\beta_n$ and
$\Delta_\delta=\widehat\delta-\delta_n$. The first-order condition for the
unpenalized coefficients implies
\[
\widehat\delta=(U'U)^{-1}U'(R-H\widehat\beta).
\]
Substituting this expression into the first-order condition for
$\widehat\beta$ yields
\[
\left(H'M_UH/N+\lambda_n I\right)\Delta_\beta
=
H'M_U(a+\varepsilon)/N-\lambda_n\beta_n.
\]
The lower eigenvalue bound therefore implies
\begin{equation}
\label{eq:partial_ridge_beta_bound}
\|\Delta_\beta\|_2
\leq
c^{-1}\left\{
\|H'M_Ua/N\|_2
+\|H'M_U\varepsilon/N\|_2
+\lambda_n\|\beta_n\|_2
\right\}.
\end{equation}

By Cauchy--Schwarz and the upper eigenvalue bound,
\[
\|H'M_Ua/N\|_2
\leq
\lambda_{\max}(H'M_UH/N)^{1/2}\|M_Ua\|_N
\leq C\|a\|_N.
\]
Because $\mathbb E\|a\|_N^2=a_n^2$, Markov's inequality gives
$\|a\|_N=O_p(a_n)$. Conditional on the regressors, independence and the bound
on $\mathbb E[\varepsilon_i^2\mid V_i]$ give
\[
\begin{aligned}
\mathbb E\!\left[
\|H'M_U\varepsilon/N\|_2^2\mid U,H
\right]
&\leq
\frac{C}{N^2}\operatorname{tr}(H'M_UH)\\
&=O_p(d_n/N).
\end{aligned}
\]
It follows from Markov's inequality that
$\|H'M_U\varepsilon/N\|_2=O_p(\sqrt{d_n/N})$. Equation
(\ref{eq:partial_ridge_beta_bound}) now bounds the coefficient error:
\[
\|\Delta_\beta\|_2
=O_p\!\left(
a_n+\sqrt{d_n/N}+\lambda_n\|\beta_n\|_2
\right).
\]

The same first-order condition implies the identity
\[
U\Delta_\delta+H\Delta_\beta
=P_U(a+\varepsilon)+M_UH\Delta_\beta.
\]
The approximation term satisfies $\|P_Ua\|_N\leq\|a\|_N=O_p(a_n)$.
Moreover, conditional on the regressors,
\[
\mathbb E[\|P_U\varepsilon\|_N^2\mid U,H]
\leq \frac{C}{N}\operatorname{rank}(P_U)
=\frac{Cq_n}{N}.
\]
Thus $\|P_U\varepsilon\|_N=O_p(\sqrt{q_n/N})$. The upper eigenvalue bound
on $H'M_UH/N$ and the rate for $\Delta_\beta$ then give
\[
\|U\Delta_\delta+H\Delta_\beta\|_N
=O_p\!\left(
a_n+\sqrt{\frac{q_n+d_n}{N}}
+\lambda_n\|\beta_n\|_2
\right).
\]
The assumed bound on the population norm therefore implies the same rate for
$u'\Delta_\delta+h'\Delta_\beta$ in the population $L^2$ norm. Adding
$\|r_{K_n}-r_0\|=a_n$ and using $N\asymp n$ proves the result.
\end{proof}

Applying Proposition~\ref{prop:partial_ridge_rate} to each of the four
regressions and then using Proposition~\ref{prop:primitive_reduction} verifies
the product condition in Assumption~\ref{ass2}. With comparable rates,
convenient sufficient conditions are
\[
q_n+d_n=o(\sqrt n),
\qquad
a_n=o(n^{-1/4}),
\qquad
\lambda_n\|\beta_n\|_2=o(n^{-1/4}).
\]
The eigenvalue bound for $H'M_UH/N$ rules out collinearity among the penalized
regressors after projecting out $U$. When $U$ contains examiner indicators,
the bound comparing the population and sample norms also requires adequate
support for each examiner, and the number of examiners contributes to $q_n$.
Partially penalized ridge does not impose sparsity, so its rate depends on the
full dimension $d_n$ of the penalized block.

\paragraph{LASSO under approximate sparsity.}
Suppose each function
$r_{0j}\in\{m_{01},m_{02},g_{01},g_{02}\}$ has an approximately sparse
representation in a basis of size $p_j$, with effective sparsity $s_j$ and
approximation error $a_{n,j}$. Under standard restricted eigenvalue conditions,
LASSO or post-LASSO estimators satisfy
\[
\|\hat{r}_{jl}-r_{0j}\|
=
O_p\!\left(a_{n,j}+\sqrt{s_j\log(p_j)/n}\right).
\]
see \citet{belloni2012sparse} and the references collected in
\citet{chernozhukov2022automatic}. Because the score has explicit representers,
no separate sparsity condition for $\alpha_1$ or $\alpha_2$ is needed:
Proposition~\ref{prop:primitive_reduction} converts these four convergence rates
directly into Assumptions~\ref{ass1} and~\ref{ass2}. A convenient sufficient
condition is
\[
a_{n,j}+\sqrt{s_j\log(p_j)/n}=o_p(n^{-1/4})
\qquad
\text{for each } r_{0j}\in\{m_{01},m_{02},g_{01},g_{02}\}.
\]

\paragraph{Other nonparametric estimators.}
Suppose estimators of the four conditional expectations attain $L^2$ rates
\[
r_{m1,n},\quad r_{m2,n},\quad r_{g1,n},\quad r_{g2,n},
\]
respectively. Assumptions~\ref{ass1} and~\ref{ass2} hold whenever
\[
\left(r_{m1,n}+r_{m2,n}+r_{g1,n}+r_{g2,n}\right)
\left(r_{g1,n}+r_{g2,n}\right)
=
o(n^{-1/2}).
\]
Estimators based on neural networks, random forests, boosting, and other
regularized nonparametric methods can attain these rates under smoothness,
complexity, and stability conditions appropriate to each method. For Assumptions~\ref{ass1} and~\ref{ass2}, only the four convergence rates
matter. Section 3.2.3 of
\citet{chernozhukov2024automatic} surveys a broad class of rate results.

\section{Saturated Linear Special Case}\label{app:saturated_example}

Under a correctly specified saturated linear first stage with finitely many values of
$(X,Z)$, UJIVE and the orthogonal estimator have the same estimand. The
example below, with two examiners and a binary covariate, also provides explicit expressions for the representers.

\paragraph{Finitely many covariate values.} Let $X$ take finitely many values
$x\in\mathcal X$, and let $Z\in\{1,\ldots,J\}$ index examiners. A saturated
first stage regresses $T$ on the complete set of cell indicators
$\{\mathbf 1\{X=x,Z=j\}\}_{x,j}$, while the saturated regression on covariates alone uses
$\{\mathbf 1\{X=x\}\}_x$. Their population least squares fitted values are the cell means,
\[
\mathbb{E}[T\mid X=x,Z=j]=\gamma_{01}(x,j),
\qquad
\mathbb{E}[T\mid X=x]=\gamma_{02}(x),
\]
so the residualized examiner propensity is the oracle instrument
$\gamma_0(x,j)=\gamma_{01}(x,j)-\gamma_{02}(x)$. Leaving an observation out
changes the finite-sample fitted values, but the long and short fits still
converge to $\mathbb E[T\mid X,Z]$ and $\mathbb E[T\mid X]$, respectively.
Thus, in the saturated correctly specified case, UJIVE and the orthogonal
estimator both estimate
\[
\theta_0=\frac{\mathbb{E}[Y\gamma_0(X,Z)]}{\mathbb{E}[T\gamma_0(X,Z)]}.
\]

More generally, Equation~(\ref{eq:projection_error_gap}) shows that the linear
UJIVE instrument equals the oracle instrument when the errors in its two
projections cancel.
Without this cancellation, the residualized UJIVE propensity can differ from
the oracle instrument, and its estimand can differ from the oracle ratio based
on conditional expectations. Saturation is one sufficient condition for equality. The orthogonal estimator
can also use nonlinear or unsaturated estimates of the conditional
expectations.

\paragraph{Two examiners and a binary covariate.} Let $X\in\{0,1\}$, let $Z\in\{0,1\}$ denote assignment to one of two examiners, and write
\[
p_{xz}=\mathbb{E}[T\mid X=x,Z=z],
\qquad
m_{xz}=\mathbb{E}[Y\mid X=x,Z=z],
\qquad
\pi_x=\Pr(Z=1\mid X=x).
\]
Then
\[
\gamma_{01}(x,z)=p_{xz},
\qquad
\gamma_{02}(x)=(1-\pi_x)p_{x0}+\pi_x p_{x1},
\]
so the residualized propensity is
\[
\gamma(x,1)=(1-\pi_x)(p_{x1}-p_{x0}),
\qquad
\gamma(x,0)=-\pi_x(p_{x1}-p_{x0}).
\]
Thus $\mathbb{E}[\gamma\mid X=x]=0$ and
\[
\mathbb{E}[T\gamma]=\mathbb{E}[\gamma^2]
=\sum_{x\in\{0,1\}}\Pr(X=x)\pi_x(1-\pi_x)(p_{x1}-p_{x0})^2.
\]
This quantity is positive whenever at least one value of $X$ has residual
first-stage variation across examiners.

The Riesz representers have the closed forms
\[
\alpha_{01}(x,z)=m_{xz}-\theta_0p_{xz},
\qquad
\alpha_{02}(x)=-\{(1-\pi_x)m_{x0}+\pi_x m_{x1}-\theta_0[(1-\pi_x)p_{x0}+\pi_x p_{x1}]\}.
\]
The negative sign in $\alpha_{02}$ reflects the subtraction of $\gamma_2$
from $\gamma_1$ in the residualized propensity.

In this model, a saturated first stage with an intercept, $X$, $Z$, and $XZ$
has population fitted value $p_{xz}=\mathbb{E}[T\mid X=x,Z=z]$. It reproduces
the general construction for finite $X$ and $Z$, and the two representers take
the forms displayed above.

\end{document}